\documentclass[12pt]{article}

\usepackage{amssymb,amsthm,undertilde}
\newcommand{\C}{{\mathbb C}}

\newcommand{\cD}{{\mathcal D}}
\newcommand{\cI}{{\mathcal I}}
\newcommand{\cM}{{\mathcal M}}

\newcommand{\SOC}{{\rm SO}(3,\C)}

\newcommand\be{\begin{eqnarray}}
\newcommand\ee{\end{eqnarray}}

\newcommand\im{{\rm i}}

\begin{document}

\vspace{1.0cm}

\begin{center}
{\Large\textbf{Effective metric Lagrangians from an underlying theory}}
\vskip 0.1cm
{\Large\textbf{with two propagating degrees of freedom}} 
\vspace{1.1cm}

\textbf{Kirill Krasnov}\footnote{kirill.krasnov@nottingham.ac.uk}  \\ \vspace{0.3cm}
\textit{School of Mathematical Sciences, University of Nottingham, Nottingham, NG7 2RD, UK}
\end{center}

\vspace{0.5cm}

\thispagestyle{empty}

\begin{center}
{\bf Abstract}
\end{center}

We describe an infinite-parametric class of effective metric Lagrangians that arise
from an underlying theory with two propagating degrees of
freedom. The Lagrangians start with the Einstein-Hilbert term, continue
with the standard $R^2, (Ricci)^2$ terms, and in the next order contain 
$(Riemann)^3$ as well as on-shell vanishing terms. This is exactly the structure
of the effective metric Lagrangian that renormalizes quantum gravity
divergences at two-loops. This shows that the theory underlying the effective field 
theory of gravity may have no more degrees of freedom than is already contained in general 
relativity. We show that the reason why an effective metric theory may describe just two 
propagating degrees of freedom is that there exists a (non-local) field redefinition that 
maps an infinitely complicated effective metric Lagrangian to the usual Einstein-Hilbert one. 
We describe this map for our class of theories and, in particular, exhibit it explicitly 
for the $(Riemann)^3$ term. 

\vspace{-0.2cm}

\newpage
\setcounter{page}{1}
\section{Introduction}

The modern effective field theory viewpoint on quantum field theory, see e.g. 
\cite{Weinberg:2009bg} for the most recent account, tells us that our cherished theories --
the Standard Model (SM) of elementary particles and General Relativity (GR) -- 
should be considered as only the first terms in an effective field theory Lagrangian 
containing all possible terms allowed by symmetries. 
This viewpoint "explains" why the Standard Model is 
renormalizable while GR is not. Indeed, the renormalizable part of an effective field 
theory Lagrangian of a given particle (and symmetry) content, if exists, is what 
describes these particles at low energies. There exists a renormalizable theory of the Standard 
Model constituents, and this is why it is the one manifesting itself as the correct theory at "low" 
energies of particle physics. However, there is no renormalizable quantum field theory of 
gravitons, as a renormalizable graviton interaction is prohibited by diffeomorphism
invariance, and so the term in the effective gravity Lagrangian that is of most 
significance at  low energies is necessarily the non-renormalizable Einstein-Hilbert one. This 
viewpoint also makes it clear that, after all terms allowed by symmetries are included in the 
Lagrangian, quoting Weinberg, "non-renormalizable theories
are just as renormalizable as renormalizable ones". Thus, within the framework of effective
field theory, gravity is renormalizable in an effective manner \cite{Weinberg:1978kz}.

From this perspective the question of quantum gravity can be reformulated as the question of what
is the theory underlying the effective field theory Lagrangian of gravity and the Standard
Model. Effective field theories (with their Lagrangians given by infinite series of all local terms
compatible with the symmetries) are easily produced from, say, renormalizable
ones (with a simple Lagrangian) by integrating out some "heavy" degrees of freedom. 
Alternatively, the field of an effective field theory may not even be present in the underlying
Lagrangian (be a composite field). Or, the underlying theory may not even be a field theory at all,
e.g. be a string theory. In all the listed possibilities there are more degrees of freedom (DOF)
in the underlying theory than in the effective field theory and, possibly, these underlying degrees 
of freedom are of a very different nature. It is a wide-spread belief 
(at least in the particle-physics community) that this is also the case with the
theory underlying the effective theory of Gravity plus the Standard Model -- this
theory should have more, and likely even different fundamental degrees of freedom than
those present in our effective field theory Lagrangian.

The purpose of this article is to point out that an alternative to this standard "more and different 
DOF" expectation may be possible. With this article being just one of the first steps of 
investigation in this direction we will certainly be unable to treat here both Gravity and the 
Standard Model, see, however, \cite{TorresGomez:2009gs}.
We shall instead concentrate on the example that is interesting by
itself -- that of pure gravity in four spacetime dimensions. 

We know that Einstein's GR is non-renormalizable and 
quantum effects require that new terms are added to the gravitational Lagrangian. 
At one-loop order these terms are the famous $R^2, R^{\mu\nu}R_{\mu\nu}$ 
counterterms \cite{'tHooft:1974bx}. They can be disposed off by a local redefinition
of the graviton field \cite{'tHooft:1974bx}. Since field redefinitions are going to play
an important role in the arguments of this paper let us briefly remind the reader how this
is done. Up to a topological term, the most general counterterm required at one loop is:
\be\label{one-loop}
\int d^4x\sqrt{-g} \left( aR^{\mu\nu}R_{\mu\nu}+bR^2\right).
\ee
This can be written as:
\be
\int d^4x \sqrt{-g} \left(R_{\mu\nu}-\frac{1}{2}g_{\mu\nu} R\right)\left( aR^{\mu\nu} -\frac{a+2b}{2}
g^{\mu\nu} R\right),
\ee
and so (\ref{one-loop}) can be removed by the following simple redefinition of the
graviton field:
\be\label{h-loc}
h_{\mu\nu}\to h_{\mu\nu}+ aR_{\mu\nu}- \frac{a+2b}{2}\eta_{\mu\nu} R.
\ee
Here $h_{\mu\nu}$ is the usual field describing a perturbation around Minkowski metric
$\eta_{\mu\nu}$, i.e., $g_{\mu\nu}=\eta_{\mu\nu}+h_{\mu\nu}$. The $R_{\mu\nu}, R$
terms in the field redefinition formula are understood as local (not containing any
inverse powers of derivatives) functions of the perturbation field $h_{\mu\nu}$,
and at this order of perturbation theory it is sufficient to keep only the linear in $h_{\mu\nu}$
terms in this formula.

At two-loop order the only term
$R_{\mu\nu}^{\quad\rho\sigma}R_{\rho\sigma}^{\quad\alpha\beta}R_{\alpha\beta}^{\quad\mu\nu}$
that cannot be disposed off by a (local) field redefinition is indeed
necessary as a counterterm, see \cite{Goroff:1985th}, which seems to remove the hope that
Einstein's GR may be an on-shell finite theory.  It is thus likely that all local terms that
are compatible with diffeomorphism invariance do arise as counterterms and
so one is in the realm of effective field theory.

Before we describe our proposal, let us note that there is one alternative to the effective 
field theory viewpoint on gravity that has been contemplated in the literature.  It 
has to do with the fact that an introduction of a higher power of the momentum 
in the propagator of the theory may make it renormalizable (in the usual sense of
a Lagrangian with only a finite number of terms being enough to absorb all the
arising divergences). This is the case with the Lagrangian \cite{Fradkin:1981hx}
quadratic in the curvature that introduces into the free graviton
action a fourth-derivative term.  If this is included in the propagator one gets a
renormalizable theory \cite{Fradkin:1981hx}, which is, however, non-unitary due
to the presence of new unphysical propagating modes (poles in the propagator). 
A more recent, but similar in spirit attempt is that of \cite{Horava:2009uw}, 
where a higher power of the momentum in
the propagator is introduced by explicitly breaking the Lorentz symmetry (at high 
energies). This proposal also turns out to introduce additional, 
not present in GR, propagating modes. These modes are strongly coupled at low-energies,
which prevents the theory to have Einstein's GR as its low-energy limit \cite{Charmousis:2009tc}.

Unlike the proposals just reviewed that make gravity renormalizable and thus remove
the need for its effective field theory interpretation (but introduce "bad" propagating
modes),  the scheme that we shall describe in this 
paper takes the conventional view on Einstein's GR as being the low-energy relevant
part of an (infinitely) complicated effective field theory Lagrangian. 
Our proposal is about the possible nature of the theory underlying this
effective Lagrangian. As we have already mentioned, it is commonly believed that
the underlying theory, at the very least, has some additional degrees of freedom on
top of those of the graviton. The purpose of this article is to point out that this
does not have to be so: what we know about the gravitational effective theory
is compatible with the possibility that the underlying theory may have just two propagating
degrees of freedom. 

More precisely, we exhibit an infinite-parameteric class of theories of metrics and
some additional fields. The theories are second-order in derivatives.  A
simple Hamiltonian analysis shows that all these theories contain just two propagating degrees
of freedom, so the additional fields are non-propagating. When the additional fields are 
integrated out  the resulting effective metric Lagrangian is the Einstein-Hilbert term plus an 
infinite set of invariants constructed from the curvature and its derivatives. 
We compute the effective Lagrangian up to terms of mass dimension six and verify that the term 
$R_{\mu\nu}^{\quad\rho\sigma}R_{\rho\sigma}^{\quad\alpha\beta}R_{\alpha\beta}^{\quad\mu\nu}$
that the two-loop analysis \cite{Goroff:1985th} requires as the counterterm 
is present in our effective metric theory. The coefficient in front of this term is a
certain combination of the (lowest-order) parameters that parametrize our theory. 
This shows that what we know about the
structure of divergences of quantum gravity is compatible with the possibility
that the underlying theory may have no more degrees of freedom than is
already present in Einstein's GR. The class of theories that we describe can thus be viewed
as the "minimal" possibility for what the underlying gravity theory may be. 

Before we describe our "underlying" theory, let us present one immediate, and 
rather interesting application. It follows from the fact  that our theory 
describes two propagating DOF and can reproduce
the $(Riemann)^3$ counterterm. Then, since our theory has only two propagating DOF,
and GR uniqueness theorems, see e.g. \cite{Hojman:1976vp}, 
tell us that the only such theory is GR, there should exist (in general non-local) field 
redefinition that maps our theory to general relativity. We shall indeed find
such a transformation below. Now, the $(Riemann)^3$ term of the effective gravity
Lagrangian cannot be removed by a local field redefinition, for it would then vanish
on-shell. However, the fact that it can be reproduced from a theory with two propagating
DOF tells us that it should be removable by a {\it non-local} field redefinition. This is indeed so,
as it is not hard to see that:
\be\label{two-loop}
a \int d^4x \, R_{\mu\nu}^{\quad\rho\sigma}R_{\rho\sigma}^{\quad\alpha\beta}
R_{\alpha\beta}^{\quad\mu\nu},
\ee
where the integrand is understood as a cubic expression in the graviton perturbation 
$h_{\mu\nu}$, can be written as:
\be
a \int d^4x \left( R^{\mu\nu} - \frac{1}{2}\eta^{\mu\nu} R\right) 
\frac{4}{\Box} \partial^\alpha \partial^\beta \left(
R_{\mu\alpha}^{\quad\gamma\delta} R_{\nu\beta\gamma\delta}
-\frac{1}{2} \eta_{\mu\nu}\eta^{\rho\sigma} 
R_{\rho\alpha}^{\quad\gamma\delta} R_{\sigma\beta\gamma\delta} \right),
\ee
where $\Box=\partial^\mu\partial_\mu$.
This is checked using the easily verifiable identity
\be\label{ddR-ident}
\partial_{[\alpha}\partial^{[\beta} R_{\mu]}^{\,\,\nu]} = \frac{1}{4} \Box R_{\alpha\mu}^{\quad\beta\nu}
\ee
that holds to first order in the perturbation field. This shows that
the field redefinition:
\be\label{h-non-loc}
h_{\mu\nu}\to h_{\mu\nu} + \frac{4a}{\Box} \partial^\alpha \partial^\beta \left( 
R_{\mu\alpha}^{\quad\gamma\delta} R_{\nu\beta\gamma\delta}
-\frac{1}{2} \eta_{\mu\nu}\eta^{\rho\sigma} 
R_{\rho\alpha}^{\quad\gamma\delta} R_{\sigma\beta\gamma\delta} \right),
\ee
where the term on the right-hand-side is viewed as being of second order in the
perturbation, removes the two-loop counterterm (\ref{two-loop}). We note that the
tensor in brackets is reminiscent of the Bel-Robinson tensor
$B_{\mu\nu\alpha\beta}=R^{\sigma\,\,\tau}_{\,\mu\,\,\,\alpha}R_{\sigma\nu\tau\beta}
+R^{\sigma\,\,\tau}_{\,\mu\,\,\,\beta}R_{\sigma\nu\tau\alpha}-(1/2)g_{\mu\nu}
R_{\alpha}^{\,\,\sigma\tau\gamma}R_{\beta\sigma\tau\gamma}$, but does not coincide with
it. An easy way to see the difference is to note that the $\alpha\beta$ trace of the tensor that 
appears in (\ref{h-non-loc}) is non-zero $-(1/4)\eta_{\mu\nu}R^{\alpha\beta\gamma\delta}
R_{\alpha\beta\gamma\delta}$, while the Bel-Robinson tensor is traceless.

The fact that (\ref{h-non-loc})
removes the $(Riemann)^3$ counterterm does not seem to have been noticed in the literature.
It is, of course, not surprising that a term that was not removable by a local field
redefinition can be removed by a non-local one. However, such a non-local transformation
typically introduces non-locality in the next, higher-order term of the resulting action. Indeed,
consider the massless free scalar with the 
Lagrangian $(1/2)(\partial_\mu \phi)^2$ and shift $\phi\to \phi+(1/\Box)\psi$, where $\psi$
is some function that depends on $\phi$ in a local way. It is obvious that the
action resulting from such a redefinition contains a non-local term $-(1/2)\psi(1/\Box)\psi$. 
Our non-local field redefinition (\ref{h-non-loc}) 
can thus be expected to introduce non-locality in higher-order terms. A non-trivial statement
then is that it is possible to complete the redefinition (\ref{h-non-loc}) by higher
powers of $1/\Box$ so that the action arising in the result of the redefinition is again local,
in the sense of not containing $1/\Box$.
The reason why such a non-local field redefinition mapping a local EH action to again a local
action must be possible at all orders is deeply related to a certain "topological symmetry" of gravity
that is not manifest in the usual metric description but reveals itself
in certain more exotic formulations such as that due to Pleba\'nski \cite{Plebanski:1977zz}.
We shall explain all this, as well as possible implications for the quantum
theory of gravity, in more details below.

Before we turn to details of our theories, let us explain why it is quite
non-trivial to have a class of metric theories with just two propagating degrees of freedom.
For this we remark that the most general effective gravity Lagrangian
containing all invariants (with arbitrary coefficients) constructed from the curvature and 
its derivatives describes more than two propagating degrees of freedom. This is well-illustrated
by e.g. the $f(R_{\mu\nu\rho\sigma})$ theories, where $f(\cdot)$ is an arbitrary
algebraic function, i.e. one that depends on the Riemann tensor but not its 
derivatives. Thus, this Lagrangian is obtained from the most general 
one by setting coefficients in front of all the derivative terms to zero. 
For a generic function $f(\cdot)$ these theories are known,
see \cite{Deruelle:2009zk},  to have six more propagating degrees of freedom in addition to 
the two present in GR. It is easy to see where the additional degrees of freedom
come from. Indeed, the action:
\be\label{f-Riemann}
S[g_{\mu\nu}]=2\int d^4x \sqrt{-g} \,\, f(R_{\mu\nu\rho\sigma}),
\ee
where we have set $32\pi G=1$,
can be rewritten in a second-derivative form by introducing auxiliary fields
$\phi_{\mu\nu\rho\sigma}$:
\be\label{S-f-phi}
S[g_{\mu\nu},\phi_{\mu\nu\rho\sigma}]=2
\int d^4x \sqrt{-g} \,\, \left( R^{\mu\nu\rho\sigma}\phi_{\mu\nu\rho\sigma} - 
\tilde{f}(\phi_{\mu\nu\rho\sigma})\right).
\ee
Here $\tilde{f}(\phi_{\mu\nu\rho\sigma}) := X^{\mu\nu\rho\sigma}\phi_{\mu\nu\rho\sigma}
-f(X_{\mu\nu\rho\sigma}), \phi^{\mu\nu\rho\sigma}=\partial f/\partial X_{\mu\nu\rho\sigma}$ 
is the Legendre transform of $f(\cdot)$. The auxiliary field $\phi_{\mu\nu\rho\sigma}$
has all the symmetries of the Riemann curvature. It is however clear that the first
term in this action contains time derivatives of the metric as well as of the auxiliary
fields, and so some of them are propagating. A careful Hamiltonian analysis
of \cite{Deruelle:2009zk} reveals that for a generic $f(\cdot)$ there are six
new modes. 

Because the class of theories that we are about to describe contains no new propagating
modes it cannot give rise to any given effective metric Lagrangian. Indeed, our 
theories are clearly unable to reproduce the $f(R_{\mu\nu\rho\sigma})$ theories 
with their additional DOF. To put it differently, our (infinite-parametric) underlying theory with its
two propagating DOF produces an effective metric theory. Even though all curvature
invariants are likely to be present in this effective theory, the coefficients in front of these
invariants are not completely arbitrary, as is illustrated by the fact that our
class of effective theories does not intersect with the $f(R_{\mu\nu\rho\sigma})$ class.
Thus, and this point is quite an important one, it is not guaranteed that the
class of theories that we shall describe is renormalizable in the sense of Weinberg
\cite{Weinberg:1978kz}, i.e. in the sense of being closed under renormalization.
We do present a reformulation of our class of theories that make it quite plausible
(to us at least) that this must be the case, but the issue of closeness of our theories under
renormalization remains open. 

A related remark is as follows. One might 
object that our claim about the existence of a large class of two propagating DOF metric
Lagrangians is trivial, since this is the property of {\it any} effective gravity Lagrangian.
Indeed, in effective field theory one is not concerned with the higher-derivative terms
of the effective Lagrangian being a source of extra propagating DOF. Such terms
are interpreted as interactions, and the counting of the propagating modes is done at
the level of the linearized action that is (typically) insensitive to higher-order terms.
However, since for the problem of quantum gravity we are interested in UV completions
of effective Lagrangians, we only understand the theory if we know the whole infinite 
series of higher-derivative terms, or, equivalently, a principle that produces such
an effective expansion. It is at this level that extra propagating DOF typically appear.
Indeed, they are those of some additional (heavy) field
that was integrated out, or of a collection of such fields. Thus, as far as we are
aware, in all known examples when the effective field theory is known completely,
i.e. with its underlying theory producing the expansion, there are more propagating
DOF in the underlying theory than is visible in the effective Lagrangian. This discussion
illustrates the non-triviality of our claim: Unlike other known examples, in the case under
consideration the underlying theory has the same number of propagating DOF as
is seen in the arising effective metric Lagrangians.

We can now turn to a description of the class of theories that is the main subject of this paper.
It is not new, and its history is briefly as follows. Gravitational theories with two
propagating DOF (distinct from GR) were first envisaged in \cite{Capovilla:1989ac}, 
\cite{Capovilla:1991kx}  by noting that the ``pure  connection'' formulation of GR described 
in these references admits a one-parameter family of 
deformations. This one-parameter family was studied by Capovilla \cite{Capovilla:1992ep} 
and by Bengtsson and Peld\'an \cite{Bengtsson:1990qh} under the
name of "neighbors of GR".  Later an infinite-parameter family of gravity theories all 
describing two propagating degrees of freedom was introduced in \cite{Bengtsson:1990qg}
and studied in a series of works 
\cite{Bengtsson:1991bq,Bengtsson:1991mg,Bengtsson:1992ds,Bengtsson:1995tx}. 
Unfortunately, in spite of providing an
infinite-parameter family of deformations of GR without changing the number of propagating
degrees of freedom, the class of theories \cite{Bengtsson:1990qg} never became widely known.
Partially, this is due to the fact that the pure connection formulation \cite{Capovilla:1991kx} 
of GR on which it was based is so far from the usual formulation in terms of spacetime metrics. 
Another problem with this class of theories was that they provided deformations
of complexified general relativity, and reality conditions that need to be imposed to
recover real Lorentzian metrics were never understood, see 
\cite{Bengtsson:1992dq,Bengtsson:1992cm}. 

The same class of theories was arrived at independently in \cite{Krasnov:2006du}, with
the author being unaware of the previous work on "neighbors of GR". This time
the starting point of the modification was the so-called Plebanski formulation of general
relativity \cite{Plebanski:1977zz}. The original paper \cite{Krasnov:2006du} obtained the class of
theories in question by studying the renormalization of GR in Plebanski
formulation. A somewhat simpler way to arrive at the same theory is
by considering what happens if one drops the so-called simplicity constraints of
Plebanski formulation; this has been described in \cite{Krasnov:2008fm}. The equivalence
of theories considered in \cite{Krasnov:2006du} to those proposed in \cite{Bengtsson:1990qg}
has been established in \cite{Bengtsson:2007zzd}. The fact that the theories in question 
have the same number of propagating degrees of freedom as GR has been established in 
\cite{Bengtsson:1990qg}, \cite{Bengtsson:2007zzd}, and in their Plebanski-like formulation in 
\cite{Krasnov:2007cq}. A metric interpretation of this class of theories has been given in
\cite{Krasnov:2008ui,Freidel:2008ku}.

As the reader may have already realized, in their easiest-to-state form the theories that
we shall consider in this paper are very remote from the usual general relativity with
its spacetime metric as the basic dynamical variable. In the simplest formulations 
the basic variable of the new theories is not a metric, and arriving at a metric formulation 
requires quite some work. A completely standard metric interpretation is nevertheless possible.
In order to make this paper accessible to as wide audience as possible, we start by
describing such a metric formulation, in spite of the fact that is not as elegant as one
might desire. Only after the basic idea of this class of deformations of GR is understood
in familiar terms do we give the most compact description.

In familiar metric terms, the basic idea of obtaining a metric Lagrangian with 
additional non-propagating scalars is to "correct" the Lagrangian in (\ref{S-f-phi}) so 
that it becomes degenerate, the field $\phi_{\mu\nu\rho\sigma}$ is a non-propagating 
auxiliary field and the theory contains exactly two
propagating modes. Integrating $\phi_{\mu\nu\rho\sigma}$ out will then produce an effective 
metric theory. To see how this might work, let us split:
\be\label{phi-g-H}
\phi_{\mu\nu\rho\sigma}= g_{\rho[\mu}g_{\nu]\sigma} + H_{\mu\nu\rho\sigma}
\ee
so that the $H_{\mu\nu\rho\sigma}=0$ theory is just GR with a cosmological constant. 
We then add to the Lagrangian a term of
the form $(D\phi)^2=(DH)^2$, where $D$ is some first order differential operator constructed
using the metric and the covariant derivative, to get:
\be\label{S-g-H}
S[g_{\mu\nu},H_{\mu\nu\rho\sigma}]=2
\int d^4x \sqrt{-g} \,\, \left( R+ R^{\mu\nu\rho\sigma}H_{\mu\nu\rho\sigma} +(DH)^2 - 
V(H_{\mu\nu\rho\sigma})\right),
\ee
where $V(\cdot)$ is just the function $\tilde{f}(\cdot)$ of the shifted argument (\ref{phi-g-H}).
One way to check whether this Lagrangian is degenerate is to linearize it
 about the Minkowski spacetime background $g_{\mu\nu}=\eta_{\mu\nu} + h_{\mu\nu}$.
Then its kinetic term takes the schematic form $(\partial h)^2 + (\partial^2 h) H + (\partial H)^2$. 
The field $H$ is non-propagating (at the linearized level) if the kinetic term here is
degenerate, i.e. if it can be written as $(\partial (h+ {\cal O}H))^2$, where $\cal O$ is some, 
possibly non-local, see below, operator acting on $H$. 

Let us see when this is possible. Our first remark is that it is quite easy to construct
degenerate Lagrangians. For example, taking the free massless field Lagrangian 
$(1/2)(\partial_{\mu}\phi)^2$ and shifting the field $\phi\to\phi+\psi$ we get a 
degenerate Lagrangian 
\be
\frac{1}{2} (\partial_{\mu}\phi)^2 + \partial_\mu \phi \partial^\mu \psi 
+ \frac{1}{2} (\partial_{\mu}\psi)^2.
\ee 
It is clear that this theory describes only one propagating field and that the second
field is an illusion. This example can be generalized to an arbitrary {\it local}
field redefinition $\phi\to\phi+{\cal O} \psi$, where $\cal O$ is a local operator,
i.e. not containing powers of $1/\Box$, where $\Box=\partial^\mu\partial_\mu$. For any such
shift the obtained theory of $\phi,\psi$ is degenerate, with only one propagating
field. At the same time, when one applies the shift $\phi\to\phi+{\cal O} \psi$ with
a non-local operator $\cal O$, one will (almost, see below) unavoidably get a "Lagrangian"
for $\phi,\psi$ that contains inverse powers of $1/\Box$. So, such non-local
transformations typically do not produce anything that can legitimately be called a $\phi,\psi$
Lagrangian.

As we shall see, however, in quite rare circumstances, applying the shift   
$\phi\to\phi+{\cal O} \psi$ with a non-local $\cal O$ of a special form, it turns out
to be possible that the $(1/2)\phi_{,\mu}\phi^{,\mu}$ Lagrangian goes into
another one for $\phi,\psi$ that is local. The underlying reason that makes it possible
is a certain "hidden" symmetry of GR, see below on this. 
In practice, for this to be possible the fields
$\phi,\psi$ cannot be scalars, and the operator $\cal O$ should have an overall
positive power of the derivatives, i.e. be of the form ${\cal O}\sim \partial^k/\Box^n$ with 
$k\geq n/2$. The simplest case is when ${\cal O}\sim \partial^2/\Box$. It is then possible,
but quite non-trivial, that the algebraic structure of the indices on the fields and the 
operator $\cal O$ is such that the resulting $\phi,\psi$ Lagrangian is still local,
and, moreover, contains only second derivatives. As we shall see in the next section,
in order for this to be possible the field $H_{\mu\nu\rho\sigma}$ must satisfy
a non-trivial algebraic conditions, and this leads to self-dual objects that are
going to play a very important role in this paper.

At the discussed linear level the non-local transformation envisaged, even though
non-trivial, still does not produce anything new - the obtained theory is still that
of a single propagating field. Things become interesting when this
non-local field redefinition idea is generalized into a map between two interacting
theories. Thus, assume that we have one theory with action $S_1[\tilde{\phi}]$, where
$\tilde{\phi}$ is some field or collection of fields, and another theory with action
$S_2[\phi,\psi]$. We assume that both actions are usual local ones, for example containing
not higher than second derivatives of the fields.  Let us now make a (non-trivial) 
assumption that there exists a non-local map $\tilde{\phi}=f(\phi,\psi)$ such that
\be
S_2[\phi,\psi]=S_1[f(\phi,\psi)].
\ee
If the transformation $f(\phi,\psi)$ were local then the theory $S_2$ would obviously
be the same as $S_1$. However, we have assumed that $f(\phi,\psi)$ is non-local.
Is it still the same theory? In order to not be comparing
apples and oranges, let us add to the action $S_2$ a potential term $V(\psi)$ for $\psi$ and
integrate this auxiliary field out. We can do it either classically, by solving its
field equations and substituting the result back into the action, or quantum mechanically,
determining the measure on the space of $\phi,\psi$ fields (i.e. computing the symplectic
form on the corresponding phase space and taking into account second-class constraints
if these are present), and then integrating over $\psi$. One gets either a classical
or quantum effective action $S^{eff}_2[\phi]$. Is the theory $S_2^{eff}$ same as $S_1$? 

The answer to this is not clear-cut. The two theories are certainly different as classical
metric theories, for a non-local field transformation is involved. Let us see this. The 
process of solving for $\psi$ gives
$\psi(\phi)$ that is some local map, in the sense that it does not contains negative
powers of derivatives (but possibly contains all positive powers). We now have
\be
S_2^{eff}[\phi]=S_2[\phi,\psi(\phi)]=S_1[f(\phi,\psi(\phi))].
\ee 
Thus, $S_2^{eff}$ is the same as $S_1$ but with a non-local 
(because of $f(\phi,\psi)$) field redefinition applied to its dynamical variable. 
Such non-local field redefinitions are forbidden in the classical theory, for they
alter content of the model. So, it is clear that these are different classical theories. 

However, the answer to the question posed above is not clear if one 
does a comparison of quantum theories. Then the
object two compare in each case is the graviton S-matrix. While in general this
is changed if a non-local field redefinition is applied, see more on this in the last
section, it is not impossible that our field redefinitions are special and that the
S-matrices of the two theories are the same. Indeed, what is most 
surprising about the field redefinitions involved is that $S_2^{eff}[\phi]$ obtained 
from $S_1[\tilde{\phi}]$ by a non-local map is still a local theory. This is by no means 
trivial and explains why we do not encounter such seemingly trivial constructions everywhere.
Thus, we do not know the answer to the question posed in the quantum case.
If the theories are quantum-equivalent, this leads to some interesting
prospects that are discussed in the last section. 

This is the scheme that is at play with our metric theories with two propagating DOF, which 
can be written in the form (\ref{S-g-H}), as we shall describe in details below.
We can now explain where the non-local transformation (\ref{h-non-loc}) comes
from. The theory (\ref{S-g-H}), having two propagating DOF, is obtained from GR by
a non-local (and quite non-trivial, see below) field redefinition. 
Once the auxiliary field $H_{\mu\nu\rho\sigma}$ is integrated out, this field
redefinition becomes a non-local transformation $\tilde{g}_{\mu\nu}=f(g_{\mu\nu})$ 
between an infinitely-parametric family of effective theories for the metric $g_{\mu\nu}$ 
and Einstein's GR for metric $\tilde{g}_{\mu\nu}$. The formula (\ref{h-non-loc}) is
just the transformation in question to lowest (second) order in the perturbation
when this map is non-local, while at first order the transformation
is a local one given by (\ref{h-loc}). This map to GR explains why all our effective metric
theories describe just two propagating DOF. At the same time, because of the non-local
nature of the map between the two theories, they are non-equivalent, at least
classically. This provides a new, and in our opinion interesting perspective on the question of 
what may be behind the effective metric theory which gravity seems to be. Moreover, and this is
an important point, our theories do not just exists as somewhat cumbersome constructs
in terms of metric and auxiliary fields -- they admit an elegant formulation that we now 
describe. As we have already said, it is seemingly quite remote from GR with its
metric as a basic variable. But we hope the reader will not be put off by an unfamiliar language.

We first give a pure connection formulation that is most compact.
Let $A^i, i=1,2,3$ be an $\SOC$ connection on the spacetime manifold $M$, and
$F^i=dA^i + (1/2)\epsilon^{ijk}A^j\wedge A^k$ be its curvature two-form. What is
a Lagrangian that can be written down for the dynamical variable $A^i$ if there is
no external metric that can be used? The simplest Lagrangian that comes to mind
is $F^i\wedge F^i$. This, however, is a surface term, and the resulting theory is known 
to be topological. Much more interesting Lagrangians can be constructed as follows.

Consider the 4-form valued matrix $F^i\wedge F^j$. This is a symmetric $3\times 3$
(complex) matrix-valued 4-form. Choosing an arbitrary volume form $(vol)$ we
can write the above matrix-valued 4-form as: $F^i\wedge F^j=\Omega^{ij} (vol)$,
where, of course, the $3\times 3$ matrix $\Omega^{ij}$ is only defined up
to rescalings: $(vol)\to \alpha (vol), \Omega^{ij}\to \alpha^{-1}\Omega^{ij}$.
Let us now introduce a (holomorphic) scalar valued function $f(\cdot)$ 
of a symmetric $3\times 3$ matrix that is in addition homogeneous of degree one in its variables:
$f(\alpha X^{ij}) = \alpha f(X^{ij})$.
The fact that this function is homogeneous degree one allows it to be 
applied to the 4-form $F^i\wedge F^j$ with the result being
again a 4-form. Indeed, $f(F^i\wedge F^j)=f((vol)\Omega^{ij})=(vol)f(\Omega^{ij})$,
and can therefore be integrated over the 4-manifold. Moreover, the homogeneity of
$f(\cdot)$ guarantees that it does not matter which background 4-form $(vol)$ is used.
With this definition of a function of a matrix-valued 4-form we can write our action as:
\be\label{action-A}
{\cal S}[A]=\int_M f(F^i\wedge F^j),
\ee
where $f(\cdot)$ is an arbitrary homogeneous order one gauge-invariant 
holomorphic function of its complex matrix-valued argument. Note that we have 
denoted the complex holomorphic action by $\cal S$ and the usual symbol $S$ is reserved 
for a real action. The action (\ref{action-A}) is not yet our class of theories, for it has to be
supplemented by certain reality conditions to give a real Lorentzian
signature gravity theory. However, as we shall see below, for a generic $f(\cdot)$ 
it does describe a class of deformations of complex general relativity in the sense that it describes 
two (complex) propagating DOF and contains GR. The clause about generic $f(\cdot)$
is important because the very special case $f(F^i\wedge F^j)\sim {\rm Tr}(F^i\wedge F^j)$ gives 
a topological theory without propagating DOF.  

The basic reason why the theory
(\ref{action-A}) describes two propagating modes is easy to see from the fact that
in its phase space is that of $\SOC$ Yang-Mills theory, i.e. is parametrized by
pairs $(A_a^i, \tilde{E}^{ai})$ where $A_a^i$ is the pull-back of the connection
$A^i$ on the spatial manifold and $\tilde{E}^{ai}$ is the conjugate momentum.
Thus, the configurational space is $3\times 3=9$-dimensional. However, the
theory is $\SOC$ and diffeomorphism-invariant, which means that there are
$3+4$ first-class constraints acting on the phase space, which reduce the
dimension of the physical configurational space down to $2$. We note that the count
given is the same as that for general relativity in terms of Ashtekar variables
\cite{Ashtekar:1987gu}.

The class of theories (\ref{action-A}) was first considered in \cite{Bengtsson:1990qg}.
This reference, however, gave a different, but equivalent formulation. Thus, 
\cite{Bengtsson:1990qg} contained, in addition to $A^i$, an extra field of density weight
minus one. The action written down in this reference is the most general one that can
be constructed from powers of $F^i\wedge F^j$ matrix-valued 4-form and the
additional density minus one field. However, the additional field is non-dynamical and
can be integrated out, with the resulting action being of the form (\ref{action-A}).

It is clear that (\ref{action-A}) is simply the most general action for a connection $A^i$ 
that can be written without any background structure such as metric. 
Such an action, to be gauge-invariant, 
can only be a functional of the curvature $F^i$ and its covariant derivatives, but the later
are zero by the Bianchi identity. Thus, it must be a function of the curvature. Then
the integrand of (\ref{action-A}) is simply the most general such function. Thus,
at least naively, the class of theories (\ref{action-A}) must be renormalizable in
the effective sense of Weinberg \cite{Weinberg:1978kz}, i.e. in the sense
that it is closed under renormalization. It is certainly non-trivial to check this,
however, and we shall comment on how this might be done below. For
now the important point for us is that there is a very compact and elegant
formulation (\ref{action-A}) of our theories, and that
our Lagrangian is the most general one given the symmetries and the field
content.

One might immediately object that (\ref{action-A}) is so remote from the usual
metric-based GR that even if it does describe deformations of GR in some formal
sense, it will never be possible to convert it into a physical theory with 
the usual matter fields coupled to it. This is a legitimate worry, but, as we have
already explained, our class of theories admits a formulation that is completely standard,
and in which the coupling to matter is straightforward. It is not completely obvious
how to go from (\ref{action-A}) to (\ref{S-g-H}) though, and this passage will take
the bulk of the paper to explain. In the last section we shall also make some comments
on how matter may be coupled to this theory directly in the pure connection
formulation. 

Let us now describe how this class of theories may be supplemented with reality conditions
that convert it into that of real Lorentzian spacetime metrics. A detailed discussion
of the reality conditions, which is to a large extent new, will be given in the main text, here we shall
just state the main claims. To obtain an action for a real Lorentzian spacetime metric 
we first need to restrict the set of connections that can appear in (\ref{action-A}). This is 
achieved with the reality conditions that already appear in \cite{Bengtsson:1992cm} and read:
\be\label{reality-A}
F^i\wedge (F^j)^* = 0.
\ee
These are nine real conditions and can be shown to guarantee that the conformal
structure of the metric that our theory describes (as will be explained below) is real and of 
Lorentzian signature. The reality of the conformal factor is subtler, and will be discussed when
we describe how the spacetime metric arises. But no extra reality conditions will be necessary. 

We now have to discuss the action. The action (\ref{action-A}) evaluated on
connections satisfying (\ref{reality-A}) can be interpreted as one for a
real Lorentzian metric constructed from $A^i$. However, in general this action will 
not be real, as we shall see. A real action that is of interest is
obtained by taking the imaginary part of (\ref{action-A}). Thus, the final action is:
\be\label{action-A-real}
S[A]={\rm Im} \int_M f(F^i\wedge F^j),
\ee
supplemented with the reality conditions (\ref{reality-A}).
As we shall demonstrate below, the class of theories so defined
describes two {\it real} propagating DOF. 

The action (\ref{action-A-real}) is reminiscent of that of the Chern-Simons formulation for
Euclidean signature gravity in 3 dimensions when the cosmological constant is negative.
In that context, one introduces an $\SOC$ connection $A^i=\omega^i+ ie^i$,
where $\omega^i$ is the spin connection and $e^i$ is the tetrad. The first-order
Einstein-Hilbert action then takes the following simple form:
\be
S_{3D}[A] = {\rm Im} \int_M {\rm Tr}(A\wedge dA+ \frac{2}{3}A\wedge A\wedge A).
\ee
Thus, it is not uncommon that one has to work with a physical theory whose Lagrangian
arises as a real (or imaginary) part of some holomorphic Lagrangian. An extra complication 
in our case as compared to 3D gravity is that one needs to impose the reality conditions 
(\ref{reality-A}). 

The described formulation (\ref{action-A-real}) 
of our class of theories, although elegant and compact,
is quite unsuited for explicit computations. Indeed, our understanding of
gravity is based on its metric description, and so is the coupling of gravity to
matter fields. It is thus absolutely necessary to develop an explicitly metric
formulation. A formulation that goes half-way towards this is
that in which this class of theories was rediscovered in \cite{Krasnov:2006du}.
In the retrospect, this formulation can be quite easily obtained from 
(\ref{action-A}). We first give a formulation of the complex theory and then
state the reality conditions. The idea is to introduce extra auxiliary fields
which remove the need to take a function of curvature 
and thus convert the theory to an explicitly first-order form. This is similar
to what is done in the passage from (\ref{f-Riemann}) to (\ref{S-f-phi})
and is achieved by the following action:
\be\label{action-AB}
{\cal S}[B,A] = \int_M B^i\wedge F^i - \frac{1}{2}V(B^i\wedge B^j),
\ee
where $A^i$ is still the $\SOC$ connection, and $B^i$ is a new field that is
an $\SOC$ Lie algebra valued two-form field. The function $V(\cdot)$ is again
a holomorphic homogeneous function of degree one of a $3\times 3$ symmetric matrix,
and its homogeneity allows it to be applied to a 4-form $B^i\wedge B^j$. The
dependence of (\ref{action-AB}) on the two-form field $B^i$ is algebraic, and
it can be integrated out with the result being the pure connection action (\ref{action-A}),
with the function $f(\cdot)$ being related to $V(\cdot)$ by an appropriate Legendre
transform. 

The price to pay for a simpler formulation (\ref{action-AB}) is that the
theory now has second-class constraints, as we shall see below. The reason for this
is that while some of the variables in $B^i$ are dynamical, the two-form field
$B^i$ also contains a subset of fields that do not propagate, and which at the level
of the Hamiltonian formulation are eliminated by certain second-class
constraints. This, however, does not appear to cause any problems, at least at
the level of the classical theory, as the second-class constraints can be solved, 
and a sufficiently simple description with only first-class constraints is possible. 
We shall return to all these points below when we describe the Hamiltonian
formulation of (\ref{action-AB}).

The best way to think about the theory (\ref{action-AB}) is to treat
$B^i$ as the main variable, and $A^i$ as only an "auxiliary" field, similar to
what happens in the first-order formulation of gravity. As in first-order gravity,
the auxiliary connection can be integrated out, and a second-order theory for
the two-form field $B^i$ only can be obtained. A way how to do this explicitly 
is described below. It then makes sense
to impose the reality conditions needed to get a real Lorentzian
signature theory directly on the two-form field $B^i$. These read:
\be\label{reality-B}
B^i \wedge (B^j)^* = 0,
\ee
A theory with two real propagating DOF is then given by the following action:
\be\label{action-AB-real}
S[B,A] = {\rm Im} \int_M B^i\wedge F^i - \frac{1}{2}V(B^i\wedge B^j),
\ee
supplemented with the reality conditions (\ref{reality-B}).

The main advantage of the formulation (\ref{action-AB-real}) is that the spacetime
metric which the theory describes is encoded in the two-form field $B^i$ in a very
simple way. Briefly, this is as follows. There exists a unique real Lorentzian signature 
conformal metric with respect to which the two-forms $B^i$ satisfying (\ref{reality-B}) 
are self-dual. This is the so-called Urbantke metric, see \cite{Schoenberg:1971yw} and
\cite{Urbantke:1984eb}, with the second reference being much better known. It is given by:
\be\label{Urb}
\sqrt{-g} g_{\mu\nu} \sim \epsilon^{ijk} B^i_{\mu\alpha} B^j_{\nu\beta} B^k_{\rho\sigma} 
\tilde{\epsilon}^{\alpha\beta\rho\sigma},
\ee
where the tilde above the Levi-Civita symbol signifies the fact that it is a densitized object
that does not need a metric for its definition. Another relevant reference on this is \cite{Samuel},
which explains why knowledge of self-dual forms (and thus the Hodge operator on two-forms)
is equivalent to knowledge of the conformal metric. 

It turns out, and this can be confirmed in
several different ways,  that (\ref{Urb}) is the (conformal) metric that the theory is about.
In terms of this metric, the following very convenient parameterization of the
two-forms $B^i$ becomes possible. One constructs the canonical triple of {\it metric} 
self-dual two-forms $\Sigma^a$, see the main text for an explicit expression for them in
terms of tetrads. The two-form field $B^i$ can then be written as:
\be\label{B-param}
B^i = b^i_a \Sigma^a,
\ee
where $b^i_a$ are complex. This is the 
parameterization used in \cite{Krasnov:2008ui,Freidel:2008ku}.
As we said, the quantities $\Sigma^a$ carry information about the metric, and
$b^i_a$ can be viewed as additional scalar fields. When the action is written
in terms of the metric and the scalars $b^i_a$ it takes the form (\ref{S-g-H}), with
$H_{\mu\nu\rho\sigma}$ being constructed from $b^i_a$ in a very simple way. 
The scalars $b^i_a$ are then non-propagating, and can be integrated out. 
This way one arrives at an effective metric theory, coming from an underlying theory with 
two propagating DOF. This briefly summarizes the logic of our construction of the 
effective Lagrangians. Most of the remainder of the paper is a detailed explanation
of how this construction works, as well as how a non-local transformation relating
two local Lagrangians is possible. 

Finally, before we turn to the main body of the paper, a word of caution about complex quantities
is in order. The described non-local field redefinition
idea, which makes possible to have a theory with two propagating DOF,
seems to require the introduction of self-dual quantities. Recall that
the Hodge operator in four dimensions splits the space $\Lambda^2$ of rank two anti-symmetric
tensors into $\Lambda^2={}^+\Lambda^2 \otimes {}^-\Lambda^2$ the self- and anti-self-dual
subspaces. When the metric is of Lorentzian signature, as is appropriate for a physical
theory, these spaces are necessarily {\it complex}. Thus, we are led to having to
work with complex objects. The usage of self-dual complex quantities and 
holomorphic Lagrangians
may be quite unfamiliar to some readers, and make it hard to follow the paper. We
have tried our best in making the paper as accessible as possible and formulated
the main ideas without referring to self-duality at all. Similarly, in the next section
we will follow the path of searching for a degenerate Lagrangian and continue
working in the usual tensor notations familiar to all readers. It is here that we shall
see that self-dual quantities are necessary. The following sections of the paper 
may require from the reader some effort in learning the basics of self-duality.
We hope this will not prove to be an unsurmountable obstacle. 

The present paper can be considered as yet another step towards understanding of
properties and interpretation of the class of theories \cite{Bengtsson:1990qg}. 
We hope that the viewpoint of effective field theory taken here will make these 
"neighbors of GR" of interest to a larger scientific community than was the case before. 

The organization of this paper is as follows. Next section
uses linearized theory to find conditions for the Lagrangian (\ref{S-g-H}) to be degenerate. 
It is here that we are led to self-dual quantities. The pure connection formulation of the
theory is described in section \ref{sec:pure-con}. We review some aspects of self-duality
in applications to gravity in section \ref{sec:self-dual}. In particular, Pleba\'nski 
formulation of GR that works with self-dual two forms instead of the metric is reviewed here.  
A formulation  of our theory in which the spacetime metric becomes the dynamical variable 
is given in section \ref{sec:2-form}. Here we give the Hamiltonian analysis, and present several 
equivalent formulations of the theory. We explain how the "physical" conformal 
factor for the metric is selected in section \ref{sec:conf-fact}. In section
\ref{sec:reality} we discuss the reality conditions. In section \ref{sec:eff}
we integrate out the non-propagating scalars and obtain the effective metric
theory. Section \ref{sec:redef} provides a key for understanding what makes
our class of theories possible. Here we explain the origin of the field redefinition that
maps our theories to GR, and work out this field redefinition to first orders in perturbation theory.
We conclude with a discussion of what results of this paper might mean
for the quantum theory of gravity. Appendix contains a summary of our conventions,
as well as some technical results relating the self-dual and usual metric-based
description of connections and curvature.

Finally, we note that a summary of the results of this paper has appeared as \cite{Krasnov:2009ip}.
Responses to \cite{Krasnov:2009ip} that we have received indicate that it is unclear whether
our work is about "usual" effective metric Lagrangians or some new theory that is being
proposed. Thus, it appears to be appropriate to stress our viewpoint from the outset:
this work is an attempt to understand what may be behind the usual effective metric Lagrangians
with their infinite number of higher-derivative terms. However, this aim is only achieved
if one understands the principle that produces such Lagrangians, or, equivalently, if
one can write all the infinite number of terms. The novelty of this work is then in the
underlying principle that is proposed, while the effective metric Lagrangians arising 
from our underlying theory are completely standard.  

\section{Degenerate Lagrangians: Linearized analysis}
\label{sec:deg}

As we have described in the Introduction, one way to arrive at the class of theories
in question is select a differential operator $D$ in (\ref{S-g-H}) so that the Lagrangian
is degenerate and the theory has only two propagating DOF. A rather complete
analysis of this is possible in the linearized theory. Thus, we just need 
to "complete the square" in the part of the Lagrangian that is metric-dependent.
The linearized Riemann curvature is given by:
\be\label{Riem-lin}
R_{\mu\nu}^{\quad\rho\sigma} = -2 \partial_{[\mu} \partial^{[\rho} h_{\nu]}^{\sigma]},
\ee
and the linearized $2(R+R^{\mu\nu\rho\sigma}H_{\mu\nu\rho\sigma})$ part of the
action takes the following form:
\be\label{R-RH}
\int d^4x\left( - \frac{1}{2} h^{\mu\nu}\Box h_{\mu\nu} 
-h^{\mu\nu}_{\,\,,\nu} h_{\mu\rho}^{\,\,,\rho} -
h^{\mu\nu}_{\,\,,\mu\nu} h + \frac{1}{2} 
h\Box h - 4 h_{\mu\nu} H^{\mu\rho\nu\sigma}_{,\rho\sigma} \right).
\ee
Here the indices are raised and lowered with the Minkowski spacetime metric $\eta_{\mu\nu}$,
the comma denotes differentiation, $h:=h^\mu_\mu$, $\Box:=\partial^\mu\partial_\mu$ and our
signature convention is $(-,+,+,+)$. The quadratic form here has the schematic form
$(1/2) h^TAh + h BH$, where $A,B$ are second-order differential operators. We 
can always (formally) complete the square by inverting the operator $A$. Thus, we 
can always write: 
\be
\frac{1}{2}h^TAh + h BH= \frac{1}{2}(h+ \frac{1}{A}BH)^TA(h+\frac{1}{A}BH) -
 (BH)^T\frac{1}{2A}(BH).
\ee
Since the theory is diffeomorphism-invariant, the operator $A$ is not invertible. However,
it is invertible on transverse functions $X^{\mu\nu}_{\,\,,\nu}=0$, and its inverse is:
\be
\frac{1}{A}X_{\mu\nu} = -\frac{1}{\Box}X_{\mu\nu} - \frac{1}{2\Box^2} 
(\partial_\mu\partial_\nu - \eta_{\mu\nu} \Box) X^\rho_\rho.
\ee
In view of
symmetries of $H_{\mu\nu\rho\sigma}$ the quantity 
$(BH)^{\mu\nu}=-4H^{\mu\rho\nu\sigma}_{\,\,,\rho\sigma}$ is transverse. 
Thus, we can easily complete the square and write (\ref{R-RH}) as the 
free graviton action with the shifted graviton field
\be\label{lin-redef}
\tilde{h}_{\mu\nu}=h_{\mu\nu} + \frac{4}{\Box} H_{\mu\rho\nu\sigma}^{\quad,\rho\sigma}
+\frac{2}{\Box^2} (\partial_\mu\partial_\nu - \eta_{\mu\nu} \Box) 
\eta^{\alpha\beta} H_{\alpha\rho\beta\sigma}^{\quad,\rho\sigma}
\ee
plus the term
\be
8\left( H_{\mu\rho\nu\sigma}^{\quad,\rho\sigma} \frac{1}{\Box} 
H^{\mu\alpha\nu\beta}_{\quad,\alpha\beta} - \eta^{\mu\nu}
H_{\mu\rho\nu\sigma}^{\quad,\rho\sigma}\frac{1}{2\Box}  
\eta^{\gamma\delta} H_{\gamma\alpha\delta\beta}^{\quad,\alpha\beta}
\right)
\ee
It is this term that we would like to be cancelled by the {\it local} term of the 
form $(DH)^2$. This is only possible if $H_{\mu\nu\rho\sigma}$ satisfies:
\be\label{HH-eqn}
\left(H_{\mu\rho\nu\sigma}H_{\gamma\alpha\delta\beta} (\eta^{\mu\gamma}\eta^{\nu\delta}-
\frac{1}{2}\eta^{\mu\nu} \eta^{\gamma\delta})\right)_{symm} \sim 
\left(g_{\alpha\beta}Y_{\rho\sigma}\right)_{symm},
\ee
where $symm$ means taking the completely symmetric part of the $\alpha\beta\rho\sigma$
tensor, and $Y_{\mu\nu}$ is some symmetric tensor that can be computed by contracting
this equation with the Minkowski metric. 

It is quite non-trivial to satisfy (\ref{HH-eqn}). Indeed, the quantity $H_{\mu\nu\rho\sigma}$
having all the symmetries of the Riemann curvature tensor has 20 independent components.
On the other hand, there are $4*5*6*7/4! - 10= 25$ equations in (\ref{HH-eqn}). We are aware
of only two solutions, but it would be of interest to analyze the equation
(\ref{HH-eqn}) in full generality. 

In order to exhibit some solutions, let us decompose $H_{\mu\nu\rho\sigma}$ into its
scalar, "Ricci" and "Weyl" parts. It is easy to check that the scalar part 
$H_{\mu\nu\rho\sigma}\sim \eta_{\mu[\rho}\eta_{\sigma]\nu}$ satisfies (\ref{HH-eqn}).
This, however, is not a very interesting solution, for it simply reproduces the well-known fact
that the linearized action
\be\label{S-lin-phi}
S[h_{\mu\nu},\phi]=2\int d^4x \left( (1+\phi)R - \frac{3}{2}\phi\Box\phi - \frac{g}{2}\phi^2\right)
\ee
that comes from the full action 
\be\label{S-full-phi}
S[g_{\mu\nu},\phi]= 2\int d^4x \sqrt{-g}\,e^\phi \left(R + \frac{3}{2}\phi^{,\mu} \phi_{,\mu} 
- V(\phi) \right)
\ee
is just Einstein-Hilbert action for the metric $\tilde{g}_{\mu\nu}=e^{\phi}g_{\mu\nu}$
plus a potential term for $\phi$. Thus,
the simplest example of (\ref{S-g-H}) with a non-propagating field is obtained
by introducing an extra scalar field $\phi$ and then writing the Einstein-Hilbert action 
for $e^{\phi}g_{\mu\nu}$ (plus a potential term), in which the kinetic term obviously
has a symmetry $\phi\to \phi-\psi, g_{\mu\nu}\to e^\psi g_{\mu\nu}$
that makes the scalar $\phi$ non-propagating. This is clearly just an uninteresting
local field redefinition, at least in the pure gravity case, see \cite{Carroll:2006jn} for a 
possible application in case when matter is present.

It is also easy to check that the "Ricci" part 
$H_{\mu\nu\rho\sigma}\sim \eta_{[\mu}^{[\rho}X_{\nu]}^{\sigma]}$ with $X^\mu_\mu=0$
does not satisfy this equation for the left-hand-side contains a term 
$X_{(\alpha\beta}X_{\rho\sigma)}$ that is not of the required form. The Weyl part
by itself does not work either, but the self-dual (or anti-self-dual) part of the Weyl
part of $H_{\mu\nu\rho\sigma}$ does satisfy (\ref{HH-eqn}). The easiest way we
know how to demonstrate this requires introduction of a triple of self-dual two-forms
$\Sigma^a_{\mu\nu}, a=1,2,3$:
\be
\frac{1}{2i} \epsilon_{\mu\nu}^{\quad\rho\sigma} \Sigma^a_{\rho\sigma}
=\Sigma^a_{\mu\nu}
\ee
normalized so that $\Sigma^{a\,\mu\nu}\Sigma^b_{\mu\nu}=4\delta^{ab}$.
Then let us write:
\be\label{H-SS}
H_{\mu\nu\rho\sigma} = \frac{1}{4}\Sigma^a_{\mu\nu}\Sigma^b_{\rho\sigma} H^{ab},
\ee
where $H^{ab}$ is a symmetric tracefree (complex) $3\times 3$ matrix. According
to this formula, the quantity $H_{\mu\nu\rho\sigma}$ is taken to be self-dual with
respect to both pairs of indices, which is what one would get by taking the "Weyl"
part of $H_{\mu\nu\rho\sigma}$ and requiring self-duality on any one pair of
indices. The basic self-dual two-forms $\Sigma^a_{\mu\nu}$ are going to
play a very important role in what follows, so it makes sense to give an explicit
formula for them already at this stage. This is easily done by introducing
an arbitrary "space plus time" split and writing:
\be\label{Sigma-def-1}
\Sigma^a_{\mu\nu} = i\, dt_\mu dx^a_\nu - i\, dt_\nu dx^a_\mu
-\epsilon^{abc} dx^b_\mu dx^c_\nu,
\ee
where $t$ and $x^a$ are the time and spatial coordinates respectively. 
Choosing a different space plus time split corresponds to making a Lorentz
transformation, and this can be seen to boil down to an $\SOC$ rotation
of the triple $\Sigma^a_{\mu\nu}$. The algebra (\ref{algebra}) of objects
(\ref{Sigma-def-1}) can be verified explicitly. Note that the basic
two-forms $\Sigma^a_{\mu\nu}$, being self-dual, are necessarily complex.
We will come back to the issue of having to work with complex objects
below.

Then using the simple identity (\ref{algebra}) we find the left-hand-side of (\ref{HH-eqn})
to be equal
\be
\frac{1}{16}\eta_{(\alpha\beta}\eta_{\rho\sigma)} H^{ab}H^{ab},
\ee
which is obviously of the form required. We thus see that the linearized theory:
\be\label{S-lin-H}
{\cal S}[h_{\mu\nu},H^{ab}]=
\int d^4x  \, \left( 2R+ \frac{1}{2}R^{\mu\nu\rho\sigma}
\Sigma^a_{\mu\nu}\Sigma^b_{\rho\sigma}H^{ab} + \frac{1}{2}H^{ab} \Box H^{ab} 
-\frac{g}{2} H^{ab}H^{ab}\right)
\ee
has just two propagating degrees of freedom and the (complex) field $H^{ab}$ is
an auxiliary, non-propagating one. The theory (\ref{S-lin-H}) is thus an analog
of (\ref{S-lin-phi}) written using a (complex) $3\times 3$ symmetric tracefree matrix 
instead of $H_{\mu\nu\rho\sigma}$. Similar to (\ref{S-lin-phi}), the action (\ref{S-lin-H}) is
obtained from the Einstein-Hilbert action by a field redefinition (\ref{lin-redef}),
which is however now non-local. As we have explained in the Introduction, it is this
non-locality of the "field redefinition" used in the construction of the action
that leads to an interesting theory (at the full non-linear level).

A remark is in order. The reader might be worried that the action (\ref{S-lin-H}) involves
a complex field $H^{ab}$ and is thus complex. This is indeed a source of some
difficulties, but a satisfactory prescription that resolves this issue is possible.
Thus, note that, as written, the action (\ref{S-lin-H}) is {\it holomorphic} in the complex
field $H^{ab}$. Such actions can be viewed as functionals of the real and imaginary
parts of their complex fields. Thus, we write: $H^{ab}=H_1^{ab}+i H_2^{ab}$
where $H_{1,2}^{ab}$ are real matrices. The action is then
\be
S[h_{\mu\nu},H^{ab}]=S_1[h_{\mu\nu},H_1^{ab},H_2^{ab}]+i
S_2[h_{\mu\nu},H_1^{ab},H_2^{ab}],
\ee
where $S_{1,2}$ are now real functionals of real fields. It can now be checked that
the field equations one obtains by varying say $S_1[h_{\mu\nu}, H_1^{ab},H_2^{ab}]$
with respect to $H_1^{ab},H_2^{ab}$ are the same as the real and imaginary parts
of the complex field equation obtained by varying the holomorphic action 
${\cal S}[h_{\mu\nu},H^{ab}]$ with respect to $H^{ab}$. This is easily checked using the
Cauchy-Riemann equations satisfied by $S_{1,2}$. Thus, as far as the equations
for $H_{1,2}$ are concerned one can work with either the complex holomorphic action
${\cal S}[h_{\mu\nu},H^{ab}]$ or with any of the two real actions 
$S_{1,2}[h_{\mu\nu}, H_1^{ab},H_2^{ab}]$ - the obtained field equations are
the same. Thus, one can view the holomorphic action 
(\ref{S-lin-H}) as just a trick that allows to work with one
complex field $H^{ab}$ instead of two real ones $H^{ab}_{1,2}$.

With these remarks being made, let us 
write down explicitly the field redefinition that takes (\ref{S-lin-H}) to the
form GR plus potential for $H$. This field redefinition is given in (\ref{lin-redef}),
and we should just apply it to the case (\ref{H-SS}) at hand. We have:
\be
\tilde{h}^{\mu\nu} = h^{\mu\nu} + \frac{1}{\Box} \Sigma^{a\,\mu\rho}\Sigma^{b\,\nu\sigma}
\partial_\rho\partial_\sigma H^{ab},
\ee
which can be compactly written as
\be\label{h-redef-H}
\tilde{h}_{\mu\nu} = h_{\mu\nu} + \frac{1}{\Box} \partial^a_\mu\partial^b_\nu H^{ab},
\ee
where we have introduced the derivative:
\be
\partial^a_\mu \equiv \Sigma^a_{\mu\nu}\partial^\nu.
\ee
The formula (\ref{h-redef-H}) was first obtained in \cite{Freidel:2008ku}, where
the action (\ref{S-lin-H}) is also contained, as well as its non-linear generalization.
The degenerate Lagrangian philosophy (at the linearized level) was first
proposed and developed in this cited paper. 

Applying (\ref{h-redef-H}) we can write (\ref{S-lin-H}) as:
\be
{\cal S}[h_{\mu\nu},H^{ab}]=\int d^4x \left( 2R(\tilde{h}) - \frac{g}{2} H^{ab}H^{ab}\right).
\ee
If now $\tilde{h}_{\mu\nu}$ and $H^{ab}$ are taken to be the fundamental variables,
which is legitimate as the transformation (\ref{h-redef-H}) is just a shift,
then our linearized theory is clearly equivalent to the linearized GR,
for the field equation for $H^{ab}$ just fixes this field to $H^{ab}=0$.

Note that the metric perturbation $\tilde{h}_{\mu\nu}$ that arises
as the result of the field redefinition (\ref{h-redef-H}) is {\it complex}. 
However, this is not a cause for concern for, as we discussed above,
one should always keep in mind the fact that $H^{ab}$ is just
an auxiliary field to be integrated out. Once this is done one should
get a real Lagrangian and a real metric perturbation. We shall see
this below.

Let us now discuss what happens if we integrate $H^{ab}$ out already
in (\ref{S-lin-H}). The field equation for $H^{ab}$ gives, formally:
\be\label{H-h}
H^{ab} = \frac{1}{2(g-\Box)}\left(\Sigma^a_{\mu\nu} \Sigma^b_{\rho\sigma} 
R^{\mu\nu\rho\sigma}\right)_{tf},
\ee
where $tf$ stands for the trace-free part:
\be
(X^{ab})_{tf}:=X^{ab}-\frac{1}{3}\delta^{ab} {\rm Tr}(X).
\ee
We can now substitute this into (\ref{h-redef-H}) to obtain the field
redefinition purely in terms of the two metric perturbations. Using
(\ref{proj-s}) we get:
\be\label{h-redef-h}
\tilde{h}_{\mu\nu} = h_{\mu\nu} + \frac{8}{\Box(g-\Box)}  \partial^\alpha \partial^\beta
\left(P_{+\,\mu\alpha\rho\sigma}P_{+\,\nu\beta\gamma\delta} 
-\frac{1}{3} P_{+\,\mu\alpha\nu\beta}P_{+\,\rho\sigma\gamma\delta} \right)
R^{\rho\sigma\gamma\delta}.
\ee
Expanding the projectors we get:
\be\label{PP-ident}
2\left(P_{+\,\mu\alpha\rho\sigma}P_{+\,\nu\beta\gamma\delta} 
-\frac{1}{3} P_{+\,\mu\alpha\nu\beta}P_{+\,\rho\sigma\gamma\delta} \right)
R^{\rho\sigma\gamma\delta}=R_{\mu\alpha\nu\beta} - \eta_{\nu[\mu} R_{\alpha]\beta} 
+ \eta_{\beta[\mu} R_{\alpha]\nu}
+\frac{R}{3} \eta_{\nu[\mu} \eta_{\alpha]\beta} \\ \nonumber
+ \frac{1}{4i}\epsilon_{\mu\alpha}^{\quad\rho\sigma}
R_{\rho\sigma\nu\beta} + 
\frac{1}{4i} R_{\mu\alpha\rho\sigma} \epsilon^{\rho\sigma}_{\quad\nu\beta}
-\frac{R}{12i}\epsilon_{\mu\alpha\nu\beta}.
\ee
The real part of the right-hand-side (the first line) is just the Weyl tensor:
\be
C_{\mu\alpha\nu\beta}:=R_{\mu\alpha\nu\beta} - \eta_{\nu[\mu} R_{\alpha]\beta} + 
\eta_{\beta[\mu} R_{\alpha]\nu}+\frac{R}{3} \eta_{\nu[\mu} \eta_{\alpha]\beta}.
\ee
It is also clear that the imaginary part does not contribute to (\ref{h-redef-h})
in view of the (differential) Bianchi identity:
\be
\epsilon_{\mu\alpha}^{\quad\rho\sigma} \partial^\alpha R_{\rho\sigma\nu\beta}=0,
\ee
which at first order in the perturbation follows directly from (\ref{Riem-lin}). Thus,
we get:
\be
\tilde{h}_{\mu\nu} = h_{\mu\nu} + \frac{4}{\Box(g-\Box)} \partial^\alpha\partial^\beta 
C_{\mu\alpha\nu\beta}.
\ee
Note that the quantity $\partial^\alpha\partial^\beta C_{\mu\alpha\nu\beta}$ that
appears here is just the so-called Bach tensor considered to first order in the
metric perturbation. We remind the reader that the importance of Bach tensor is
in the fact that this tensor must vanish for a metric to be conformal to an Einstein metric.
The right-hand-side in the above formula is explicitly real, and illustrates well the nature 
of the non-local field redefinition involved. Expanding the $1/(g-\Box)$ in powers of $\Box$ we see
that only the first term contains $1/\Box$ with other terms being local. 

We can now also compute the effective metric action. Substituting (\ref{H-h}) 
into (\ref{S-lin-H}) and again using (\ref{proj-s}) we get the following
effective theory:
\be
{\cal S}^{eff}[h_{\mu\nu}] = \int d^4x\left( 2R + 2 \left( P^{+\,\alpha\beta\gamma\delta}
P^{+\,\mu\nu\rho\sigma} -\frac{1}{3} P^{+\,\mu\nu\gamma\delta}
P^{+\,\alpha\beta\rho\sigma}\right)
R_{\gamma\delta\mu\nu} \frac{1}{g-\Box} R_{\rho\sigma\alpha\beta}
\right),
\ee
where, as before, only the terms quadratic in the perturbation $h_{\mu\nu}$ should be
kept. The denominator $g-\Box$ here should be interpreted as an expansion in
powers of $\Box$, so that the effective action contains all powers of the derivatives.
Note that the effective action does not contain 
powers of the $1/\Box$ operator. Now using (\ref{PP-ident}) we notice
that the imaginary parts do not contribute to the action. Indeed, the quantity
\be
\int d^4x \, R_{\mu\nu}^{\quad\rho\sigma} \epsilon_{\rho\sigma}^{\quad\alpha\beta} 
\frac{1}{g-\Box} R_{\alpha\beta}^{\quad\mu\nu} 
\ee
can be seen to be zero (modulo a surface term) by integration 
by parts using the explicit form (\ref{Riem-lin})
of the Riemann curvature to first order in perturbation. Thus, overall we obtain
the following {\it real} effective metric action:
\be\label{eff-lin}
S^{eff}[h_{\mu\nu}] = \int d^4x\left( 2R + C^{\mu\nu\rho\sigma}\frac{1}{g-\Box}
C_{\mu\nu\rho\sigma} \right),
\ee
where only the terms quadratic in the perturbation are to be retained. Note that
we have replaced the symbol ${\cal S}$ for the action by $S$, because the action
is real and no extra step of taking the real part is necessary.

The obtained effective action (\ref{eff-lin}) looks non-trivial, but using (\ref{Riem-lin})
it can be shown to reduce to the Einstein-Hilbert term plus a set of terms that
vanish on-shell. This is well-known in the case of the $(Weyl)^2$ term, but can
be checked by a similar integration by parts argument for the full action (\ref{eff-lin}).
Another way to reach the same conclusion is to analyze the field redefinition 
(\ref{h-redef-h}). Indeed, an elementary computation using (\ref{Riem-lin}) gives:
\be\label{dd-C}
2\partial^\rho\partial^\sigma C_{\mu\rho\nu\sigma}=\Box \left( R_{\mu\nu}
-\frac{1}{6} \eta_{\mu\nu} \right) - \frac{1}{3} \partial_\mu\partial_\nu R,
\ee
where only the linear in perturbation terms are to be kept. Thus, the only 
non-local term in (\ref{h-redef-h}) is seen to be $(1/\Box)\partial_\mu\partial_\nu R$,
which, however, is just an infinitesimal diffeomorphism corresponding to the vector field
$\xi^\mu = \partial^\mu R$. Thus, modulo a non-local diffeomorphism, the
transformation (\ref{h-redef-h}) that maps the effective linearized 
metric theory (\ref{eff-lin}) into the linearized GR is {\it local}, and so
(\ref{eff-lin}) is just GR in disguise. Thus, at the described linearized
level the construction of an effective metric theory via a degenerate
Lagrangian does not produce anything new. But, as we shall see below,
things become much more interesting when we consider interactions.

The construction described above was, although not completely trivial, quite 
elementary. The natural question that now arises if there exists a non-linear
completion of the action (\ref{S-lin-H}), still of the general form (\ref{S-g-H}),
so that the property of theory to have only two propagating degrees of freedom
is preserved at the full non-liear level. The answer to this is yes, and the
corresponding theory is the one that is obtained from (\ref{action-A-real})
as explained in the Introduction. We now turn to more details of these
constructions.

\section{The class of theories: Pure connection formulation}
\label{sec:pure-con}

In the previous sections we have followed a down-up approach and attempted to
construct a degenerate Lagraingian with additional non-propagating fields. We have
see how this works at the linearized level, but we have also concluded that
at this level one does not obtain anything new - the theory is general relativity
with a rather complicated (and local) field redefinition applied to the metric 
variable. There is no guarantee that the same non-propagating fields idea can
be extended to an interacting theory, and there is no guarantee that the
resulting field redefinitions are non-local so that one gets something inequivalent to GR.
We do not know how to arrive at any such construction starting from
the linearized considerations we have given so far. Thus, we shall now switch
gears completely and describe the class of theories in question in the form
it was discovered. Only after a detailed investigation of the properties of these
theories will we be able to see that they indeed extend to non-linear level the ideas we were
describing so far. 

We start with the pure connection formulation in which our
theories were first discovered in \cite{Capovilla:1992ep}, \cite{Bengtsson:1990qg}.
Our analysis is not going to be very detailed, for the formulation with additional
two-form fields is more convenient and will be paid much more attention to below.
However, some points, e.g. the fact that the theories describe just two propagating
DOF, can be seen already at this level.

\subsection{Action}

We have defined our class of theories via the action (\ref{action-A}) in the Introduction.
This action involves a function of a matrix-valued 4-form, which is quite unconventional
and needs time to get used to. So, some alternative definitions might be helpful.

Thus, consider the trace of the matrix $F^i\wedge F^j$. If $F^i\wedge F^i\not = 0$ we can define a
$3\times 3$ matrix $\Omega^{ij}: F^i\wedge F^j = \Omega^{ij} F^k\wedge F^k$,
or, with the understanding that a common 4-form prefactor is introduced and then
cancelled:
\be\label{Omega}
\Omega^{ij} = \frac{F^i\wedge F^j}{F^k\wedge F^k}.
\ee
The field $\Omega^{ij}$ is a function of the connection $A^i$, and has the mass dimension zero,
where our convention is that the mass dimension of the connection is equal to one $[A^i]=1$.
Note that by definition ${\rm Tr}(\Omega)=1$. Since the mass dimension 
$[\Omega^{ij}]=0$ we can introduce in the Lagrangian any possible power of
this field (as long as a gauge-invariant combination is used). We are then led
to consider the following class of theories:
\be\label{action-A1}
S[A]= \int_M F^i\wedge F^i f(\Omega^{ij}),
\ee
where $f(\cdot)$ is an arbitrary dimensionless $[f]=0$ gauge-invariant function of its
matrix argument. 
When $f=const$ we get back the topological theory. Since ${\rm Tr}(\Omega)=1$ the 
function $f(\cdot)$ in (\ref{action-A1}) needs only be defined on this surface. When
it is extended off this surface as a homogeneous function of degree one, one gets
back the formulation (\ref{action-A}) we are already familiar with.

An alternative convenient description of the function $f(\Omega^{ij})$ in 
(\ref{action-A1}) is as follows. This function being $\SOC$-invariant,
it can only depend on the invariants of the (symmetric) matrix $\Omega^{ij}$. 
There are in general 3 such independent invariants, but since ${\rm Tr}\Omega=1$ we
are left with only two independent invariants. These can be taken to be
\be
{\rm Tr}\Omega^2, \qquad {\rm Tr}\Omega^3,
\ee
and so the action can be written as 
\be\label{action-A2}
S[A]= \int_M F^i\wedge F^i \rho({\rm Tr}\Omega^2, {\rm Tr}\Omega^3),
\ee
where $\rho(\cdot,\cdot)$ is now an arbitrary (holomorphic) function of its
two arguments. In this form the function that appears in the action is 
a simple complex-valued function of two complex-valued arguments,
and can be dealt with in the usual fashion. 

The question that the reader must now be asking is what (\ref{action-A1}), (\ref{action-A2}) has
to do with general relativity. An answer to this was given in \cite{Capovilla:1989ac},
\cite{Capovilla:1991kx} where it is shown that (complex) GR can be put into the form 
(\ref{action-A1}), (\ref{action-A2}) provided one chooses the defining function to be of 
a very special form. Namely, the function that turns out to reproduce GR is the delta-function 
that imposes the constraint:
\be\label{cond-GR}
{\rm Tr}\Omega^2 =\frac{1}{2},
\ee
or, equivalently ${\rm Tr}\Omega^{1/2}=0$, see \cite{Capovilla:1991kx} for
more details. If one allows the number on the right-hand-side of (\ref{cond-GR}) 
to be different from $1/2$ one obtains
the class of "neighbors of GR" studied in \cite{Capovilla:1992ep} and still
describing two DOF. Moreover, our effective Lagrangians analysis below
establishes that for a generic $f(\cdot)$ the low-energy limit of theory (\ref{action-A1})
is given by GR. Thus, theories (\ref{action-A1}) with varying $f(\cdot)$
provide a particular family of UV completions of general relativity, and
are indistinguishable from GR at energies much smaller than
the Planck energy. To put it stronger, our real world gravity theory
may be one of the theories (\ref{action-A1}) and we would not have noticed
this at low energies that we have access to.

Below we shall sketch the Hamiltonian analysis of a theory with
general $f(\cdot)$ to see why it still describes two (complex) DOF. But first,
let us write down the field equations.

\subsection{Field equations}

First, we need to understand how to deal with the function $f(\cdot)$ of a 4-form 
when variations are taken, e.g. for the purpose of deriving the field equations.
It is easiest to work this out if one puts $f(F^i\wedge F^j)$ into the form
$F^k\wedge F^k f(\Omega^{ij})$ with $\Omega^{ij}$ given by (\ref{Omega}).
When extended off the surface ${\rm Tr}\Omega=1$ as a homogenous function,
the function $f(\Omega^{ij})$ is a usual function of a matrix, and can be
differentiated with respect to its argument in the normal way. Then the first
variation of the Lagrangian is given by:
\be
2F^k\wedge \delta F^k f(\Omega) + F^k\wedge F^k \frac{\partial f}{\partial \Omega^{ij}}
\left( \frac{2F^i\wedge \delta F^j}{F^k\wedge F^k} - \frac{F^i\wedge F^j}{F^k\wedge F^k}
\frac{2F^l\wedge \delta F^l}{F^m\wedge F^m}\right) = 2 \frac{\partial f}{\partial \Omega^{ij}}
F^i\wedge \delta F^j,
\ee
where we have used $f-(\partial f/\partial \Omega^{ij})\Omega^{ij}=0$ that follows
from the homogeneity of $f(\cdot)$. The derivative $(\partial f/\partial \Omega^{ij})$
is a matrix-valued homogeneous function of degree zero in its argument, and so
one can instead write $(\partial f/\partial X^{ij})$, where $X^{ij}=F^i\wedge F^j$.
This derivative is then a matrix-valued zero-form. The field equations take
the following simple form:
\be\label{conn-EOM}
D_A \left( \frac{\partial f(X)}{\partial X^{ij}} F^j\right) = 0.
\ee
This is a set of $3\times 4$ equations for $3\times 4$ components of the connection $A^i$,
so the number of equations matches the number of unknowns. The combination 
that appears in (\ref{conn-EOM}) plays a very special role in this theory, so it deserves
a separate name to be given to it. So, we define a two-form:
\be\label{B-def}
B^i:= \frac{\partial f(X)}{\partial X^{ij}} F^j.
\ee
The field equations (\ref{conn-EOM}) then simply state that the two-form $B^i$ is
covariantly constant with respect to the connection $A^i$. 

\subsection{Hamiltonian analysis}

Our treatment here is analogous to that in e.g. \cite{Bengtsson:1990qg}, with
the main difference being that a compact notation using a function of a matrix-valued
4-form is employed. 

The two-form (\ref{B-def}) plays a special role in the Hamiltonian analysis
of the theory (\ref{action-A}) as giving the momentum conjugate to $A^i$. 
Indeed, introducing the time plus space split we can write the action (\ref{action-A}) in
the following form (modulo unimportant at this stage numerical factors):
\be
S=\int dt d^3x\, f(F_{0a}^{(i} F^{j)}_{bc} \tilde{\epsilon}^{abc}),
\ee
where $a,b,c$ are spatial indices and $\tilde{\epsilon}^{abc}$ is the 3-dimensional
Levi-Civita symbol, densitized so that it does not need any spatial metric for its definition.
It is now easy to determine the momentum conjugate to $A^i$. Indeed, we have:
\be
\frac{\delta S}{\delta \dot{A}^i_a} = \frac{\partial f(X)}{\partial X^{ij}}  \tilde{\epsilon}^{abc}
F^j_{bc} \equiv \tilde{\epsilon}^{abc} B^i_{bc}\equiv \tilde{E}^{ai},
\ee
where we have introduced a new notation $\tilde{E}^{ai}$ for the 
the (spatial) dual of the pull-back of the two-form $B^i$ onto the spatial slice, which
plays the role of the "electric" field in this formulation. Indeed, recall that in the
usual Yang-Mills theory the quantity canonically conjugate to the connection
is precisely the electric field.

To write the action in the Hamiltonian form we now have to solve for the velocities
$\dot{A}^i_a$ in terms of the momenta $\tilde{E}^{ai}$. We can however expect that
not all the components of $\dot{A}^i_a$ can be solved for. Indeed, the theory 
(\ref{action-A}) is diffeomorphism and $\SOC$ invariant, so we should at the
very least to have constraints that generate these symmetries. Some constraints
are very easy to see. Indeed, the spatial projection of the field equations $D_A B^i=0$
do not contain time derivatives of the basic variables and are thus constraints. Thus,
we get:
\be\label{Gauss}
D_a \tilde{E}^{ai}=0
\ee
as a set of constraints. These generate $\SOC$ rotations of the phase space variables
$A_a^i, \tilde{E}^{ai}$, as is not hard to check. 

Another set of constraints is that obtained from the identity:
\be\label{diffeo}
\frac{\partial f(X)}{\partial X^{ij}} F^i_{ad} \tilde{\epsilon}^{abc} F^j_{bc} =
\tilde{E}^{ai} F^i_{ad}=  0.
\ee
Indeed, it is not hard to see that the matrix $Y^{ij}=F^i_{ad} \tilde{\epsilon}^{abc} F^j_{bc}$
is anti-symmetric $Y^{ij}=-Y^{ji}$, while it is contracted with a symmetric matrix of first derivatives.
It is not hard to check that these constraints generate spatial diffeomorphisms. 

In addition to (\ref{Gauss}), (\ref{diffeo}) there is also the Hamiltonian constraint, which we
will refrain from exhibiting here in view of some algebra needed for this. We will describe
the Hamiltonian constraint explicitly in an equivalent version of the theory that works
with additional two-form fields. 

We shall ask the reader to believe us (for now) that the only constraints that arise are
3 Gauss constraint (\ref{Gauss}), 3 diffeomorphism constraints (\ref{diffeo}) and
one Hamiltonian constraints, and that these form a first-class algebra, i.e. the
Poisson bracket of constraints is again a constraint. All this can be verified explicitly,
but involves some algebra, especially when one deals with the Hamiltonian 
constraint. Thus, we shall refrain from giving the calculations here and send
the reader to references \cite{Bengtsson:1990qg}, \cite{Krasnov:2007cq} where
the computations are done. These results immediately allow us to count the
number of propagating DOF described by the theory. The configurational space
is that of $\SOC$ connections on the spatial manifold, and is thus 9-dimensional.
In addition, we have $3+3+1$ first class constraints that reduce one to a two-dimensional
physical configurational space. Thus, as promised, the theory describes two
(complex) DOF. To get a theory that describes two real DOF we will need to
impose reality conditions, and these shall be dealt with below.

\subsection{Remarks about the pure connection formulation}

Ideally, we would like to be able to work with the theory in its pure connection
formulation (\ref{action-A}) and e.g. perform the quantization already at this level.
Indeed, an important point about this class of theories is that they do not
contain any dimensionful parameters, and so one could expect a reasonably nice perturbative
behavior. However, this is not easy. The immediate problem with the connection
formulation is that its "vacuum" solution is not obvious. Indeed, we would e.g. like
to see the two propagating DOF appearing as propagating modes of the theory linearized 
around some good vacuum solution. What should be taken as such a vacuum?

To discuss this, it is convenient to introduce a notation:
\be
\frac{\partial f(X)}{\partial X^{ij}} F^j \equiv \frac{\partial f}{\partial F^i}.
\ee
Thus, we take a derivative of a function of a 4-form with respect to a 2-form to
obtain a 2-form, which is the one (\ref{B-def}) that already played an important 
role above. Using this notation, the first variation of the action reads:
\be
\delta S=\int \frac{\partial f}{\partial F^i} \wedge D_A \delta A^i,
\ee
and the second variation is given by:
\be\label{A-sec-var}
\delta^2 S= \int \frac{1}{2}\frac{\partial f}{\partial F^i} \epsilon^{ijk}\wedge 
\delta A^j \wedge \delta A^k + \frac{\partial^2 f}{\partial F^i\partial F^j} 
D_A \delta A^i \wedge D_A \delta A^j.
\ee

The most natural "vacuum" of the theory then seems to be 
\be\label{A-vac}
F^i = 0, \qquad \frac{\partial f}{\partial F^i} = 0, \qquad 
\frac{\partial^2 f}{\partial F^i\partial F^j} \not =0.
\ee
Indeed, this is indeed a "vacuum" of the theory in the sense that the first derivative of 
our "potential" function vanishes, which then automatically satisfies the field equation
$D_A B^i=0$, and only the second derivative is non-trivial. From (\ref{A-sec-var})
we see that the first "mass" term is absent, and there is only the "kinetic" term
for the connection. Thus, it seems like the perfect vacuum to expand about.
However, an immediate problem with this vacuum is that in the absence of any 
background structure the second derivative in (\ref{A-vac}) can only be proportional to 
$\delta^{ij}$, which gives a degenerate kinetic term. So, there does not seem to be any 
way to build a meaningful perturbation theory around (\ref{A-vac}). What might
be possible, however, is to keep the background connection $A^i$ general and simply
perform the one-loop computation with the action (\ref{A-sec-var}) using the
background field method. This might shed light on the conjectured closeness of this 
class of theories under the renormalization. We do not attempt this computation in 
the present paper, but hope to return to it in a separate publication.

A conventional perturbative treatment for our theory is possible, but requires
a rather strange, at least from the pure connection point of view, choice of
vacuum. Thus, as we shall see in details in the following sections, the usual 
perturbative expansion around a flat metric corresponds in the pure connection language 
to an expansion around the following point:
\be
F^i = 0, \qquad \frac{\partial f}{\partial F^i} \not = 0.
\ee
This is a very strange point to expand the theory about, but the non-zero right-hand-side
of the first derivative of the potential receives the interpretation of essentially the Minkowski
metric, and a usual expansion results. It might seem that this choice introduces a "mass"
term for the connection, but this is not so. In fact, the second "kinetic" term is still
a total derivative and plays no role, and there is only the "mass" term. However, as
we shall see, the connection is no longer a natural variable in this case, and one works with
the two-form field $B^i$ via which the connection is expressed as $A^i\sim \partial B^i$,
so what appears as a mass term is in fact the usual kinetic one but for the two-form field. 

The purpose of the above discussion was to motivate the need to rewrite the theory in
terms of some other variables using which e.g. a conventional perturbative treatment 
is possible. These are also the variables in which the metric description of the theory
becomes transparent. We give such a formulation in the following sections. We first
treat the complex theory, and discuss the reality conditions once the complex case
is understood. 

\section{Self-dual two-forms}
\label{sec:self-dual}

Before we describe a formulation that is based on an ${\mathfrak su}(2,\C)$-valued
two-form field we need to remind the reader how one can describe the usual
general relativity in terms of self-dual two-forms rather than the metric. This
formulation of GR became known in the literature as that due to Pleba\'nski.
\cite{Plebanski:1977zz}. We first describe how the self-dual
two-forms are constructed once the metric is given.

\subsection{Metric self-dual two-forms}

Let us start with a metric spacetime, and choose a tetrad $\theta^I$ corresponding to the 
metric, i.e. represent the metric as $g_{\mu\nu}=\theta^I\otimes \theta^J \eta_{IJ}$, 
where $\eta_{IJ}$ is the Minkowski metric. Then, making an arbitrary time-space split 
$\theta^I=(\theta^0,\theta^a), a=1,2,3$, consider the following triple of two-forms:
\be\label{sigma}
\Sigma^a := \im\, \theta^0 \wedge \theta^a - \frac{1}{2} \epsilon^{abc} \theta^b\wedge \theta^c.
\ee
Here $\im=\sqrt{-1}$ is the imaginary unit.
By construction, these two-forms are self-dual with respect to $g_{\mu\nu}$, and moreover
\be\label{sigma-prop}
(1/4) \Sigma^a_{\mu\nu} \Sigma^b_{\rho\sigma} g^{\mu\rho} g^{\nu\sigma} = \delta^{ab},
\qquad 
(1/4) \Sigma^a_{\mu\nu} \Sigma^a_{\rho\sigma} = P^+_{\mu\nu\,\,\rho\sigma},
\ee
where
\be
P^{+\,\mu\nu}_{\quad\rho\sigma} = \frac{1}{2}\left( \delta^{[\mu}_\rho \delta^{\nu]}_\sigma + 
\frac{1}{2i}\epsilon^{\mu\nu}_{\quad\rho\sigma} \right)
\ee
is the projector on the self-dual two-forms. Another important property of
two-forms $\Sigma^i_{\mu\nu}$ is their algebra:
\be\label{sigma-algebra}
\Sigma^{a\,\,\, \nu}_{\mu} \Sigma^b_{\nu\rho} = -\delta^{ab} g_{\mu\rho} + \epsilon^{abc} \Sigma^c_{\mu\rho},
\ee
where the spacetime index is raised using the metric. Note the similarity to the Pauli matrix algebra. 
We also note that the two-forms $\Sigma^a_{\mu\nu}$ are ``orthogonal''
\be
\Sigma^a_{\mu\nu} (\Sigma^b_{\rho\sigma})^* g^{\mu\rho} g^{\nu\sigma} = 0
\ee
to the complex conjugate (and anti-self-dual) two forms $(\Sigma^a_{\mu\nu})^*$.
It is a simple exercise to verify all these properties. For convenience of the
reader they are listen again in the Appendix, for we shall use them quite often.

It is very convenient to think of the triple of two-forms $\Sigma^a$ as being a two-form
taking values in a vector bundle with fibers ${\mathfrak su}(2,\C)$ 
over the spacetime $M$. Let us denote this bundle by $\cM$, for ``metric'', so we 
have: ${\mathfrak su}(2,\C)\to \cM\to M$. Note the the fibers ${\mathfrak su}(2,\C)$
are equipped with a canonical metric $\delta^{ab}$. It turns out that, given $\Sigma^a$, there is
a canonical $\SOC$ fibre metric $\delta^{ab}$ preserving connection $\gamma^a$ in 
$\cM$ with respect to which the two-forms $\Sigma^a$ are covariantly constant:
\be\label{metric-comp}
d\Sigma^a + \epsilon^{abc}\gamma^b \wedge \Sigma^c=0.
\ee
An explicit formula for this connection (in terms of $\Sigma^a$) 
can be obtained as follows. Let us apply the Hodge
dual to the 3-forms present in equation (\ref{metric-comp}). If we replace the operator of exterior 
derivative $d$ by the metric-compatible one $\nabla$, we can interchange the operator of Hodge
dual with the covariant derivative one, and then use the self-duality of $\Sigma^a_{\mu\nu}$ to 
obtain: $\nabla^\mu \Sigma^a_{\mu\nu} + \epsilon^{abc} \gamma^{b\,\mu}\Sigma^c_{\mu\nu}=0$. 
We can now multiply this equation by $\Sigma^a_{\alpha\beta} \Sigma^{d\,\beta\nu}$ and
use the identity:
\be\label{ident-sigma}
\epsilon^{abc}\Sigma^a_{\mu\nu}\Sigma^b_{\rho\sigma}\Sigma^{d\,\nu\sigma}=-2\delta^{cd}g_{\mu\rho}
\ee
that follows from (\ref{sigma-algebra}) to get:
\be\label{conn-gamma}
\gamma^a_{\mu} = \frac{1}{2} \Sigma^{a\,\rho\nu} \Sigma^b_{\mu\nu} \nabla^\alpha \Sigma^b_{\alpha\rho}.
\ee
In this expression the metric-compatible covariant derivative $\nabla^\alpha$ acts only
on the spacetime indices of the two-form $\Sigma^b_{\alpha\rho}$. We note that this expression
has a structure analogous to that of the well-known expression 
\be
\Gamma_\mu^{IJ} = \theta^I_{\nu} \nabla_\mu \theta^{J\nu}
\ee
for the Ricci rotation coefficients in terms of the covariant derivative of the tetrad.
It can moreover be shown by an explicit computation, see more on this in the Appendix, that the 
connection $\gamma^a$ given by (\ref{conn-gamma}) is 
just the self-dual part $\im \Gamma^{0a}-(1/2)\epsilon^{abc}\Gamma^{bc}$ of the 
tetrad-compatible ${\rm SO}(1,3)$ connection $\Gamma^{IJ}_\mu$.

Let us discuss the issue of gauge transformations. 
In choosing the split $I=(0,a)$ we had to select a time direction. However, we can apply to the
tetrad $\theta^I$ a Lorentz transformation, and use the resulting new tetrad (corresponding
to the same metric) to construct $\Sigma^a$ via (\ref{sigma}). It is not hard to show
that the new $\Sigma^a$ are related to the old ones by a $\SOC$
transformation which is just the self-dual part of the Lorentz transformation we have
applied. Thus, similar to the tetrads $\theta^I$ being defined only up to a Lorentz
${\rm SO}(1,3)$ transformation, our metric two-forms $\Sigma^a$ are defined only modulo
a $\SOC$ transformation. The infinitesimal $\SOC$ action
on $\Sigma^a$ is given by $\delta_\omega \Sigma^a = \epsilon^{abc} \omega^b \Sigma^c$,
where $\omega^a$ are the gauge transformation parameters. The corresponding action on
the canonical $\Sigma$-compatible connection $\gamma$ can be easily computed from 
(\ref{conn-gamma}) using (\ref{ident-sigma}) and reads: $\delta_\omega \gamma^a = 
\epsilon^{abc}\omega^b \gamma^c - \partial_\mu \omega^a$, exactly as is appropriate
for an $\SOC$ gauge transformation. 

\subsection{Plebanski formulation}

The importance of the self-dual objects introduced lies in the fact that 
general relativity can be described very simply in these terms. Thus, consider
the curvature of the $\Sigma$-compatible connection $\gamma^a$, i.e. 
$F^a=d\gamma^a + (1/2)\epsilon^{abc}\gamma^b\wedge \gamma^c$.
It is not hard to show, see the Appendix, that this coincides with the self-dual part 
of the curvature $F^{IJ}$ of the tetrad-compatible connection $\Gamma^{IJ}$.
This, together with the fact that Einstein condition can be reformulated as
the requirement that the self-dual part of the Riemann curvature tensor with
respect to one pair of indices must also be self-dual with respect to the other pair,
allows one to reformulate Einstein equations as:
\be\label{eq-Pleb}
F^a =\Phi^{ab} \Sigma^b,
\ee
where $\Phi^{ab}$ is an arbitrary $3\times 3$ matrix. Indeed, (\ref{eq-Pleb})
just states that the curvature of the self-dual part of the spin connection
is self-dual, which is equivalent to the Einstein condition. 

Another way to obtain a reformulation of GR that uses self-dual two-forms instead
of the metric is at the action level. For this we note that the quantity 
$\Sigma^{a\mu\nu} F^a_{\mu\nu}$ coincides with the Ricci scalar of the metric. 
This means that the action for general relativity can be rewritten using
the metric self-dual two-forms $\Sigma^a$, which is the formulation discovered
by Pleba\'nski \cite{Plebanski:1977zz}.

\subsection{Variations}

In this subsection, to gain more intuition about how the self-dual two-form capture the
familiar from GR concepts, we develop the calculus of variations for these objects. 
We do not need the technology developed here till section \ref{sec:redef} where
the non-local field redefinition mapping our theory into GR is worked out, so 
it can be skipped on the first reading. 

Let us first obtain a formula for variation of the metric two-forms $\Sigma^a$ when the metric is
varied. A metric variation $\dot{g}_{\mu\nu}$ can be described by a tetrad variation 
$\dot{\theta}^I_\mu$, and this can always be decomposed into the background tetrads: 
$\dot{\theta}^I_\mu = M^{IJ} \theta^J_\mu$
for some matrix $M^{IJ}$. Without loss of generality we can assume this matrix to be symmetric, for
its anti-symmetic part represents just a Lorentz rotation of the tetrad. With this assumption it
can easily be related to the metric variation 
$M^{IJ}=(1/2) \theta^{I\mu}\theta^{J\nu} \dot{g}_{\mu\nu}$,
where $\theta^{I\mu}$ is the inverse tetrad such that $\theta^I_\mu \theta^{J\mu}=\eta^{IJ}$ and  
$\theta^I_\mu \theta^{I\nu}=\delta_\mu^\nu$. Thus, we have:
\be
\dot{\theta}^I_\mu = \frac{1}{2} \theta^{I\nu} \dot{g}_{\mu\nu}.
\ee
Then, using the definition (\ref{sigma}) of the metric two-forms we get:
\be\label{dot-sigma}
\dot{\Sigma}^a_{\mu\nu} = \Sigma^{a\,\,\rho}_{[\mu} \dot{g}_{\nu]\rho},
\ee
where the square brackets denote anti-symmetrization. Using this formula we can get a 
formula for the mixed object with one of its spacetime indices raised, as well as for
the quantity with both indices raised:
\be\label{sigma-var}
\dot{\Sigma}^{a\,\mu}_{\quad\rho}:= (g^{\mu\nu}\Sigma^{a}_{\nu\rho})^{\cdot} 
= g^{\mu\nu} \Sigma^{a\,\,\sigma}_{(\nu} \dot{g}_{\rho)\sigma},
\qquad 
\dot{\Sigma}^{a\,\mu\rho}:= (g^{\mu\nu}g^{\rho\sigma}\Sigma^{a}_{\nu\sigma})^{\cdot} = 
g^{\mu\nu} g^{\rho\sigma} \Sigma^{a\,\,\alpha}_{[\sigma} \dot{g}_{\nu]\alpha}\, .
\ee
Note that these quantities are {\it not} equal to $\dot{\Sigma}^a_{\mu\nu}$ with its indices raised.

The inverse relation can also be found. Thus, we have, by an explicit computation:
\be
\Sigma^{a\,\rho}_{\quad\mu} \dot{\Sigma}^a_{\rho\nu} =  \dot{g}_{\mu\nu} + \frac{1}{2}g_{\mu\nu}
g^{\alpha\beta} \dot{g}_{\alpha\beta}.
\ee
Note that the quantity on the left-hand-side here is the one from (\ref{dot-sigma}).
This remark is important for, for a general $\dot{\Sigma}^a_{\rho\nu}$ on the left-hand-side
the result would not be symmetric.
From this we get:
\be\label{dot-g}
g^{\alpha\beta} \dot{g}_{\alpha\beta} = \frac{1}{3} \Sigma^{a\,\mu\nu} \dot{\Sigma}^a_{\mu\nu},
\qquad
\dot{g}_{\mu\nu} = \Sigma^{a\,\rho}_{\quad\mu} \dot{\Sigma}^a_{\rho\nu} -
\frac{1}{6} g_{\mu\nu} \Sigma^{a\,\alpha\beta} \dot{\Sigma}^a_{\alpha\beta}.
\ee
Let us also discuss the meaning of this formula for a general quantity $\dot{\Sigma}^a_{\rho\nu}$
substituted on the left-hand-side. Any two-form can be decomposed into the self- and 
anti-self-dual basic ones, so we can write:
\be
\dot{\Sigma}^a_{\mu\nu} = M^{ab} \Sigma^b_{\mu\nu} + N^{ab}\bar{\Sigma}^b_{\mu\nu}.
\ee
Let us now see what the projection in (\ref{dot-g}) does to such a general two-form variation.
We have:
\be
\Sigma^{a\,\rho}_{\quad\mu} \dot{\Sigma}^a_{\rho\nu}={\rm Tr}(M) \eta_{\mu\nu}
+ \Sigma^{a\,\rho}_{\quad\mu}\bar{\Sigma}^b_{\rho\nu} N^{ab} -
\epsilon^{abc} M^{ab} \Sigma^c_{\mu\nu}.
\ee
We note the the tensor $\Sigma^{a\,\rho}_{\quad\mu}\bar{\Sigma}^b_{\rho\nu}$ in the second
term is automatically symmetric, see (\ref{id-sd-asd}) and traceless. The trace-free property
follows from the orthogonality of self- and anti-self-dual forms. Thus, we see that the trace part of
the result picks up the trace ${\rm Tr}(M)$ of the matrix $M^{ab}$ of the self-dual coefficients.
The tracefree symmetric part picks up the matrix $N^{ab}$ of the anti-self-dual coefficients.
Finally, the anti-symmetric part of this tensor picks up the anti-symmetric part of $M^{ab}$.
The symmetric tracefree part of $M^{ab}$ has been projected out by this operation. 
As we shall see below when we discuss a more general two-form geometry, the trace part
of $M^{ab}$ as well as the matrix of anti-self-dual coefficients $N^{ab}$ is the part of
a general two-form perturbation that describes the metric part of this perturbation. Thus,
for future reference, we shall rewrite the formula (\ref{dot-g}) in a form applicable to
any two-form variation $\dot{B}^a_{\mu\nu}$. The "metric" part of such a perturbation is 
given by:
\be
\dot{g}_{\mu\nu} = \Sigma^{a\,\rho}_{\quad(\mu} \dot{B}^a_{|\rho|\nu)} - 
\frac{1}{6} g_{\mu\nu} \Sigma^{a\,\alpha\beta} \dot{B}^a_{\alpha\beta}.
\ee

Equipped with understanding of how the metric geometry can be described in
the language of self-dual two-forms, we are ready to study the two-form field formulation of our
class of theories. 

\section{Two-form field formulation}
\label{sec:2-form}

The action (\ref{action-A1}) contains derivatives of the basic connection field in the
denominator of the matrix $\Omega^{ij}$. This is extremely inconvenient, for
it makes e.g. the canonical analysis of the theory difficult. As we have seen,
in the connection formulation it is also difficult to choose the right "vacuum" for the theory 
to be expanded about. Moreover, the theory is very far from the usual metric formulation
of gravity, and, in particular, it is not clear how to impose the reality conditions to
make sure it describes two real degrees of freedom. All these problems are solved
by performing a Legendre transform and introducing a new basic variable 
$B^i = \partial f/\partial F^i$. The action then takes the form (\ref{action-AB}),
which, upon integrating $B^i$ out gives back the original action (\ref{action-A}).

One immediate advantage of the formulation (\ref{action-AB}) is that it does not involve
arbitrary powers of the derivatives. Indeed, the formulation is explicitly first-order.
As we shall soon see, the two-form $B^i$ now becomes our basic variable.
In particular, it describes the metric which our theory is about in a direct way. 
We start our description of the theory in this formulation by deriving the field equations.

\subsection{Field equations}

We start from the action (\ref{action-AB}) which, for the convenience of the reader, is 
\be\label{action-Pleb}
{\cal S}[B,A]=\int B^i \wedge F^i(A) - \frac{1}{2}V(B^i\wedge B^j),
\ee
where $V$ is a homogeneous function of degree one in its 4-form arguments, 
and $F^i(A)$ is the curvature of an $\SOC$ connection $A^i$.

Field equations that result from (\ref{action-Pleb}) are as follows. Varying the action
with respect to the connection $A^i$ one gets the compatibility equation: 
\be
dA^i + \epsilon^{ijk}A^j\wedge B^k=0,
\ee 
which can be viewed as an algebraic equation for the components of the connection. 
This equation can be explicitly solved by introducing a metric in the conformal class 
determined by $B^i_{\mu\nu}$, as we shall explain in details below. Once this is
done one obtains a second-order theory for the two-form field, which will
be our main object of study below. 

Varying the action with respect to $B^i$ we get:
\be\label{field-eqs}
F^i = \frac{1}{2}\frac{\partial V}{\partial B^i},
\ee
where the meaning of the derivative on the right-hand-side was explained
in the previous section. After the connection $A^i$ is solved for in terms of $B^i$,
this is a second-order differential equation for the components of the two-form field. 

Already at this stage we can see how general relativity is contained in 
the class of theories (\ref{action-Pleb}). Namely, one gets GR in its Plebanski
formulation if one takes the two-form field $B^i$ to be given by the metric
self-dual two-forms $\Sigma^a$ introduced in the previous subsection. 
Indeed, in this case, the $\Sigma^a$ compatible connection is given by
(\ref{conn-gamma}), and its curvature is can be written in terms of
the Riemann tensor as (\ref{F-Riemm}). Then, using (\ref{proj-s})
we have:
\be
i\,\Sigma^a\wedge F^a = R_{\mu\nu\rho\sigma} P^{+\,\,\mu\nu\rho\sigma} \sqrt{-g} \, d^4 x = 
(1/2) R \sqrt{-g} \, d^4 x,
\ee
and thus the first term in (\ref{action-Pleb}) is a multiple of the Einstein-Hilbert action.
The second term is obtained by noting that 
\be
\frac{i}{2}  \,\Sigma^a\wedge \Sigma^b = \delta^{ab} \sqrt{-g} \, d^4 x,
\ee
and so the potential function $V(\cdot)$ is evaluated on the identity matrix $\delta^{ab}$,
which produces a constant -- a multiple of the cosmological constant. As we shall see
below, one can also obtain GR from (\ref{action-Pleb}) by taking a limit when
the potential $V(\cdot)$ becomes infinitely steep in the sense explained below.
This effectively constraints $B^i$ to be of the metric form $\Sigma^a$ and gives
rise to GR in the way we have just observed. Yet another way to obtain GR
is to simply take the low-energy limit of the theory. As we have already mentioned,
for generic potentials $V(\cdot)$ the low-energy limit is given precisely by GR.

\subsection{Hamiltonian analysis}

The material in this subsection is along the lines of the treatment given in
\cite{Krasnov:2007cq}, even though the starting point is a slightly different
(but equivalent) action. In this Hamiltonian analysis section the lower
case Latin letters from the beginning of the alphabet denote spatial indices,
not internal $\SOC$ indices.  We hope this will not lead
to any confusion. 

The formulation in terms of two-forms allows to complete the Hamiltonian analysis
that was only sketched in the previous section. Thus, we introduce a space plus time
split and write the action as (modulo an overall numerical factor):
\be
{\cal S} = \int dt d^3x \left( \tilde{\epsilon}^{abc}( B^i_{0a} F^i_{bc} + B^i_{ab} F^i_{0c}) 
- 2V(\tilde{\epsilon}^{abc} B^{(i}_{0a} B^{j)}_{bc})\right).
\ee
Noting that $F^i_{0a}=\dot{A}_a^i - D_a A_0^i$, where $D_a$ is the covariant
derivative with respect to the spatial connection $A_a^i$, we see that the momentum
conjugate to $A_a^i$ is:
\be
\frac{\partial {\cal S}}{\partial \dot{A}_a^i} = \tilde{\epsilon}^{abc} B^i_{bc}\equiv
\tilde{E}^{ai}.
\ee 
This is exactly as we have observed in the pure connection formulation of the
previous section. Written in terms of the momentum variable $\tilde{E}^{ai}$
the action takes the following form:
\be\label{action-Ham1}
{\cal S}= \int dtd^3x\left( \tilde{E}^{ai} \dot{A}_a^i + A_0^i D_a \tilde{E}^{ai} 
+ \tilde{\epsilon}^{abc} B^i_{0a} F^i_{bc} - 2V(B^{(i}_{0a} \tilde{E}^{j)a}) \right),
\ee
where we have integrated by parts in the term containing the $A_0^i$ component
of the connection. It is clear that $A_0^i$ is a Lagrange multiplier that imposes the
Gauss constraint $D_a \tilde{E}^{ai}=0$.

Let us now discuss other constraints. If not for the last potential term our action would
be that of the so-called BF theory, which is a topological theory without any degrees
of freedom. Indeed, if not for the last term, all quantities $B^i_{0a}$ would be Lagrange
multipliers enforcing the constraints $F^i_{ab}=0$. Thus, our theory would be about
flat connections and thus void of any interesting physics. The potential term changes
this by making the action depend on most of the would-be Lagrange multipliers 
$B^i_{0a}$ in a non-linear way and thus removing most of the gauge symmetries present
in BF theory. 

To see this, let us use the momentum variables $\tilde{E}^{ai}$ to build a spatial
metric ${\rm det}(q) q^{ab}:=\tilde{E}^{ai}\tilde{E}^{bi}$. This can then be used
to raise and lower the spatial indices. For instance, we can introduced a matrix
$E_a^i$ inverse to $\tilde{E}^{ai}$ and having moreover density weight zero.
This can then be used to expand the quantities $B^i_{0a}$ as:
\be\label{M-Ham}
B^i_{0a} = M^{ij} E^j_a.
\ee
Now the argument of the potential function in (\ref{action-Ham1}) can be written as
\be\label{BB-Ham}
B^{(i}_{0a} \tilde{E}^{j)a}= \sqrt{{\rm det}(q)} M^{(ij)}.
\ee
Thus, the potential function only depends on the symmetric part of the matrix $M^{ij}$.
Moreover, since $V(\cdot)$ is a homogeneous function of degree one it only depends
non-linearly on the tracefree part of $M^{ij}$, for the trace part can be pulled out. 
In other words, we can always decompose:
\be
M^{(ij)} = \frac{{\rm Tr}(M)}{3}(\delta^{ij}+H^{ij}),
\ee
where $H^{ij}$ is tracefree. Then defining the lapse and shift functions
$N,N^a$ and a shifted potential via:
\be
\frac{{\rm Tr}(M)}{3} :=N, \qquad M^{[ij]} = \frac{1}{2} \epsilon^{ijk} E^k_a N^a, \qquad
V(\delta^{ij}+H^{ij}):=3U(H^{ij})
\ee
we can write the last two terms in (\ref{action-Ham1}) as
\be
\frac{1}{2} \tilde{\epsilon}^{abc} \epsilon^{ijk} E^j_a E^k_d N^d F^i_{bc} + N\left( \tilde{\epsilon}^{abc} E_a^i (\delta^{ij}+H^{ij}) F^j_{bc}
- 6\sqrt{{\rm det}(q)}U(H) \right).
\ee
We can rewrite this in terms of the momentum variable $\tilde{E}^{ai}$ using the following
simple identities: 
\be
\sqrt{{\rm det}(q)} E^i_a \tilde{\epsilon}^{abc} = \epsilon^{ijk}\tilde{E}^{jb} \tilde{E}^{kc},
\qquad
\epsilon^{ijk} E^j_a E^k_b = \utilde{\epsilon}_{abc} \tilde{E}^{ci},
\qquad
3! {\rm det}(q) = \epsilon^{ijk}\utilde{\epsilon}_{abc}\tilde{E}^{ai}\tilde{E}^{bj}\tilde{E}^{ck}.
\ee
In all these formulas the density weight is explicitly indicated. After some elementary
transformations we get for the following lapse and shift parts of the Hamiltonian:
\be
N^a \tilde{E}^{bi} F^i_{ab} + \utilde{N}\left( \epsilon^{ijk} \tilde{E}^{aj}\tilde{E}^{bk} 
(\delta^{il}+H^{il}) F^l_{ab} - 
\epsilon^{ijk}\utilde{\epsilon}_{abc}\tilde{E}^{ai}\tilde{E}^{bj}\tilde{E}^{ck} U(H)
\right).
\ee
It is now clear that the quantity $H^{ij}$, which is just the appropriate symmetric 
tracefree part of the would-be Lagrange multipliers $B^i_{0a}$, is non-dynamical.
It is no longer a Lagrange multiplier it used to be in BF theory, since the potential
depends on it non-trivially. However, there is still no time derivatives of this
quantity in the action, so its conjugate momentum is zero $\pi_{ij}=0$. 
Commuting the Hamiltonian with this primary constraint we get a secondary
constraint:
\be\label{H-constr}
\left(\frac{F^{(i}_{ab} \epsilon^{j)kl}\tilde{E}^{ak} \tilde{E}^{bl}}{\epsilon^{pqr}\utilde{\epsilon}_{abc}
\tilde{E}^{ap}\tilde{E}^{bq}\tilde{E}^{cr}}\right)_{tf} = \frac{\partial U}{\partial H^{ij}},
\ee
where $tf$ stands for the tracefree part of the matrix.
The Poisson bracket of this constraint with $\pi_{ij}=0$ gives 
$\partial^2 U/\partial H^{ij}\partial H^{kl}$ and if this matrix is non-degenerate, which is
the case for a generic $U(\cdot)$ at a generic point $H^{ij}$, then
the $H$-sector constraints are second class. This means that the field $H^{ij}$
is auxiliary and needs to be eliminated. At the level of the classical theory
this is done by simply solving for $H^{ij}$ from (\ref{H-constr}) and
substituting the resulting $H^{ij}$ expressed in terms of other phase space
variables into the action. This results in a Hamiltonian system with only
first class constraints, with the Hamiltonian given by:
\be
N^a \tilde{E}^{bi} F^i_{ab} + \utilde{N}\left( \epsilon^{ijk} \tilde{E}^{aj}\tilde{E}^{bk} 
F^i_{ab} + \epsilon^{ijk}\utilde{\epsilon}_{abc}\tilde{E}^{ai}\tilde{E}^{bj}\tilde{E}^{ck} 
\Lambda(\Psi)
\right),
\ee
where we have introduced a notation:
\be\label{Psi}
\Psi^{ij}=\left(\frac{F^{(i}_{ab} \epsilon^{j)kl}\tilde{E}^{ak} \tilde{E}^{bl}}{\epsilon^{pqr}
\utilde{\epsilon}_{abc}\tilde{E}^{ap}\tilde{E}^{bq}\tilde{E}^{cr}}\right)_{tf}
\ee
and $\Lambda(\cdot)$ is the Legendre transform of $U(H)$:
\be
\Lambda(\Psi)=\Psi^{ij} H^{ij}- U(H(\Psi)).
\ee

Thus, the final result is the following set of constraints:
\be\label{constraints}
D_a \tilde{E}^{ai}\approx 0, \qquad F^i_{ab} \tilde{E}^{bi}\approx 0, \qquad 
\epsilon^{ijk} \tilde{E}^{aj}\tilde{E}^{bk}F^i_{ab} + \epsilon^{ijk}\utilde{\epsilon}_{abc}\tilde{E}^{ai}
\tilde{E}^{bj}\tilde{E}^{ck} \Lambda(\Psi)\approx 0.
\ee
When $\Lambda(\Psi)=const$ we recognize in this the Hamiltonian formulation 
of GR due to Ashtekar \cite{Ashtekar:1987gu}, while a non-trivial function
$\Lambda(\Psi)$ corresponds to a theory distinct from GR. Thus, the Hamiltonian
formulation given allows to see how the described class of theories provide
deformations of GR in a particularly clean way. Indeed, one can recognize in
the quantity $\Psi^{ij}$ as given by (\ref{Psi}) the self-dual part of the
Weyl curvature. Thus, at the level of the Hamiltonian formulation the
deformation is obtained by making the cosmological constant of the theory
dependent on the Weyl curvature. 

It can be verified explicitly that the algebra of the constraints (\ref{constraints}) is
first class. The only non-trivial computation is that of the Poisson bracket of
the Hamiltonian constraint with itself, and this is performed in \cite{Krasnov:2007cq}.
We will only state the result. Let us introduce the smeared Hamiltonian constraint
\be
C_{\utilde{N}} = \int d^3x\, \utilde{N} \,(Hamiltonian).
\ee
Then we get:
\be
\{ C_{\utilde{N}_1}, C_{\utilde{N}_2} \} = 4 \int d^3x\, \tilde{\tilde{Q}}^{ab} \utilde{\utilde{N}}_b
F^i_{ac} \tilde{E}^{ci},
\ee
where
\be
\utilde{\utilde{N}}_a = (\partial_a \utilde{N}_1) \utilde{N}_2 -  (\partial_a \utilde{N}_2) \utilde{N}_1
\ee
and
\be
\tilde{\tilde{Q}}^{ab} = \frac{1}{2} \tilde{E}^{ai}\tilde{E}^{bl} \epsilon^{ijk}\epsilon^{lmn}
m^{jm} m^{kn}, 
\qquad m^{ij}=\delta^{ij} + H^{ij}(\Psi)=M^{(ij)},
\ee
where the matrix $M^{ij}$ has been introduced above in (\ref{M-Ham}).
Note that the internal $\SOC$ indices are tacitly assumed to be contracted using the
canonical metric $\delta^{ij}$, so it does not matter whether such an index is a
subscript or a superscript. Thus, the result of the Poisson bracket of two
Hamiltonian constraints is a diffeomorphism constraint, and the algebra closes.
This immediately allows us a count of the number of propagating DOF of
the theory. We see that the phase space and the character of the algebra of the constraints
are unmodified as compared to GR, so the number of DOF is unmodified as well.

According to \cite{Hojman:1976vp} it is the metric
that appears in the result for the Poisson bracket of two Hamiltonian constraints that must be
interpreted as the spatial metric. Thus, the physical spatial metric that the theory
is about is $Q^{ab}$, not the auxiliary metric $q^{ab}$ constructed from the 
momentum variable $\tilde{E}^{ai}$. It is thus important to understand this 
metric better. We can rewrite it as 
\be\label{metric-phys}
\tilde{\tilde{Q}}^{ab} =  \tilde{E}^{ai}\tilde{E}^{bj} {\rm det}(m) (m^{-1})^{ij}.
\ee
Thus, an important difference with the case of GR is that the physical spatial 
metric does no longer coincide with the "naive" metric build from the momentum
variable, and the matrix $(m^{-1})^{ij}$ should be used to contract the
indices instead. When $\Lambda(\Psi)=const$ then $H^{ij}=0$ and we
are back to the case of GR. We will soon see that the metric (\ref{metric-phys})
coincides with the conformal Urbantke metric that is defined by the
two-form field $B^i$. Thus, the Hamiltonian analysis gives one way
to identify which physical metric the theory is about. A different procedure
that leads to the same conclusion is to consider a motion of a "small body"
in this class of theories. This has been done in \cite{Krasnov:2008ui}
with the result being again that the physical metric in which bodies
move along geodesics is in the conformal class of the Urbantke metric.
The issue of the conformal factor of the physical metric is subtle, 
and will be discussed below. 

To summarize the results of this subsection, we have performed the Hamiltonian
analysis of our theory in its formulation that uses the two-form field $B^i$.
We have found that there are 5 second-class constraints, absent in the
pure connection formulation, but these can be easily dealt with and
the auxiliary fields they describe eliminated. So, the price to pay for the
first-order formulation is presence of second-class constraints. However,
these do not appear to be problematic. 

We have not considered the
question of the arising determinant of the Dirac bracket that would need
to be included in the measure if one is to quantize the theory. This 
determinant is quite easy to compute and the result is given by the
square root of the determinant of the matrix of second derivatives of the potential.
But we shall not dwell of these points any longer as they are only relevant for
quantum theory and can be safely ignored in this purely
classical paper. 

After the elimination of the auxiliary variables one obtains a
pure first-class algebra that deforms the algebra of constraints of general
relativity in Ashtekar formulation \cite{Ashtekar:1987gu} in the sense
that the canonical phase space variables $(A^i_a, \tilde{E}^{ai})$ are
no longer related in an elementary way to the physical spatial metric
that appears in the result of a commutator of two Hamiltonian constraints. 
Note that the algebra itself is unmodified, and, of course, it cannot be,
for the algebra in question is just that of diffeomorphisms. What is
modified as compared to GR is a {\it realization} of this algebra in
terms of a pair of canonically conjugate variables $(A^i_a, \tilde{E}^{ai})$,
see \cite{Peldan:1991as} where the same conclusion has been reached
in the (equivalent) pure connection formulation.
This is how the class of theories under consideration avoids the
uniqueness theorem of \cite{Hojman:1976vp}, for the starting
point of the analysis in this paper is an assumption that the spatial
metric is one of the conjugate variables. This is clearly not the case in our realization of the diffeomorphisms algebra, see \cite{Peldan:1991as} for further
discussion of this point.

However, one might ask a question about what happens if one decides to
use the variable $Q^{ab}$ on our phase space as one of the canonical variables. 
Then one has to find the canonically conjugate variable and express all
the constraints in terms of the new conjugate pair. It is at this step that
we expect that non-locality will enter, and the Hamiltonian will not be
a local function of the momentum conjugate to $Q^{ab}$. This means
that the analysis of  \cite{Hojman:1976vp}, which makes an assumption about locality, is 
inapplicable, which explains why a different realization of the constraint algebra is possible. 
It would be interesting to see all this explicitly, but we shall not attempt such a calculation in 
the present paper.

We now have two -- modified for our theory and the standard one for GR -- different realizations
of the constraint algebra of diffeomorphisms on the same phase space manifold. There must exist
a (presumably non-canonical) transformation relating these realizations. In section \ref{sec:redef} 
we will see that this is indeed the case by exhibiting this transformation as a 
non-local field redefinition at the level of the action. 

\subsection{Metric plus self-dual forms formulation}

Now that we have understood the canonical formulation of the theory and count
of the DOF that it describes, let us continue with its formal development and 
exhibit several different reformulations of it, with the aim being to get closer
to an explicitly metric formulation. Some of the formulations below may
appear not very suited for any practical computations, but it is useful to
have as many different perspectives on the same theory as possible, and
this motivates the analysis below. Note that we are still at the level of
working with a theory providing deformations of complex GR, for no reality
conditions have been imposed yet. These will be dealt with in a separate section.

The logic of the following developments is to first learn how the connection $A^i$
can be integrated out, and then learn how to have an explicitly metric parameterization
of the resulting theory of the two-form field $B^i$.

Let us start with the connection. We would like to solve the
``compatibility'' equation: $dB^i + \epsilon^{ijk} A^j\wedge B^k=0$ for
the components of the connection $A^i$. To do this it is very convenient
to introduce a (conformal) metric defined by $B^i$. Indeed, as we
have already discussed in the Introduction, a triple of two-forms $B^i$
defines a (conformal) metric via the condition that $B^i$ is self-dual
with respect to this metric. This is the Urbantke metric (\ref{Urb}).
We will not need an explicit expression for it in terms of $B^i$, but
it is important that it exists and is unique, up to a conformal factor. 
Using this metric we can raise and lower the spacetime indices of the objects.
$B^i_{\mu\nu}$. Then a set of identities analogous to those derived
for the metric two-forms $\Sigma^a$ can be obtained. A particularly useful
identity is given by:
\be\label{ident}
- \frac{1}{2\,{\rm det}(B)} \epsilon^{ijk} B^i_{\mu\nu} B^j_{\rho\sigma} B^{l\,\nu\sigma}
 = \delta^{kl} g_{\mu\rho},
\ee
where $g_{\mu\nu}$ here is the Urbantke conformal metric, and 
\be
{\rm det}(B):=- \frac{1}{24} \epsilon^{ijk} B^{i\,\,\nu}_\mu B^{j\,\,\rho}_\nu B^{k\,\,\mu}_\rho.
\ee
It can be verified explicitly that (\ref{ident}) is invariant under conformal rescalings
of the the metric, as it should, since only a conformal class of $g_{\mu\nu}$ is
well-defined. 

Using the introduced conformal metric we can easily solve the equation for $A^i$.
Indeed, using the self-duality of $B^i_{\mu\nu}$ 
one rewrites the compatibility equation as: 
\be\label{compat}
\nabla_\nu B^{i\, \mu\nu} + \epsilon^{ijk} A^j_\nu B^{\, \mu\nu}=0,
\ee
where $\nabla_\mu$ is the metric-compatible derivative operator that
acts only on the spacetime indices,
multiplies this equation by $B^{i\, \alpha\beta} B^l_{\alpha\mu}$, and uses
(\ref{ident}) to get 
\be\label{A-forms}
A^i_{\mu}(B) := \frac{1}{2\, {\rm det}(B)} 
B^{i\,\rho\sigma} B^j_{\rho\mu} \nabla^\nu B^j_{\nu\sigma}.
\ee
Note that this connection is conformally-invariant and has the correct
transformation properties of an ${\rm SO}(3)$ connection. That is, when the two-form
field $B^i_{\mu\nu}$ transforms as $\delta B^i_{\mu\nu}=\epsilon^{ijk} \omega^j B^k_{\mu\nu}$ 
the connection transforms as: $\delta_\omega A^i_\mu = 
\epsilon^{ijk}\omega^j A^k_\mu - \partial_\mu \omega^i$.
A demonstration of this is a simple exercise involving (\ref{ident}). 
Note also that our expression (\ref{A-forms}) is essentially the same
as the expression (\ref{conn-gamma}) obtained earlier for the
metric-compatible connection. The two coincide when the two-forms
$B^i$ are taken to be the metric two-forms $\Sigma^a$. A linearized version 
of the formula (\ref{A-forms}) was used in \cite{TorresGomez:2009gs}.

We can now substitute the expression (\ref{A-forms}) for the connection into 
the action (\ref{action-Pleb}) to obtain a second-order theory for the two-form field. 
Thus, using the compatibility equation, and switching from form to index notations,
we can rewrite (\ref{action-Pleb}) evaluated on $A^i(B)$ as:
\be\label{action-HH}
{\cal S}[B,A(B)]= \frac{1}{2\im}\int d^4x\,\sqrt{-g} \left( - \frac{1}{2} 
B^{i\,\mu\nu} \epsilon^{ijk} A^j_\mu(B) A^k_\nu(B) - 2V(m) \right),
\ee
where
\be
m^{ij} = \frac{1}{4} B^{i\,\mu\nu} B^j_{\mu\nu}.
\ee
We have used the self-duality of $B^i_{\mu\nu}$ to simplify expressions. Our conventions
for the self-duality are $(B^i)^* = \im \, B^i$, where $\im=\sqrt{-1}$ is the imaginary unit, 
and the Hodge operator defined by the metric is: 
$X^*_{\mu\nu} := (1/2)\epsilon_{\mu\nu}^{\quad\rho\sigma}X_{\rho\sigma}$,
where $\epsilon_{\mu\nu\rho\sigma}$ is the volume form of $g_{\mu\nu}$ and the indices are
raised-lowered with the metric. 

We can now substitute (\ref{A-forms}) into (\ref{action-HH}) and apply the 
identity (\ref{ident}) to obtain the following sigma-model-like action:
\be\label{action}
{\cal S}[g,B] = \frac{1}{2i} \int d^4x\,\sqrt{-g} \left( \frac{1}{4\, {\rm det}(B)} 
(B^{i\,\mu\nu}\nabla_\alpha B^{i\,\alpha\rho})(B^j_{\rho\mu} \nabla^\beta B^j_{\beta\nu}) - 2V(m) \right).
\ee

This formulation of the theory could be taken as its definition. However, it is not the
most useful one for practical computations, for the fields $B^i_{\mu\nu}$ and $g_{\mu\nu}$
on which this action depends are not completely independent, for $B^i_{\mu\nu}$
being self-dual with respect to $g_{\mu\nu}$ varies when the metric varies. Below
we will give some alternative formulations that are more convenient.

Properties of the theory in this formulation are as follows. 
First, it is conformally-invariant, i.e. invariant under the transformation 
$g_{\mu\nu}\to \Omega^2 g_{\mu\nu}$, with the two-form field $B^i_{\mu\nu}$ not
transformed. Indeed, this must be so because we have obtained (\ref{action})
from a metric independent theory (\ref{action-Pleb}), and the two-form
field $B^i$ present in our original formulation determines metric only modulo
conformal transformations. This invariance is interesting to verify explicitly.
To this end one observes that the quantity
$\nabla_\nu B^{i\,\mu\nu}$ appearing in the action, in view of self-duality of $B^i_{\mu\nu}$,
can be written as a multiple of $\epsilon^{\mu\nu\rho\sigma}\nabla_\nu B^i_{\rho\sigma}$,
which is essentially the Hodge dual of the 3-form $\nabla B^i$. When written this way it is obvious
that it does not matter which derivative operator is used, and one can use the metric-independent
derivative operator $\partial_\mu$ instead of $\nabla_\mu$. It follows that the 
quantity $\nabla_\nu B^{i\,\mu\nu}$ transforms under conformal transformations as 
$\nabla_\nu B^{i\,\mu\nu}\to \Omega^{-4} \nabla_\nu B^{i\,\mu\nu}$. 
The other quantities transform as $h^{ij}\to \Omega^{-4} h^{ij},
{\rm det}(B)\to \Omega^{-6} {\rm det}(B)$ and the invariance is obvious. 

Second, the theory (\ref{action}) is invariant under $\SOC$ rotations
of the two-form field: $B^i_{\mu\nu} \to M^{ij} B^j_{\mu\nu}$, where $M^{ij}$ is an
orthogonal matrix $M M^T=1$. This can be verified using (\ref{ident}). Indeed,
the variation of the action under an infinitesimal $\SOC$ transformation 
$\delta B^i_{\mu\nu}=\epsilon^{ijk} \omega^j B^k_{\mu\nu}$ takes the following form:
\be
\delta_\omega {\cal S} = \frac{1}{2i} \int d^4x\,\sqrt{-g} \, (\nabla^\nu \omega^i)
(\nabla^\beta B^i_{\beta\nu}) = \frac{1}{i} \int d^4x\,\sqrt{-g} \,\omega^i R^{\mu\nu} B^i_{\mu\nu} = 0,
\ee
where we have integrated by parts and used the anti-symmetry 
of $B^j_{\beta\nu}$ to convert $\nabla^{[\nu}\nabla^{\beta]}$ into Riemann curvature 
which then gives the equality with $R^{\mu\nu} B^i_{\mu\nu}$. The later is zero, 
because the Ricci tensor $R_{\mu\nu}$ is symmetric while $B^i_{\mu\nu}$ is anti-symmetric.

We now give a formulation in which the basic fields are the spacetime metric and
a set of scalar fields.

\subsection{Sigma-model-like formulation}

A particularly inconvenient feature of the formulation (\ref{action}) is that we
can not freely vary with respect to its dynamical fields, as a variation of the conformal
structure induces a variation in the self-dual forms. A formulation that is quite
similar in spirit to (\ref{action}) but which works with fields that can be varied
independently is obtained as follows. The two-forms $B^i_{\mu\nu}$ that are 
required to be self-dual with respect to the metric $g_{\mu\nu}$ can always
be decomposed into a basis of metric self-dual two-forms $\Sigma^a$, whose
construction and properties were explained above. 

Thus, for any triple $B^i_{\mu\nu}$ we can write:
\be\label{B-Sigma}
B^i_{\mu\nu}= b^{i}_a \Sigma^{a}_{\mu\nu},
\ee
where $\Sigma^{a}_{\mu\nu}$ are as in (\ref{sigma}), and a new type of index (lower case 
latin letters from the beginning of the alphabet) has been introduced in order to distinguish 
between the ${\rm SO}(3)$ indices originally present in the action (latin letters from the middle
of the alphabet) and the ones that appear when (\ref{sigma}) are introduced. The relation
(\ref{B-Sigma}) can be understood in more abstract terms by introducing two $\SOC$ bundles
over the spacetime, both with fibers being the Lie algebra ${\mathfrak su}(2)$ (complexified).
We shall denote the first of these bundles as $\cI$ for ``internal''. This is where the original
two-forms $B^i$ take values. The other bundle will be referred to as $\cM$ for ``metric''. This is
where the ``metric'' two-forms $\Sigma^a$ take values. Then the quantities $b^i_a$ is just a map
between these two bundles:
\be\label{b-map}
b: \cM \to \cI
\ee 

We also note, for future use, the behavior of $b^i_a$ under conformal rescalings.
With the metric $g_{\mu\nu}$ at this
stage being defined only modulo conformal transformations, so are the metric
two-forms $\Sigma^a$, which transform as $\Sigma^a_{\mu\nu}\to \Omega^2 \Sigma^a_{\mu\nu}$
when $g_{\mu\nu}\to \Omega^2 g_{\mu\nu}$. For $B^i_{\mu\nu}$ to remain invariant
the scalars $b^i_a$ must transform as $b^i_a \to \Omega^{-2} b^i_a$. 

The quantities
$b^{i}_a$ are nine scalars, so, after the substitution (\ref{B-Sigma}), the theory (\ref{action}) 
becomes that of metrics plus the scalars
$b^{i}_a$, and is invariant under the simultaneous rescaling $g_{\mu\nu}\to \Omega^2 g_{\mu\nu},
b^{i}_a\to \Omega^{-2} b^{i}_a$. The action is also invariant under two independent 
$\SOC$ rotations $b^{i}_a \to M^{i}_j b^{j}_a$ and $b^{i}_a\to b^{i}_b (N^{-1})^{b}_a, 
\Sigma^{a} \to N^{a}_b\Sigma^{b}$ where $M,N\in{\rm SO}(3)$. The second of these 
transformations is just the $\SOC$ freedom in choosing the forms (\ref{sigma}). 

We would like to characterize the theory arising this way in more details. To this end, let us 
recall that above we have introduced an $\SOC$ connection $\gamma^a$ on the bundle 
$\cM$ with respect to which the metric two-forms $\Sigma^a$ are covariantly constant:
$d\Sigma^a + \epsilon^{abc} \gamma^b \wedge \Sigma^c = 0$. This connection, we
recall, is just the self-dual part of the Levi-Civita connection. Then, using $\gamma^a$ we have:
\be
d B^i = (\cD b^i_a) \wedge \Sigma^a,
\ee
where $\cD b^i_a = d b^i_a + \epsilon_{ab}^{\,\,\, c} \gamma^b b^i_c$ is the covariant
derivative that acts only on the bundle $\cM$ index of the matrix $b^i_a$. Taking now the Hodge
dual of the above three-form and using the self-duality of $B^i$ we get: 
\be\label{D-gamma-b}
\nabla^\mu B^i_{\mu\nu} = (\cD^\mu b^i_a) \Sigma^a_{\mu\nu}.
\ee 
Substituting this expression into (\ref{action}) we get a new formulation of
our theory:
\be\label{action-sigma-mod}
{\cal S}[g,b] = - \frac{1}{2i} \int d^4x\,\sqrt{-g} \left( \frac{1}{4\, {\rm det}(b)} 
(\Sigma^{a\,\,\mu}_\alpha \Sigma^{b\,\,\nu}_{\mu}\Sigma^{c\,\, \rho}_\nu \Sigma^{d}_{\rho\beta})
(b^{i}_c \cD^\alpha b^{i}_a)(b^j_b \cD^\beta b^j_d) + 2V(m) \right),
\ee
where now ${\rm det}(b)=(1/3!) \epsilon_{ijk}\epsilon^{abc} b^i_a b^j_b b^k_c$ is the
determinant of the $3\times 3$ matrix $b^i_a$ and $m^{ij}=b^i_a b^j_b \delta^{ab}$.
The product of 4 $\Sigma$-matrices in the first term can be expanded using their
algebra (\ref{algebra}), but the arising expression is not elegant, so we decided
to keep the action in the form given.

A useful feature of the formulation (\ref{action-sigma-mod}) is that one can vary
with respect to the metric and the scalars $b^i_a$ independently. The theory is
still conformally invariant, so we get 9 equations by varying with respect to the
conformal metric, and 9 equations when varying with respect to the scalars. 
The resulting field equations are easiest to derive by going back to the 
original formulation in terms of forms and then expressing (\ref{field-eqs})
in terms of the metric and the scalars $b^i_a$. The same field equations
can of course be derived directly from the action, using the calculus of variations 
for the metric self-dual two-forms that was developed in the previous section. 
We found this formulation of the theory most suited for practical computations
with it for the purpose e.g. of finding explicit solutions. 

The formulation (\ref{action-sigma-mod}) is still not the most economical,
as it turns out to be possible to eliminate some of the gauge freedom present
in this formulation and write the theory in terms of only the internal metric
$m_{ab} = b^i_a b^i_b$. This formulation of the theory that arises is 
not as elegant as (\ref{action-sigma-mod}), but is more suited for our
purposes in this paper because it shows the theory to be general relativity
plus other fields. It was obtained and studied in \cite{Freidel:2008ku}. 
Here we will rederive the results of this reference keeping, however, the choice 
of the conformal factor for the metric arbitrary.

\subsection{Metric plus internal metric formulation}

We start by noting that the map (\ref{b-map}) is not ${\rm SO}(3)$ equivariant. However, it is 
convenient to extend it to an equivariant map between bundles. To this end, we introduce certain
enlarged internal and metric bundles, for which the structure groups are ${\rm GL}(3)$.
We shall denote this enlarged bundles by $\cI'$ and $\cM'$ respectively. Then $A^i, B^i$ become
the connection and a Lie-algebra valued two-form on $\cI'$ correspondingly, with an
explicit index description of these objects being:
\be
A^i_{\,\, j} = \epsilon^i_{\,\, kj} A^k, \qquad
D X^i = dX^i + A^i_{\,\, j} X^j, \qquad D X_i = dX_i - A^j_{\,\, i} X_j
\ee
for the connection and
\be
B^{i}_{\,\, j} = \epsilon^i_{\,\, kj} B^k
\ee
for the two-form field. In these ${\rm GL}(3)$ notations the compatibility equation takes
the following form:
\be
dB^{i}_{\,\, j} + A^{i}_{\,\, k} \wedge B^{k}_{\,\, j} - A^{k}_{\,\, j} \wedge B^{i}_{\,\, k} = 0.
\ee
Of course, $A^{i}_{\,\, j}$ and $B^i_{\,\, j}$ take values in the ${\rm SO}(3)$ subbundle $\cI\subset \cI'$
of the ${\rm GL}(3)$ bundle $\cI'$. We can perform a similar procedure on the connections $A^a$ and
the two-form field $\Sigma^a$ in $\cM$ to get a ${\rm GL}(3)$ connection $A^a_{\,\, b}$ and
a ${\mathfrak gl}(3)$-valued two-form field $\Sigma^{a}_{\,\, b}$.

Now, following \cite{Freidel:2008ku}, we note that the problem of finding the
$B^i_{\,\, j}$-compatible ${\rm GL}(3)$ connection in $\cI'$ is the same as the problem of finding
a certain $\Sigma^a$-compatible ${\rm GL}(3)$ connection on $\cM'$. Indeed, if one pulls
the $B^i_{\,\, j}$-compatible connection $A^i_{\,\, j}$ on $\cI'$ to a connection on $\cM'$
using the map (\ref{B-Sigma}) one gets:
\be\label{omega}
\omega^{a}_{\,\, b} = (b^{-1})^{a}_i A^i_{\,\, j} b^j_b + (b^{-1})^{a}_i d b^i_b.
\ee
where we have introduced the
inverse matrix $(b^{-1})^{a}_i$ satisfying:
\be
(b^{-1})^{a}_i b^{ja} = \delta_{i}^{j}, \qquad (b^{-1})^{a}_i b^{i}_b=\delta^{a}_b.
\ee
Now it is easy to check that, for any section $X^a$ of the ``metric'' bundle, the $A$-covariant
derivative of its $b$-image in the original bundle is just the $b$-image of the $\omega$-covariant
derivative in the ``metric'' bundle: 
\be
D (b^i_a X^a) = b^i_a \cD_\omega X^a,
\ee
where
\be
\cD_\omega X^a = d X^a + \omega^a_{\,\, b} X^b
\ee 
is the covariant derivative operator for the connection $\omega^a_{\,\, b}$. This immediately
implies that the connection $\omega^a_{\,\, b}$ on $\cM'$ that arises as the pull-back (\ref{omega})
is $\Sigma^a_{\,\, b}$-compatible:
\be\label{w-compat}
\cD_\omega \Sigma^a_{\,\, b} = 0.
\ee
However, unlike the connection $A^i_{\,\, j}$ on $\cI'$ that is an ${\rm SO}(3)$ connection
preserving the ``trivial'' metric $\delta_{ij}$:
\be
D \delta^{ij}=0,
\ee
the connection $\omega^a_{\,\, b}$ preserves 
\be\label{w-m}
\cD_\omega m_{ab}=0
\ee 
the metric 
\be
m_{ab}= b^i_a b^j_b \, \delta_{ij},
\ee
which is just the pull-back of $\delta_{ij}$ on fibers of $\cI'$ to a metric on fibers of $\cM'$.
The ${\rm GL}(3)$ connection $\omega^a_{\,\, b}$ can be explicitly determined from the conditions
(\ref{w-compat}) and (\ref{w-m}) as was done in \cite{Freidel:2008ku}. Let us
repeat this calculation.

To find $\omega^a_{\,\, b}$ let us decompose it into a part that preserves $\delta_{ab}$ and
the remainder:
\be
\omega^a_{\,\, b} = \gamma^a_{\,\, b} + \rho^a_{\,\, b},
\ee
where the metric-compatible connection $\gamma^a_{\,\, b}$ is such that:
\be
\cD_\gamma\Sigma^a=d\Sigma^a + \gamma^a_{\,\, b}\wedge \Sigma^b = 0, \qquad 
\cD_\gamma\eta_{ab}=d\eta_{ab} - \gamma^c_{\,\, a}\eta_{cb} - \gamma^c_{\,\, b}\eta_{ac}=0.
\ee
This connection is given by $\gamma^a_{\,\, b}=\epsilon^a_{\,\,cb}\gamma^c$, with
$\gamma^a$ given by (\ref{conn-gamma}). In what follows we shall omit the subscript
$\gamma$ next to the symbol $\cal D$, with understanding that such a symbol
always means the $\Sigma^a$-compatible derivative operator.
From the condition that $\Sigma^a$ is covariantly constant with respect to the derivative defined 
by $\omega^a_{\,\, b}$ we obtain:
\be\label{rho-2}
\rho^a_{\,\, b}\wedge \Sigma^b = 0.
\ee
The condition that the $\omega$-covariant derivative preserves $m_{ab}$ gives:
\be\label{rho-1}
\cD m_{ab} - \rho^c_{\,\, a}m_{cb} - \rho^c_{\,\, b}m_{ac}=0.
\ee
If we now introduce 
\be
\rho_{ab}:=m_{ac}\rho^c_{\,\,b}
\ee
the equation (\ref{rho-1}) gives:
\be
\rho_{(ab)}= \frac{1}{2}\cD m_{ab}.
\ee
It remains to find the anti-symmetric $\rho_{[ab]}$ part of $\rho_{ab}$. To this end
we use the equation (\ref{rho-2}) and write:
\be
(\rho_{[ab]}+ \frac{1}{2}\cD m_{ab} )\wedge \Sigma^b = 0.
\ee
Introducing $\rho_{[ab]}=(1/2)\epsilon_{abc}\rho_c$, and rewriting the resulting equation
in component notations (using the self-duality of $\Sigma^a$) we get:
\be
\epsilon_{abc} \Sigma^{b\,\mu\nu} \rho^c_\nu +  \Sigma^{b\,\mu\nu} \cD_{\nu} m_{ab} = 0.
\ee
We can solve this equation in exactly the same way as we solved the compatibility 
equation for the connection previously. Thus, we multiply it by 
$\Sigma^{a\,\alpha\beta}\Sigma^d_{\alpha\mu}$ and use (\ref{ident-s}) to get:
\be
\rho_\mu^a = \frac{1}{2} \Sigma^{a\,\alpha\beta}\Sigma^b_{\alpha\mu}
\Sigma^c_{\beta\gamma} \cD^\gamma m_{bc}.
\ee
Bringing the symmetric and anti-symmetric parts together, and using some algebra
of the $\Sigma$-matrices, we can write the answer for the $\rho$-part of the connection as:
\be\label{rho}
\rho_{\mu\,ab} = \frac{1}{2}\cD_{\mu} m_{ab} + \frac{1}{4}\epsilon_{abc}
(\delta^{ce}\Sigma^f_{\mu\nu} + \delta^{cf}\Sigma^e_{\mu\nu} - \delta^{ef}\Sigma^c_{\mu\nu})
\cD^\nu m_{ef},
\ee
which agrees with  \cite{Freidel:2008ku}. It can also be checked explicitly
that the corresponding connection $A^i$ coincides with the one obtained
earlier in (\ref{A-forms}) by a different method.

We can now compute the BF-part of the action in this formalism. We have:
\be\label{temp-act-1}
\int B^i\wedge F^i = -\frac{1}{2}\int b^i_a \Sigma^a \wedge \epsilon^k_{\,\, ij} F^j_{\,\, k} 
= -\frac{1}{2}\int b^i_a \Sigma^a \wedge \epsilon^k_{\,\, ij} b^j_b F^b_{\,\, c} (b^{-1})^c_k
\\ \nonumber
=-\frac{1}{2}\int {\rm det}(b) \epsilon_{abd}(m^{-1})^{dc} \Sigma^a\wedge F^b_{\,\,c} 
\\ \nonumber
=-\frac{1}{2}\int {\rm det}(b) \epsilon_{abd}(m^{-1})^{dc} \Sigma^a\wedge 
(F^b_{\,\,c}(\gamma) + \cD \rho^b_{\,\, c} + \rho^b_{\,\, e}\wedge \rho^e_{\,\, c}).
\ee
Here ${\rm det}(b)=(1/3!)\epsilon^{abc}\epsilon_{ijk}b^i_a b^j_b b^k_c$. Let us now
simplify the resulting expression. To this end, let us integrate by parts in the second term
in brackets. We have:
\be
\cD\left( {\rm det}(b) (m^{-1})^{dc}\right)= \cD_\omega \left( {\rm det}(b) (m^{-1})^{dc}
\right)- {\rm det}(b) \rho^d_{\,\, e} (m^{-1})^{ec} -  {\rm det}(b) \rho^c_{\,\, e} (m^{-1})^{de}.
\ee
The first term on the right-hand-side is zero since the $\omega$-derivative preserves $b^i_a$.
The other two terms combine with the last term in the brackets in (\ref{temp-act-1})
to give:
\be
-\frac{1}{2}\int {\rm det}(b) \epsilon_{abd}(m^{-1})^{dc} \Sigma^a\wedge F^b_{\,\,c}(\gamma)
+ \frac{1}{2} \int {\rm det}(b) \epsilon_{abc} \Sigma^a\wedge \rho^b_{\,\, e}\wedge \rho^c_{\,\, f}
(m^{-1})^{ef}.
\ee
The two terms here can be further rewritten. Thus, let us rewrite the ${\rm GL}(3)$ curvature
$F^b_{\,\, c}(\gamma)$ via the ${\rm SO}(3)$ curvature: $F^b_{\,\,c}=\epsilon^b_{\,\, ec}F^e$
and then expand the product of two $\epsilon$-tensors. Let us also rewrite the second term
in terms of the quantities $\rho_{ab}$. We get:
\be
\frac{1}{2}\int {\rm det}(b) \left( \delta^{ab} (m^{-1})^{cd}\delta_{cd}-(m^{-1})^{ab}\right)  
\Sigma^a\wedge F^b
+ \frac{1}{2} \int {\rm det}(b^{-1}) \Sigma^a m_{ad} \epsilon^{dbc} \wedge \rho_{be}
\wedge \rho_{cf} (m^{-1})^{ef}.
\ee

Writing everything in component notations, and adding the potential term, we get the full
action:
\be\label{action-m}
i{\cal S} = \frac{1}{4}\int d^4x\, \sqrt{g} \,\sqrt{{\rm det}(m)}
\left( \delta^{ab} (m^{-1})^{cd}\delta_{cd}-(m^{-1})^{ab}\right)  
\Sigma^{a\,\mu\nu}F^b_{\mu\nu} \\ \nonumber
+ \frac{1}{2} \int d^4x \, \sqrt{g} \,\sqrt{{\rm det}(m^{-1})}
\Sigma^{a\,\mu\nu} m_{ad} \epsilon^{dbc} \rho_{\mu\,be}
\rho_{\nu\,cf} (m^{-1})^{ef} - \int d^4x \sqrt{g}\, V(m).
\ee
The merit of this formulation is that the action is explicitly a functional of the metric (via the
$\Sigma$-forms) and a symmetric internal tensor $m_{ab}$ (as well as its inverse). 
It is considerably more involved than the actions we have encountered before. 
Thus, it is not the best starting point for, say, finding explicit solutions of the theory,
as it is a pain to even write down the arising field equations.
The most compact original formulation in terms of forms appears to be more
suited for explicit calculation. However, the formulation just derived 
will be a useful starting point for integrating the non-propagating modes out
of the theory, which we will come to below. Indeed, the obtained theory is
of the form (\ref{S-g-H}), and so the logic outlined in the Introduction is applicable. 
However, we first need to discuss reality conditions, as the theory is still complex. 
Another important issue that we have not yet discussed is that of the conformal factor of the
physical metric. Indeed, at this stage the metric formulations we have presented
are all conformally-invariant (even if this is not obvious from e.g. (\ref{action-m})), 
so it is not clear what is the physical metric among all the conformally-equivalent ones.

It is quite easy to see the usual GR arise from the action (\ref{action-m}). Indeed, one should just
take $m_{ab}=\delta_{ab}$. Then $\rho_{ab}=0$ and the second term is absent. The first
term then gives an integral of $(1/2)\Sigma^{a\,\mu\nu}F^a_{\mu\nu}$, which is equal to
$(1/2)R$, where $R$ is the Ricci scalar for the metric. We thus get the Einstein-Hilbert action
(in units $8\pi G=1$) with the cosmological constant $\Lambda=V(\delta)$.

\section{The physical metric}
\label{sec:conf-fact}

As we have seen in the previous section, the theory (\ref{action-Pleb}) can be viewed as that 
of metrics (modulo conformal transformations) plus either 9 or 6 scalars 
(modulo $\SOC$ rotations). The metric in terms of which the action was written
has been introduced in a natural way (as the metric that makes the two-forms $B^i$
of the original formulation self-dual). But we still have not verified that this
is the metric that the theory is about. Also, the theory is invariant under a simultaneous rescaling
if the metric and the scalars. We need to understand what is the "physical"
conformal factor that determines e.g. the metric that matter couples to. 
We are still at the stage of working with a complex theory,
so the question of coupling to matter is not completely physical, but it will come
so when the reality conditions are imposed. 

There are (at least) two ways to arrive to a conclusion about what the "physical" metric is. One
is by looking at the constraint algebra, where the Poisson bracket of two Hamiltonian
constraints gives the physical spatial metric (up to rescalings). We have seen that
this physical spatial metric is given by (\ref{metric-phys}). Now the metric $m^{ij}$
that appears in this formula is just a multiple of the matrix $B^i\wedge B^j$, 
see (\ref{BB-Ham}). Thus, it is the same matrix $m^{ij}=b^i_a b^j_a$ that we introduced when
we parameterized the two-form field by metric two-forms $\Sigma^a$ and the
scalars $b^i_a$. But this immediately shows that the physical spatial metric is that
constructed from the projection of the forms $\Sigma^a$ on the spatial slice. 
Thus, if we define $\tilde{\sigma}^{ap}=\tilde{\epsilon}^{abc}\Sigma^p_{bc}$,
where $p$ is an $\SOC$ index, then $\tilde{E}^{ai} = b^i_p \tilde{\sigma}^{ap}$, and
\be
\tilde{\tilde{Q}}^{ab} = \tilde{E}^{ai}\tilde{E}^{bj} {\rm det}(m) (m^{-1})^{ij} = 
{\rm det}(m) b^i_p b^j_q  \tilde{\sigma}^{ap}\tilde{\sigma}^{bq} (m^{-1})^{ij}=
{\rm det}(m) \tilde{\sigma}^{ap}\tilde{\sigma}^{bq}\delta_{pq}.
\ee
In other words, the physical spatial metric is the spatial projection of the metric used
to construct $\Sigma^a$. The same conclusion has also been reached in 
\cite{Bengtsson:1990qh} for the one-parameter family of deformations of GR
\cite{Capovilla:1992ep}.

Another way to get the physical metric is to look at the motion of a "small body" in the 
theory under consideration. This gives more information, because one determines
not just the (conformal) spatial metric, but the complete information about how any
type of matter may be consistently coupled to our theory. Such an analysis was
performed in \cite{Krasnov:2008ui}. The main results of this analysis are as
follows. While it can be confirmed that the physical conformal metric is that
defined by $B^i$ two-forms, and thus the one in terms of which we have written
down the actions above, the conformal factor of the physical metric cannot be
fixed by considering the pure gravity theory. Thus, only after a specific coupling
to matter is given, one obtains a preferred physical metric (along geodesics of which 
matter moves) in the conformal class defined by the two-forms. In principle, the 
theory is consistent with matter coupling to any physical metric in this conformal class. 
The information about which metric is physical is supplied by the matter part of the action, 
not the gravity part.

One way to understand this is to imagine that the material action couples to the
gravity theory in question solely via the two-forms $B^i$. After the parameterization
(\ref{B-Sigma}) is introduced, the coupling is that to the metric and
scalars $b^i_a$, and is invariant under conformal transformations
since the original $B^i_a$ is invariant. Such conformally-invariant couplings
of matter to gravity with extra scalars are easy to construct, and they have
been considered in the literature on conformal gravity, see e.g. \cite{Barabash:1999bj}. Thus,
in this reference, the authors have considered the conformal gravity theory whose action is
given by the square of the Weyl tensor coupled to conformally-invariant matter described
by the following action:
\be\label{action-yura}
S_M = \int d^4x \sqrt{-g}\left[\partial^\mu \phi \partial_\mu \phi +\lambda \phi^4 - \phi^2 R/12
+ i\bar{\psi}\gamma^\mu(x)\nabla_\mu \psi-\zeta \phi\bar{\psi}\psi\right].
\ee
Here $\lambda, \zeta$ are dimensionless parameters, $\phi$ is a scalar field 
(Higgs field) that transforms as $\phi\to \Omega^{-1}\phi$, and $\psi$ is a fermionic field 
describing matter. In a solution of the theory in which $\phi$ is nowhere zero one can use 
the conformal freedom to put it to a constant
$\phi=1$. After this is done the last term in (\ref{action-yura}) becomes the
usual mass term for the fermion. Thus, when $\phi\not=0$ it is always possible to
go to a gauge in which all matter is described in a standard way as
moving along spacetime geodesics of a certain spacetime metric. When working
in this gauge, the field equation for $\phi$ obtained by varying (\ref{action-yura}) becomes
an equation for the conformal factor of the metric. 

Our case is analogous,  with the additional scalar fields $m_{ab}$ playing the
role of the Higgs field $\phi$. Indeed, our conformally invariant action (\ref{action-m}) for the 
metric and the fields $m_{ab}$ is an analog of the Weyl-squared action 
plus the first three terms of (\ref{action-yura}). Thus, it remains to insert
a certain combination of the $m_{ab}$ fields into the matter mass terms to make them
conformally invariant, as in the last term in (\ref{action-yura}). However, since we
now have not one, but a multiplet of scalar fields, there are many combinations with 
right transformation properties that can be constructed. It is convenient to
parametrize such combinations with an arbitrary function of the matrix $m_{ab}$
that is homogeneous of degree one in its components. Thus, we introduce yet
another potential function $R(m): R(\alpha m)=\alpha R(m)$, similar to the
potential $V(m)$ we already have in the gravity sector. This potential should be
thought of as supplied by the matter part of the action. It has transformation properties 
$R(m)\to \Omega^{-4}R(m)$, and so $(R(m))^{1/4}$ has the properties of the scalar field 
$\phi$ in (\ref{action-yura}). Therefore, it can be inserted in the fermionic mass term(s)
in place of the usual Higgs field $\phi$ in (\ref{action-yura}). 

An alternative, and easier to deal with prescription, is to introduce a potential $R(m)$,
which is an arbitrary gauge-invariant homogeneous of order one function of
the "internal" metric $m_{ab}$, which should be thought of as supplied by
the matter part of the Lagrnagian, and then "fix the gauge" in which 
\be\label{mat-pot}
R(m)=1.
\ee
This makes all the material couplings to be the standard metric  ones,
and also fixes the physical metric that the theory couples to. This is the prescription
that we will use below, when we derive an effective metric action. We note that
the work \cite{Freidel:2008ku} used the prescription (\ref{mat-pot}) with
$R(m)=({\rm det}(m))^{1/3}$. This is convenient, for it eliminates factors
of ${\rm det}(m)$ from the formulas. But, as the analysis of \cite{Krasnov:2008ui}
shows, this is not the most general prescription allowed by the consistency of the
theory.

Thus, to summarize, the question of which physical metric all matter fields couples to
cannot be decided at the level of pure gravity. Any metric from the conformal class 
determined by $B^i$ can be such a metric. The input from the material sector that
is necessary to settle this question can be parameterized by the matter sector
potential $R(m)$. The the physical metric is selected by (\ref{mat-pot}). We shall
see how this prescription works below. 

\section{Reality conditions}
\label{sec:reality}

Now that we have obtained several equivalent formulations of the theory
providing deformations of complex GR we would like to explain how a consistent
real section can be selected. The idea is to impose reality conditions that
guarantee that the dynamical variable of the theory is a real Lorentzian signature
metric. In addition, since one is working with holomorphic Lagrangians,
one needs to choose either a real or imaginary part of the Lagrangian
as the physical real Lagrangian of the theory. The main ideas can be
explained using an example of a usual finite-dimensional dynamical system,
to which we now turn.

\subsection{Holomorphic Hamiltonians}

Let $(p,q)$ be complex-valued momentum and position, and $H(p,q)$
be a Hamiltonian that we assume depends on $p,q$ holomorphically:
$\partial_{\bar{q}}H=\partial_{\bar{p}} H=0$. The "holomorphic" Lagrangian is:
\be
{\cal L}=\int dt( p\dot{q} - H(p,q)),
\ee
where the time variable is assumed to be usual real, and the 
Poisson brackets are given by:
\be
\{f,g\} = \frac{\partial f}{\partial p}\frac{\partial g}{\partial q} - \frac{\partial f}{\partial q}
\frac{\partial g}{\partial p}.
\ee

We can then treat this system as
that with real phase space variables $(p_1,p_2,q_1,q_2)$, where
$q=q_i+\im q_2, p=p_1+\im p_2$, and two real "Hamiltonians" $H_{1,2}(p_{1,2},q_{1,2})$
arising as $H=H_1+\im H_2$. Expanding the Lagrangian we get:
${\cal L}=L_1+\im L_2$ where
\be
L_1= \int dt( p_1 \dot{q}_1 - p_2 \dot{q}_2 - H_1(p_{1,2},q_{1,2})), \qquad
L_2=\int dt( p_1 \dot{q}_2 + p_2 \dot{q}_1 - H_2(p_{1,2},q_{1,2})).
\ee
Thus, the holomorphic system gives rise to two real Hamiltonian systems with
two different Hamiltonians $H_{1,2}$ and with Poisson brackets:
\be
\{F,G\}_1 = \frac{\partial F}{\partial p_1}\frac{\partial G}{\partial q_1} 
- \frac{\partial F}{\partial q_1}\frac{\partial G}{\partial p_1}
-\frac{\partial F}{\partial p_2}\frac{\partial G}{\partial q_2} 
+\frac{\partial F}{\partial q_2}\frac{\partial G}{\partial p_2}
\ee
and
\be
\{F,G\}_2 = \frac{\partial F}{\partial p_1}\frac{\partial G}{\partial q_2} 
- \frac{\partial F}{\partial q_2}\frac{\partial G}{\partial p_1}
+\frac{\partial F}{\partial p_2}\frac{\partial G}{\partial q_1} 
-\frac{\partial F}{\partial q_1}\frac{\partial G}{\partial p_2}
\ee
respectively. Here $F,G$ are assumed to be real functions of phase space variables. 

We can now derive a set of relations between the evolution in these two systems using
the Cauchy-Riemann equations $\partial_{\bar{q}}H=\partial_{\bar{p}} H=0$. These read:
\be
\frac{\partial H_1}{\partial p_1}=\frac{\partial H_2}{\partial p_2}, 
\qquad \frac{\partial H_1}{\partial p_2}=-\frac{\partial H_2}{\partial p_1},
\qquad \frac{\partial H_1}{\partial q_1}=\frac{\partial H_2}{\partial q_2}, 
\qquad \frac{\partial H_1}{\partial q_2}=-\frac{\partial H_2}{\partial q_1}.
\ee
Consider now a (real) observable -- function on the phase space $Q(p_{1,2},q_{1,2})$.  
Using the Cauchy-Riemann equations for $H$ we have:
\be
\{ H_1,Q\}_1=\{H_2,Q\}_2.
\ee
Thus, the time derivative of
this observable with respect to the first Hamiltonian (using the first symplectic structure)
is the same as the time derivative with respect to the second Hamiltonian
(using the second symplectic structure). This means that the real and imaginary parts
of our original holomorphic Hamiltonian system describe the same dynamics and
are equivalent.

Let us now consider a question of how a real Hamiltonian
system can be consistently constructed from a given holomorphic one. 
For this purpose, let us assume
that we are given a holomorphic function $Q(p,q)$ on our phase space
and we would like to select a section on which this function is real. Our
holomorphic observable gives rise to two real ones $Q=Q_1+\im Q_2$,
and we are thus interested in a surface in our $(p_{1,2},q_{1,2})$ phase
space on which
\be\label{constr-1}
Q_2(p_{1,2},q_{1,2})=0.
\ee
Since the $H_{1,2}$ Hamiltonians generate equivalent evolution (if taken with
their respective symplectic structures), we can work at the level of e.g. the
real part of our Hamiltonian system with $H_1$ and the Poisson bracket
$\{\cdot,\cdot\}_1$. Then the required half-dimensional surface in our phase
space is obtained by enforcing the condition that $Q_2$ remains zero under
evolution. Thus, we need to require:
\be\label{constr-2}
\{H_1,Q_2\}_1 = 0.
\ee
We can now restrict our system to the constraint surface (\ref{constr-1}), (\ref{constr-2})
to obtain a dynamical system whose configuration space is one-dimensional.

Thus, the idea here is analogous to what happens in a system with second-class
constraints, where these constraints are used to eliminate some non-dynamical
phase space variables and arrive at a smaller phase space. As in the second-class
constraints case it is thus important that the symplectic structure reduced to
the constraint surface (\ref{constr-1}), (\ref{constr-2}) is non-degenerate. If this is
the case then we get a usual Hamiltonian system with the Hamiltonian being the
restriction of $H_1$ to the constraint surface and serving as a generator of evolution 
tangential to the surface. 

Let us illustrate this procedure on some simple examples. As the first example, let
us take a "holomorphic" harmonic oscillator with the Hamiltonian $H=(1/2)p^2 + (1/2)q^2$, or 
\be
H_1=\frac{1}{2}(p_1^2-p_2^2)+\frac{1}{2}(q_1^2-q_2^2), \qquad
H_2=p_1p_2 + q_1 q_2.
\ee
Let us assume that we would like the complex configurational variable $q$ to be real, i.e.
$q_2=0$. Taking the Poisson bracket of this constraint with the first Hamiltonian we
get: $\{H_1,q_2\}_1=p_2$. Thus, both $q_2$ and $p_2$ need to be set to zero and
we obtain the usual real harmonic oscillator. Note that in this example $H_2$ becomes
zero on the constraint surface, even though the evolution it generates is the same as
the real part $H_1$ of the holomorphic Hamiltonian $H$.

A bit more non-trivial example is that illustrating Ashtekar's Hamiltonian formulation
of GR, see \cite{Ashtekar:1991hf}, section 8. In this example the holomorphic
phase space is coordinatized by pairs $(q,z)$, with the symplectic structure being
$\{z,q\}=\im$. The holomorphic Hamiltonian is $H=zq-(1/2)z^2$, or
\be
H_1=z_1q_1-z_2q_2-(1/2)(z_1^2-z_2^2), \qquad
H_2=z_1q_2+z_2q_1-z_1z_2.
\ee
We now wish $q$ to be real, or $q_2=0$. However, our relevant symplectic structure
is now $\{\cdot,\cdot\}_2$ with the Hamiltonian $H_1$, since there is an extra factor
of $\im$ in the defining symplectic structure. Thus, we get the secondary constraint
to be: $\{H_1,q_2\}_2=q_1-z_1$. The symplectic structure induced on the 
corresponding constraint surface is that $\{z_2,q_1\}=1$, and the induced
Hamiltonian is: $H_1=(1/2)z_2^2+(1/2)q_1^2$, which is the usual real harmonic
oscillator with momentum $p=z_2$.

We now turn to the more non-trivial example of gravity.

\subsection{Reality conditions for  gravity: Hamiltonian treatment}

The case of gravity will be treated along the lines of the previous subsection. However,
there are some complications. First, as we have seen above, our gravitational
system has some non-dynamical fields and the associated second-class constraints.
Second, there are also first-class constraints that generate symmetries. 

We start by formulating the theory in terms of real variables. We get two real symplectic
structures as the real and imaginary parts of the original holomorphic one. 
The real and imaginary parts of the constraints give twice the number of
the constrains of the holomorphic formulation. 

It is convenient to deal with the
second-class constraints from the outset, as we have done it in the complex case.
We get twice the number of the holomorphic second-class constraints (as real
and imaginary parts of the later). As we have already discussed, these real
second-class constraints can be obtained from either the real or imaginary
parts of the action, so there is no loss of information if one chooses to work
with say only the real part. The non-dynamical variables can then be eliminated
in exactly the same way as they were in the complex theory, for the real constraints
are just the real and imaginary parts of the holomorphic equation (\ref{H-constr}) we were
solving in the complex case. Eliminating $H_{1,2}^{ab}$ we end up with
a real phase space coordinatized by $\tilde{E}^{ia}_{1,2}, A^i_{a\,1,2}$
(with two different symplectic structures on it), and a set of real first-class
constraints on it. 

The first-class constraints arising are twice in number as compared to the
complex case. Indeed, considering e.g. the Hamiltonian
constraint $H=H_1+\im H_2$, and decomposing the lapse function $N=N_1+\im N_2$
we get: $N_1 H_1-N_2 H_2$ as a contribution to the real part of the Lagrangian,
and $N_1 H_2 + N_2 H_1$ as that to the imaginary part. Varying with respect to
$N_1,N_2$ we get $H_1=0, H_2=0$ as the constraints. The algebra of the
constraints evaluated using either real or imaginary part of the symplectic structure
closes so the real constraints are still first-class. At this stage we have a theory
with 4 real propagating DOF, even though the extra pair of DOF has the wrong
sign in front of its kinetic term and is thus an unphysical ghost. 

We now wish to impose the reality conditions that eliminate the unphysical
pair of DOF. The most natural condition to impose is that the physical
spatial metric $Q^{ab}$ is real. This gives us 6 constraints $Q_2^{ab}=0$.
To reduce the dimension of the phase space by two we also need the
conjugate constraints that arise by requiring that the time evolution
preserves $Q_2^{ab}=0$. The time "evolution" in our completely constrained
case is given by commuting the constraints $Q^{ab}_2$ with the first-class
constraints. Let us first discuss the easy Gauss and (spatial) diffeomorphism
constraints. The Gauss constraint only acts on the internal indices and so
leaves the metric $Q^{ab}$ invariant. It is convenient though to introduce a
triad $\tilde{\sigma}^{ai}$ for the spatial metric, on which gauge-transformations
act by $\SOC$ rotations, and then require this triad to be real, as is the
case in Ashtekar Hamiltonian formulation of GR. This reduces the 
gauge group from $\SOC$ to ${\rm SO}(3)$. The holomorphic diffeomorphism constraint
generates infinitesimal diffeos along a complex-valued vector field $N^a$.
If we restrict the shift function to be real, this will remove the "imaginary"
diffeos. Thus, overall, restricting the Lagrange multipliers generating
the Gauss and diffeomorphism constraints to be real, we get a real
first-class constraint algebra of gauge transformations and spatial diffeomorphisms, 
which preserves the condition that $Q^{ab}$ is real. 

It remains to discuss the Hamiltonian constraint. As the first step we need
to require the lapse function to be real. One can then see that the Poisson
bracket (computed using say the real part of the symplectic structure)
of two smeared constraints $\int d^3x N_1 H_1$ and $\int d^3x M_1 H_1$,
where $N_1,M_1$ are real lapse functions and $H_1$ is the real
part of the Hamiltonian constraint, is given by $Q^{ab}$ times
$N_1\partial_a M_1 - M_1\partial_a N_1$, which are both real,
times the real part of the diffeomorphism constraint. Indeed, this can be
deduced just by taking the real part of the complex Poisson bracket of
two Hamiltonians smeared with real lapse functions. Thus, the real algebra
still closes. 

Computing the Poisson bracket of $H_1$ with $Q_2^{ab}$
we get the secondary constraints that together with $Q_2^{ab}$ eliminate half 
of the phase space variables. Alternatively, since our phase space is
extended by ${\rm SO}(3)$ gauge variables, we can compute the 
$H_1$ time derivative of the condition that $\tilde{\sigma}^{ai}$ are real,
which gives 9 conjugate constraints. Reducing the symplectic
structure on the constraint surface we get a real $9+9$ dimensional
phase space with a set of $1+3+3$ real first-class constraints on it,
which gives us 2 real propagating DOF as the result. 

An alternative, but easier to work in practice prescription is to work
with the holomorhic constraints and holomorphic symplectic structure,
and then require the triad of the physical metric to be real, as well
as the time derivative of this condition to be zero. This gives $9+9$
constraints that half the dimension of the phase space. When using
this prescription it is of course important to keep the lapse and shift
functions (as well as the functions that generate the ${\rm SO}(3)$ rotations)
real. The resulting description is exactly analogous to that discussed
in \cite{Ashtekar:1991hf}, section 8, with the only difference being that
the reality conditions on the triad now become more complicated --
it is the physical triad $\tilde{\sigma}^{ai}$, not the phase space variable $\tilde{E}^{ai}$
that is now required to be real. The rest of the treatment is the same,
in that the condition that the time derivative of the primary constraints
vanishes gives secondary constraints, and the whole set is used
to eliminate half of the phase space. 

The above discussion shows that it is possible to impose reality conditions
on our complex theory in a consistent way with the result being a
first-class constrained Hamiltonian system with two real propagating
DOF. What is missing in our case as compared to GR in Ashtekar
formulation is an explicitly real description, an analog of
$A^i_a = \Gamma^i_a + i K^i_a$, where $K^i_a$ is the extrinsic
curvature. It would be very interesting to develop such a
description, but an attempt to do it in this paper would take us too far.

\subsection{Reality conditions: Action}

We would now like to translate the above Hamiltonian-level
prescription into one for the Lagrangian of the theory. As we saw
in the previous section, the physical spatial metric, on which
we imposed our reality conditions, is just the spatial restriction of
the Urbantke metric constructed from the two-form field. Now
having required the lapse and shift functions to be real,
as well as the time derivative of the spatial metric to be
real, we have required the full spacetime metric to be real.
Thus, all the above reality conditions can be encoded in
the condition that the physical metric described by our theory
is real. These are exactly the 9 conditions $B^i\wedge (B^j)^*=0$
plus a condition on the conformal factor. This last condition 
is essentially the "gauge-fixing" condition (\ref{mat-pot}), which uses
the available conformal freedom in choosing the metric to make
it real, and then fixes this conformal ambiguity.

Thus, once our complex theory is written for the real physical metric
and the non-dynamical fields are integrated out, we get two
copies of a real theory (as real and imaginary parts). As we have
seen before, even though these two copies have different symplectic
structure and different Hamiltonians, they have the same content.
It is then sufficient to restrict one's attention to say only the real part
of the arising action. This is done in (\ref{action-A-real}),
where the imaginary part is taken so
that the usual Einstein-Hilbert action can arise when the
non-propagating scalars are set to zero. The reason why
imaginary, not real part needs to be taken here is that the
wedge product of two-forms rewritten using the self-duality
of $\Sigma^a_{\mu\nu}$, gives rise to a factor of $\im$,
and it is this extra imaginary unit that leads to ${\rm Im}$
part of the action being taken. 

\section{Effective action}
\label{sec:eff}

In this section we compute the effective metric action by integrating out the non-propagating
scalars present in our theory. This computation can only be done perturbatively
in powers of what can be called "non-metricity", which is just a collection of fields that measure
the departure of our theory from GR. Thus, our starting point is to expand the action
in powers of "non-metricity". 

We assume that the reality conditions on the (conformal) metric
are imposed, that the full physical metric is real, and that the "gauge-fixing"
condition $R(m)=1$  is imposed. We then write the
resulting (complex) action for a real metric and (complex) fields $m_{ab}$. This action depends on
the complex fields $m_{ab}$ holomorphically, and an appropriate real part
will later have to be taken to obtain a real action. We will continue to denote
the holomorphic action by ${\cal S}$ so that it is clear whether the final projection
to a real theory has been taken or not. 

\subsection{Expansion in "non-metricity"}

In this subsection we would like to use the formulation (\ref{action-m}) to obtain
an expansion of the action around the "metric" point with $m_{ab}=\delta_{ab}$.
We have already seen that the zeroth-order action one gets is just the Einstein-Hilbert one,
so we would like to compute the higher-order terms. We take:
\be\label{m-expans}
m_{ab}=\delta_{ab} + H_{ab} + \kappa\delta_{ab} {\rm Tr}(H^2)+O(H^3),
\ee
where $H_{ab}$ is a symmetric, traceless matrix, and $\kappa$ is a parameter that
comes from the "gauge-fixing" condition $R(m)=1$. To see how this comes about,
we note that we can always parameterize
\be
R(m)=\frac{{\rm Tr}(m)}{3} U_m(H), \qquad m_{ab}=\frac{{\rm Tr}(m)}{3} (\delta_{ab}+H_{ab}),
\ee
where $H_{ab}$ is the tracefree part of $m_{ab}$ and $U_m(\cdot)$ is
an arbitrary function that must be thought of as being supplied by the matter sector action. 
The condition $R(m)=1$ then means that the trace part
of $m_{ab}$ is the function of the tracefree part $H_{ab}$: ${\rm Tr}(m)=3/U_m(H)$.
The function $U_m(H)$ starts with the term proportional to ${\rm Tr}H^2$, and this
is where the parameter $\kappa$ in (\ref{m-expans}) comes from.

We would like to
expand the action up to order $H^2$ terms. We work with the holomorphic action,
and the "non-metricity" field $H_{ab}$ is complex. 
Let us first derive the linear terms. To this end we note that the one-forms $\rho_{ab}$ are
order $H$, and so the second term in (\ref{action-m}) only contributes at the second order.
Thus, to first order we only need to work out the expansion of the first term (below we shall see
that there is no contribution at order $H$ from the potential either). We have:
\be
{\rm det}(b)=\sqrt{{\rm det}(m)}= 1- \frac{1}{4}{\rm Tr}(H^2) + O(H^3), 
\\ \nonumber
(m^{-1})^{ab} = \delta^{ab} - H^{ab} + H^{ac}H^{cb} -  \kappa\delta_{ab} {\rm Tr}(H^2)+O(H^3),
\ee
and thus
\be
{\rm det}(b) 
\left( \delta^{ab} (m^{-1})^{cd}\delta_{cd}-(m^{-1})^{ab}\right)  =
2\delta^{ab} + H^{ab} - H^{ac}H^{cb} + \frac{1-4\kappa}{2}\delta^{ab}
{\rm tr}(H^2)+O(H^3).
\ee
The potential can be expanded as follows:
\be
V(m)= \frac{g_2}{2l^2} {\rm Tr}(H^2) + \frac{g_3}{3l^2}{\rm Tr}(H^3) + O(4),
\ee
where $g_2,g_3$ are numerical coefficients that can, in principle, be complex, 
$l^2$ is a length scale (real) required to give the
potential the correct ($1/L^2$) dimension, and the reason why we have kept the cubic term 
will become clear below. Note that we have set the cosmological constant to zero
for simplicity.

The formulas above allow to expand all but the second term in (\ref{action-m}). The expansion
of this term to its first non-trivial -- second -- order in $H^2$ is as follows. 
As we have already noted each $\rho_{ab}$
is order $H$, and so each occurrence of the metric $m_{ab}$ and its inverse 
(apart from that in $\rho_{ab})$ may be replaced by the zeroth-order metric $\delta_{ab}$.
Thus, we need to compute
\be
\frac{1}{2}\int d^4x\sqrt{g}\, \Sigma^{a\,\mu\nu} \epsilon^{abc} 
\rho_{\mu\,be}\rho_{\nu\,cf} \delta^{ef},
\ee
with the $\rho_{ab}$ one forms given by:
\be
\rho_{\mu\,ab} = \frac{1}{2}\cD_\mu H_{ab} + \frac{1}{4}\epsilon_{abc}
\Sigma^{c\,\alpha\beta}\Sigma^e_{\alpha\mu}
\Sigma^f_{\beta\gamma} \cD^\gamma H_{ef},
\ee
where we have rewritten the answer in a form convenient for computations.

There are three terms to compute. The first one is:
\be
\frac{1}{8}\int d^4x\sqrt{g}\, \Sigma^{a\,\mu\nu} \epsilon^{abc} 
\cD_{\mu}H_{be}\cD_{\nu}H_{ce}.
\ee
We integrate by parts and use the fact that $\cD_\mu\Sigma^{a\,\mu\nu}=0$ to
write the result as:
\be
-\frac{1}{8}\int d^4x\sqrt{g}\, \Sigma^{a\,\mu\nu} \epsilon^{abc} 
H_{be}\cD_\mu\cD_{\nu}H_{ce}= -\frac{1}{8}\int d^4x\sqrt{g}\, \Sigma^{a\,\mu\nu} 
\epsilon^{abc} H_{be} (\epsilon_{cdf}F^d_{\mu\nu}H_{fe} +\epsilon_{edf}F^d_{\mu\nu}H_{cf}),
\ee
where we have used the definition of the curvature. Simple algebra gives:
\be\label{term-1}
-\frac{1}{8}\int d^4x\sqrt{g}\, \Sigma^{a\,\mu\nu}F^b_{\mu\nu}(2\delta^{ab}{\rm Tr}(H^2)
-3H^{ac}H^{cb}).
\ee

The second term to compute is:
\be
\frac{1}{8}\int d^4x\sqrt{g}\, \Sigma^{a\,\mu\nu} \epsilon^{abc} 
\cD_{\mu}H_{be}\epsilon_{ced}\Sigma^{d\,\alpha\beta}\Sigma^m_{\alpha\nu}
\Sigma^n_{\beta\gamma}\cD^\gamma H_{mn}.
\ee
Integrating by parts and expanding the product of two $\epsilon$'s we get:
\be\label{term-2}
-\frac{1}{8}\int d^4x\sqrt{g}\, \Sigma^{a\,\mu\nu}H_{ab}\cD_\mu \Sigma^{b\,\alpha\beta}
\Sigma^m_{\alpha\nu} \Sigma^n_{\beta\gamma}\cD^\gamma H_{mn}.
\ee
We can further simplify this term by first expanding the three $\Sigma$'s under
the first derivative operator. The result is available in (\ref{rho}). The term proportional to 
$\delta^{mn}$ gets contracted with a traceless $H_{mn}$ and gives no contribution and we get:
\be
-\frac{1}{4}\int d^4x\sqrt{g}\, \Sigma^{a\,\mu\nu}H_{ab}\cD_\mu 
\Sigma^c_{\nu\rho} \cD^\rho H_{bc}.
\ee
We can now take the first $\Sigma$ under the operator of covariant derivative and 
expand the product of two $\Sigma$'s using their algebra. We get:
\be
-\frac{1}{4}\int d^4x\sqrt{g}\, H_{ab}\cD_\mu 
(-\delta^{ac}g^\mu_\rho + \epsilon^{acd}\Sigma^{d\,\mu}_{\quad\rho}) \cD^\rho H_{bc}.
\ee
The second term here can again be reduced to the curvature. Overall we get for
this term:
\be\label{term-2-final}
-\frac{1}{4}\int d^4x\sqrt{g}\, \left((\cD_\mu H_{ab})^2 + \Sigma^{a\,\mu\nu}F^b_{\mu\nu}
(2\delta^{ab} {\rm Tr}(H^2) - 3H^{ac}H^{cb})\right).
\ee

The third term is:
\be
\frac{1}{32}\int d^4x\sqrt{g}\, \Sigma^{a\,\mu\nu} \epsilon^{abc} 
\epsilon_{bed}\Sigma^{d\,\alpha\beta}\Sigma^m_{\alpha\mu}
\Sigma^n_{\beta\gamma}\cD^\gamma H_{mn}
\epsilon_{cef}\Sigma^{f\,\rho\sigma}\Sigma^p_{\rho\nu}
\Sigma^q_{\sigma\delta}\cD^\delta H_{pq}.
\ee
Expanding the product of two $\epsilon$'s we get:
\be
\frac{1}{32}\int d^4x\sqrt{g}\, \Sigma^{a\,\mu\nu} \epsilon^{abc} 
\Sigma^{b\,\alpha\beta}\Sigma^m_{\alpha\mu}
\Sigma^n_{\beta\gamma}\cD^\gamma H_{mn}
\Sigma^{c\,\rho\sigma}\Sigma^p_{\rho\nu}
\Sigma^q_{\sigma\delta}\cD^\delta H_{pq}.
\ee
We can now use the identity (\ref{ident-s}) for instance for the first three $\Sigma$'s.
Let us also integrate by parts in the result. We get:
\be\label{term-3}
\frac{1}{16}\int d^4x\sqrt{g}\, \Sigma^{a\,\mu\nu}H_{ab}\cD_\mu \Sigma^{b\,\alpha\beta}
\Sigma^m_{\alpha\nu} \Sigma^n_{\beta\gamma}\cD^\gamma H_{mn},
\ee
which is precisely of the same form as (\ref{term-2}), but with a different numerical coefficient.
We have already simplified this term in (\ref{term-2-final}).

We can now combine everything together and write the result for the expansion of the
action in powers of $H$:
\be\label{action-H2}
S = \frac{1}{4}\int d^4x \sqrt{g}\, \Sigma^{a\,\mu\nu} F^b_{\mu\nu} 
\left( 2\delta^{ab} + H^{ab} + 2H^{ac}H^{cb} - \frac{3+4\kappa}{2}\delta^{ab}
{\rm tr}(H^2)\right)
\\ \nonumber
-\frac{1}{8}\int d^4x\sqrt{g}\, (\cD_\mu H_{ab})^2+O(H^3)
\\ \nonumber
- \int d^4x \sqrt{g}\left( \frac{g_2}{2l^2} {\rm Tr}(H^2) + \frac{g_3}{3l^2}{\rm Tr}(H^3)
+O(H^4)\right),
\ee
where we have kept different powers of $H$ in the "kinetic" and "potential" terms.
The quadratic part of this action is the already familiar to us action (\ref{S-lin-H}).
It can already be anticipated that the limit to GR can be obtained by taking
the length scale $l\to 0$. This makes the potential for the "non-metricity" fields
$H^{ab}$ infinitely steep, and thus effectively sets them to zero, giving GR.
We will see this explicitly when we compute the effective action. Alternatively,
to get GR one can simply pass to the low energy limit $E\ll (1/l)$, where $E$ is
the energy of a typical field configuration. In this limit the fields $H^{ab}$ are
infinitely massive, and should be set to zero, which gives GR.

\subsection{Effective action}

Let us now write down the equation that one obtains by varying (\ref{action-H2})
with respect to $H^{ab}$. It is convenient to introduce:
\be\label{F-ab}
F^{ab}:=\Sigma^{a\,\mu\nu}F^b_{\mu\nu}=\frac{1}{2}\Sigma^{a\,\mu\nu}R_{\mu\nu\rho\sigma}
\Sigma^{b\,\rho\sigma},
\ee
where to write the second equality we have used (\ref{F-Riemm}). This, in particular,
shows that $F^{ab}$ is a symmetric matrix. Then the equation for $H^{ab}$ takes the 
following form:
\be
\left( F^{ab} + 2F^{ac} H^{cb} + 2H^{ac}F^{cb} \right)_{tf}-(3+4\kappa) {\rm Tr}(F) H^{ab}
+ \cD^\mu\cD_\mu H^{ab} \\ \nonumber
= \frac{4g_2}{l^2} H^{ab} + \frac{4g_3}{l^2}
\left( H^{ac}H^{cb}\right)_{tf},
\ee
plus higher-order terms.
We can now solve for $H^{ab}$ in terms of $F^{ab}$. To first order in curvature we get:
\be\label{H-1*}
H_{(1)}^{ab} = \frac{l^2}{4g_2}(F^{ab})_{tf}.
\ee
It is thus clear that we are solving for the "non-metricity" $H^{ab}$ in terms of
an expansion in the small parameter, which is the product of the length scale
$l^2$ times the typical (sectional) curvature of our metric. In the approximation when
this dimensionless quantity is small the terms $l^2F$ are first order, and so are the terms
$H$. This is why it was consistent to keep only the terms of order $H^2$ in the "kinetic"
part of the action and terms $H^3$ in the potential part, for the kinetic part of the action
contains an additional factor of the curvature that makes it the same order as the
$H^3$ term in the potential part. 

Solving to the second order in curvature we find:
\be\label{H-2*}
H_{(2)}^{ab} = \frac{l^4}{4g_2^2}\left( F^{ac}F^{cb} \left( 1- \frac{g_3}{4g_2}\right) 
- F^{ab} {\rm Tr}(F) \left( \frac{1}{3} - \frac{g_3}{6g_2} + \frac{3+4\kappa}{4}\right) \right)_{tf}
+\frac{l^4}{16g_2^2} \cD^\mu\cD_\mu F^{ab}\Big|_{tf}.
\ee
We could now also solve for the order $H^{ab}_{(3)}$, but we do not need it as our aim
is to obtain the action only to third order in curvature. The corresponding Lagrangian is:
\be
\frac{1}{4} {\rm Tr}\left( F(2\,{\rm Id}+H_{(1)} + H_{(2)} + 2H_{(1)}^2
-\frac{3+4\kappa}{2}{\rm Id} \,{\rm Tr}(H_{(1)})^2 )\right) 
\\ \nonumber
-\frac{1}{8}{\rm Tr}(\cD_\mu H_{(1)})^2 - \frac{g_2}{2l^2}{\rm Tr}(H_{(1)}^2
+ 2H_{(1)}H_{(2)}) - \frac{g_3}{3l^2}{\rm Tr}(H_{(1)})^3.
\ee
To first order in curvature this gives $(1/2){\rm Tr}(F)$, which, as we already discussed,
is just the Einstein-Hilbert Lagrangian. To second order (or "one-loop") in curvature we have:
\be\label{act-1-loop}
{\cal L}^{(1)} = \frac{1}{4}{\rm Tr}(FH_{(1)}) - \frac{g_2}{2l^2} {\rm Tr}(H_{(1)})^2 = 
\frac{l^2}{32g_2}{\rm Tr}(F|_{tf})^2.
\ee
The Lagrangian to third order in curvature (two-loop) is given by:
\be\label{act-2-loop}
{\cal L}^{(2)} = \frac{1}{4}{\rm Tr}(F(H_{(2)}+2H_{(1)}^2)) - \frac{3+4\kappa}{8}{\rm Tr}(F)
{\rm Tr}(H_{(1)})^2 - \frac{1}{8}{\rm Tr}(\cD_\mu H_{(1)})^2
\\ \nonumber
 - \frac{g_2}{l^2}{\rm Tr}(H_{(1)}H_{(2)}) - \frac{g_3}{3l^2}{\rm Tr}(H_{(1)})^3
 \\ \nonumber
 = \frac{l^4}{32g_2^2}\left[ \left(1-\frac{g_3}{6g_2}\right) {\rm Tr}(F|_{tf})^3 
 -\frac{3+4\kappa}{4}{\rm Tr}(F){\rm Tr}(F|_{tf})^2 
 -\frac{1}{4} {\rm Tr}(\cD_\mu F|_{tf})^2\right].
\ee

The effective metric Lagrangian obtained is still complex, as it depends on the
complex self-dual part of the Riemann curvture. Its real part needs
to be taken to obtain a real metric theory. We shall rewrite everything in
explicitly metric terms below.

\subsection{Curvature computations}

In this subsection we will express the curvature invariants that appeared above 
in the usual spacetime index notations. 

Let us start with the invariant that appears at "one-loop" level ${\rm Tr}(F|_{tf})^2$. Note
that the tracefree part of the matrix $F^{ab}$ given by (\ref{F-ab}) 
is just the self-dual part of the Weyl curvature tensor that is unconstrained by the
Einstein equations. We now use the relation (\ref{F-ab}) to the usual Riemann curvature
to compute the invariant of interest. We have:
\be
{\rm Tr}(F|_{tf})^2 = {\rm Tr}(F(F-(1/3){\rm Id}\,{\rm Tr}(F))) = {\rm Tr}F^2 -\frac{1}{3}({\rm Tr}F)^2.
\ee
Recall now, see (\ref{R-scalar}), that ${\rm Tr}(F)=R$, the Ricci scalar. The other quantity
can be computed using (\ref{proj-s}). We have:
\be\label{PRPR}
{\rm Tr}F^2= 4P^{+\,\gamma\delta\mu\nu}R_{\mu\nu\rho\sigma}P^{+\,\rho\sigma\alpha\beta}
R_{\alpha\beta\gamma\delta}= 
\\ \nonumber
R^{\mu\nu\rho\sigma}R_{\mu\nu\rho\sigma}
+\frac{1}{i} R^{\mu\nu}_{\quad\rho\sigma}\epsilon^{\rho\sigma\alpha\beta}R_{\alpha\beta\mu\nu}
-\frac{1}{4}\epsilon^{\gamma\delta\mu\nu}R_{\mu\nu\rho\sigma}\epsilon^{\rho\sigma\alpha\beta}
R_{\alpha\beta\gamma\delta}.
\ee
Expanding the product of two $\epsilon$'s we get:
\be\label{ee-RR}
\epsilon^{\gamma\delta\mu\nu}R_{\mu\nu\rho\sigma}\epsilon^{\rho\sigma\alpha\beta}
R_{\alpha\beta\gamma\delta}= -4R^{\mu\nu\rho\sigma}R_{\mu\nu\rho\sigma}
+16R^{\mu\nu}R_{\mu\nu}-4R^2,
\ee
and so overall
\be\label{F2}
{\rm Tr}F^2=2R^{\mu\nu\rho\sigma}R_{\mu\nu\rho\sigma}-4R^{\mu\nu}R_{\mu\nu}+R^2 
+\frac{1}{i} R^{\mu\nu}_{\quad\rho\sigma}\epsilon^{\rho\sigma\alpha\beta}R_{\alpha\beta\mu\nu}.
\ee

We now note that the quantity $R_{\alpha\beta\gamma\delta}
\epsilon^{\gamma\delta\mu\nu}R_{\mu\nu\rho\sigma}\epsilon^{\rho\sigma\alpha\beta}$ 
is a total derivative. So, modulo a total derivative we can trade the invariant 
$R^{\mu\nu\rho\sigma}R_{\mu\nu\rho\sigma}$
for other curvature invariants. Thus, from (\ref{ee-RR}) it follows that modulo a topological term:
\be
R^{\mu\nu\rho\sigma}R_{\mu\nu\rho\sigma}\approx 4R^{\mu\nu}R_{\mu\nu}-R^2.
\ee
The quantity 
$R^{\mu\nu}_{\quad\rho\sigma}\epsilon^{\rho\sigma\alpha\beta}R_{\alpha\beta\mu\nu}$,
which the imaginary part of ${\rm Tr}F^2$ is proportional to, is also a total derivative.
Using this, we can finally write our answer for the "one-loop" term (modulo a
surface term):
\be\label{curv-1-loop}
{\rm Tr}(F|_{tf})^2\approx 4R^{\mu\nu}R_{\mu\nu}-\frac{4}{3}R^2.
\ee
Thus, assuming that the length parameter $l$ is real, we
get for our effective metric action at order $l^2$:
\be
{\cal L}^{(1)} = \frac{l^2}{8{\rm Re}(g_2)} (R^{\mu\nu}R_{\mu\nu}-\frac{1}{3}R^2).
\ee

At "two-loop" order we need to compute two terms. One of them is:
\be
{\rm Tr}(F|_{tf})^3 = {\rm Tr}((F-(1/3){\rm Id}\,{\rm Tr}(F))
(F-(1/3){\rm Id}\,{\rm Tr}(F))(F-(1/3){\rm Id}\,{\rm Tr}(F))) 
\\ \nonumber
={\rm Tr}F^3 - {\rm Tr}F^2{\rm Tr}(F)+ \frac{2}{9}({\rm Tr}(F))^3.
\ee 
We thus need to compute:
\be
{\rm Tr}F^3= 8P^{+\,\gamma\delta\mu\nu}R_{\mu\nu\rho\sigma}P^{+\,\rho\sigma\alpha\beta}
R_{\alpha\beta\eta\xi}P^{+\,\eta\xi\lambda\tau}R_{\lambda\tau\gamma\delta}
\\ \nonumber
= R^{\gamma\delta}_{\quad\rho\sigma}R^{\rho\sigma}_{\quad\alpha\beta}
R^{\alpha\beta}_{\quad\gamma\delta} + \frac{3}{2i}\epsilon^{\gamma\delta}_{\quad\mu\nu}
R^{\mu\nu}_{\quad\rho\sigma}
R^{\rho\sigma}_{\quad\alpha\beta}R^{\alpha\beta}_{\quad\gamma\delta}
-\frac{3}{4}R^{\gamma\delta}_{\quad\rho\sigma}\epsilon^{\rho\sigma\alpha\beta}
R_{\alpha\beta\eta\xi}\epsilon^{\eta\xi\lambda\tau}R_{\lambda\tau\gamma\delta}
\\ \nonumber
-\frac{1}{8i}\epsilon^{\gamma\delta\mu\nu}R_{\mu\nu\rho\sigma}\epsilon^{\rho\sigma\alpha\beta}
R_{\alpha\beta\eta\xi}\epsilon^{\eta\xi\lambda\tau}R_{\lambda\tau\gamma\delta}.
\ee
The product of two $\epsilon$'s in the third and fourth terms can be expanded. We have:
\be
\epsilon^{\rho\sigma\alpha\beta}
R_{\alpha\beta\eta\xi}\epsilon^{\eta\xi\lambda\tau}R_{\lambda\tau\gamma\delta}=
-4\left(R^{\rho\sigma\alpha\beta}R_{\alpha\beta\gamma\delta} 
+4R^{\alpha[\rho}R^{\sigma]}_{\,\,\,\alpha\gamma\delta}+R R^{\rho\sigma}_{\quad\gamma\delta}
\right).
\ee
Thus, overall,
\be\label{F3}
{\rm Tr}F^3=4R^{\gamma\delta}_{\quad\rho\sigma}R^{\rho\sigma}_{\quad\alpha\beta}
R^{\alpha\beta}_{\quad\gamma\delta} - 12 R^{\gamma\delta\rho\sigma}
R_\rho^{\,\,\alpha}R_{\alpha\sigma\gamma\delta}+3R R^{\rho\sigma\gamma\delta}
R_{\rho\sigma\gamma\delta}
\\ \nonumber
+2i \epsilon^{\gamma\delta}_{\quad\mu\nu} R^{\mu\nu}_{\quad\rho\sigma}
R^{\rho\sigma}_{\quad\alpha\beta}R^{\alpha\beta}_{\quad\gamma\delta}
-2i \epsilon^{\gamma\delta}_{\quad\mu\nu} R^{\mu\nu}_{\quad\rho\sigma}
R_\rho^{\,\,\alpha}R_{\alpha\sigma\gamma\delta}
+\frac{1}{2i}R\epsilon^{\gamma\delta}_{\quad\mu\nu} R^{\mu\nu}_{\quad\rho\sigma}
R^{\rho\sigma}_{\quad\gamma\delta},
\ee
and
\be
{\rm Tr}(F|_{tf})^3=4R^{\gamma\delta}_{\quad\rho\sigma}R^{\rho\sigma}_{\quad\alpha\beta}
R^{\alpha\beta}_{\quad\gamma\delta} - 12 R^{\gamma\delta\rho\sigma}
R_\rho^{\,\,\alpha}R_{\alpha\sigma\gamma\delta}+R R^{\rho\sigma\gamma\delta}
R_{\rho\sigma\gamma\delta}+4RR^{\mu\nu}R_{\mu\nu} - \frac{7}{9}R^3
\\ \nonumber
+2i \epsilon^{\gamma\delta}_{\quad\mu\nu} R^{\mu\nu}_{\quad\rho\sigma}
R^{\rho\sigma}_{\quad\alpha\beta}R^{\alpha\beta}_{\quad\gamma\delta}
-2i \epsilon^{\gamma\delta}_{\quad\mu\nu} R^{\mu\nu}_{\quad\rho\sigma}
R_\rho^{\,\,\alpha}R_{\alpha\sigma\gamma\delta}
-\frac{1}{2i}R\epsilon^{\gamma\delta}_{\quad\mu\nu} R^{\mu\nu}_{\quad\rho\sigma}
R^{\rho\sigma}_{\quad\gamma\delta},
\ee
This contains a non-trivial imaginary part, which is odd under orientation reversal. 

We can also compute the other invariant. To this end we extend the $\Sigma^a$-compatible
derivative operator $\cD$ into one $\tilde{\cD}$ that acts on both the internal and spacetime 
indices and such that $\tilde{\cD}_\mu \Sigma^a_{\rho\sigma}=0$. Since its action on
quantities without spacetime indices is the same as that of $\cD$, we can freely replace
$\cD$ by $\tilde{\cD}$ in ${\rm Tr}(\cD_\mu F\Big|_{tf})^2$. We can then rewrite
$F^{ab}$ as in (\ref{F-ab}) and then pull the $\Sigma^a$ matrices out of the
derivative operators. These then act only on spacetime indices, where their action
is that of the metric-compatible ones. Thus, we get:
\be
{\rm Tr}(\cD_\mu F\Big|_{tf})^2= 4P^{+\,\gamma\delta\mu\nu}\nabla^\tau 
R_{\mu\nu\rho\sigma}P^{+\,\rho\sigma\alpha\beta}\nabla_\tau R_{\alpha\beta\gamma\delta}
- \frac{1}{3}\nabla^\mu R \nabla_\mu R.
\ee
This has essentially been computed in (\ref{F2}), so the result is:
\be
{\rm Tr}(\nabla_\mu F\Big|_{tf})^2=2\nabla^\tau R^{\mu\nu\rho\sigma}
\nabla_\tau R_{\mu\nu\rho\sigma}-4\nabla^\tau R^{\mu\nu}\nabla_\tau R_{\mu\nu}+
\frac{2}{3}\nabla^\mu R\nabla_\mu R \\ \nonumber +\frac{1}{i} \nabla^\tau 
R^{\mu\nu}_{\quad\rho\sigma}\epsilon^{\rho\sigma\alpha\beta}
\nabla_\tau R_{\alpha\beta\mu\nu}.
\ee
We can now use the Lichnorowicz identity that says that modulo a surface term
\be
2\nabla^\tau R^{\mu\nu\rho\sigma}
\nabla_\tau R_{\mu\nu\rho\sigma} \approx -
R^{\gamma\delta}_{\quad\rho\sigma}R^{\rho\sigma}_{\quad\alpha\beta}
R^{\alpha\beta}_{\quad\gamma\delta}
\ee
to put the quantity in question into the form $(Riemann)^3$ plus terms that
vanish on shell.

Ignoring the terms containing $R_{\mu\nu}$ and $R$ that vanish on-shell,
taking the real part of the action, and assuming for simplicity that $g_2,g_3$ are
real, we get the following order $l^4$ effective action:
\be
{\cal L}^{(2)} = \frac{l^4}{32 g_2^2} \left( 4(1-\frac{g_3}{6g_2})+\frac{1}{4}\right)
R^{\gamma\delta}_{\quad\rho\sigma}R^{\rho\sigma}_{\quad\alpha\beta}
R^{\alpha\beta}_{\quad\gamma\delta}.
\ee
If either of the constants $g_2,g_3$ is complex, then the action also picks up
the parity-odd terms containing the $\epsilon$-tensor. This finishes our demonstration
of the fact that the $(Riemann)^3$ term is contained in our effective metric
theory. Note that, as we have expected, all the corrections to GR come with
an appropriate power of the length scale $l$ in front. Thus, the limit to GR
is obtained by taking this length scale to zero, which corresponds to the
original potential $V(\cdot)$ being infinitely steep, or, equivalently, by considering the
low-energy limit of the theory.

\section{Field redefinitions}
\label{sec:redef}

In this section we would like to show that the higher-derivative effective
metric Lagrangian obtained above can
be brought into the standard EH form by a certain redefinition of the metric field. 
We do this in two steps. First, recall that the higher-derivative metric action was obtained
from a BF-type theory with extra scalars by integrating out the scalars. So, we go back
to the formulation with extra scalars and study some available field redefinitions at
the BF-level. After the scalars are integrated out the BF-level transformation
becomes a field redefinition that acts on spacetime metrics, and we compute it
using perturbation theory around Minkowski background.

\subsection{Topological shift symmetry}

The availability of the field redefinition that makes a class of metric theories
with two propagating DOF possible has its origins in the fact that GR in
Pleba\'nski formulation takes the form of BF theory, and the BF term 
has the "topological" symmetry $B^i\to B^i+d_A \eta^i$. This underlying
reason for the field redefinition mapping our theory into GR has been
identified already in \cite{Freidel:2008ku}. However, this reference only
treated the linearized theory. We extend this result to the full non-linear
theory. As before, we first deal with the complex theory, and later take the
real projection.

The topological symmetry of the BF term makes two different
parameterizations of the two-form field possible. One parameterization is
the already discussed $B^i = b^i_a \Sigma^a$, where $\Sigma^a$ are the metric
two-forms in the conformal class defined by $B^i$, and $b^i_a$ are
the non-propagating scalars that we have learned how to integrate out. 
The other parameterization arises by looking for $\eta^i$ such that
\be\label{B-split}
B^i = \tilde{\Sigma}^i + d_{\tilde{\gamma}} \eta^i,
\ee
where $\tilde{\Sigma}^i$ are some other metric two-forms, $\tilde{\gamma}^i$ are
the associated metric-compatible $\SOC$-connection one-forms, and
$\eta^i$ is some one-form valued in the Lie algebra. As we shall see, 
the representation (\ref{B-split}) is possible for any $B^i$, but is not unique,
for one can always perform a diffeomorphism on $\tilde{\Sigma}^i$ and correct
its effect by changing $\eta^i$ without changing $B^i$. However, modulo diffeomorphisms,
the split is unique, at least in the Riemannian signature when the associated
differential equation is elliptic. Thus, at least in the Riemannian signature,
given any two-form field $B^i$, there exists
a unique (modulo diffeomorphisms) metric $\tilde{g}$ such that the corresponding
metric two-forms $\tilde{\Sigma}^i$ can be obtained from $B^i$ by shifting it
with a derivative of a Lie-algebra valued one-form. In the case of Lorentzian signature
that is of more physical significance we will be able to find a unique formal solution
that involves a $1/\Box$ operator. 

Before we discuss this statement further, let us 
note that, unlike the process that we have used to deduce a conformal metric from $B^i$ 
(looking at the subspace in the space of two-forms spanned by $B^i$), the representation
(\ref{B-split}) allows us to deduce not just a conformal class, but a full metric.  
Note also that the metric $\tilde{g}$ that appears 
via (\ref{B-split}) from $B^i$ is, in general, different from the one whose conformal class
is deduced directly from $B^i$. Thus, a general $B^i$ can be said to carry information
about two different metrics, or rather about a natural conformal metric arising via
$B^i=b^i_a\Sigma^a$ and a natural metric arising via (\ref{B-split}). 

To convince oneself that there is enough parameters in $\eta^i$ to achieve the
decomposition (\ref{B-split}), let us recall that a general $B^i$ is characterized by
18 parameters, of which, however, only 15 are "physical", with 3 others being 
${\rm SO}(3)$-gauge. We would like to see how 5 of these parameters can be "killed"
to obtain a metric two-form via a shift (\ref{B-split}). Let us see how many
parameters are there at our disposal. We have a Lie-algebra-valued one-form field,
which has 12 components. However, 3 of these correspond to a gauge freedom in
choosing $\eta^i$. Indeed, $\eta^i\to \eta^i+D_{\tilde{\gamma}}\phi^i$ affects
$D_{\tilde{\gamma}}\eta^i$ only by shifting it with a two-form
$\epsilon^{ijk} F^j(\tilde{\gamma}) \phi^k$ proportional to the curvature of $\tilde{\gamma}$. 
This is zero for a flat $\tilde{\gamma}$ and is thus invisible at least in the lowest
order of a perturbative expansion around a flat background. Thus, there is only 
9 "physical" parameters in $\eta^i$. However,
4 more of them correspond to the possibility of performing diffeomorphisms on
$\tilde{\Sigma}^i$ and shifting $\eta^i$ at the same time (below we shall see
how this is described in details). Thus, only 5 parameters of $\eta^i$ correspond
to those that really change the two-form field, and this is exactly enough to 
kill all 5 "non-metric" components of a general two-form field $B^i$. Below we shall 
see all this in details in perturbation theory. 

Let us give a more detailed treatment of the problem of finding a decomposition
(\ref{B-split}) for a given $B^i$. This is a problem of finding simultaneously a
metric $\tilde{g}_{\mu\nu}$ as well as a Lie-algebra-valued one-form $\eta^i$
so that (\ref{B-split}) holds. The fact that $\tilde{\Sigma}^i$ is metric implies
that the following 5 differential equations on $\eta^i$:
\be\label{eta-eqs-gen}
(B^i - D_{\tilde{\gamma}}\eta^i)\wedge (B^j - D_{\tilde{\gamma}}\eta^j) \sim\delta^{ij}.
\ee
These must be supplemented with some gauge-fixing conditions on $\eta^i$. 
Convenient conditions fixing the diffeomorphism freedom can be taken to be:
\be\label{eta-gf-2}
\epsilon^{ijk} \tilde{\Sigma}^{j\,\mu\nu} D_{\tilde{\gamma}\,\mu}\eta^k_\nu=0,
\qquad \mathrm{and} \qquad 
\tilde{\Sigma}^{i\,\mu\nu} D_{\tilde{\gamma}\,\mu}\eta^i_\nu=0.
\ee
A condition that fixes the "gauge" freedom in choosing $\eta^i$ can be taken to be:
\be\label{eta-gf-1}
\tilde{\nabla}^\mu \eta_\mu^i = 0.
\ee
Here $\tilde{\nabla}^\mu$ is the covariant derivative operator $D^\mu_{\tilde{\gamma}}$
extended to act not just on internal, but also on the spacetime indices and preserve the
metric $\tilde{g}_{\mu\nu}$. Note that the conditions (\ref{eta-gf-2}) can be written using
$\tilde{\nabla}^\mu$ instead of $D^\mu_{\tilde{\gamma}}$.

It is obvious that (\ref{eta-gf-1}) fixes the $\eta^i\to\eta^i+D_{\tilde{\gamma}}\phi^i$
freedom, while to convince oneself that (\ref{eta-gf-2}) are good gauge-fixing conditions
for the diffeomorphisms it is enough to note that these act on $\tilde{\Sigma}^i$ by
shifting it with $D_{\tilde{\gamma}} \iota_\xi \tilde{\Sigma}^i$, where $\iota_\xi$ is
the interior multiplication with a vector field $\xi$. One can then see that the gauge
(\ref{eta-gf-2}) can indeed be achieved by finding a diffeomorphism generated by 
a vector field with given $\tilde{\Sigma}^{i\,\mu\nu} D_{\tilde{\gamma}\,\mu} \xi_\nu$ as well as
$D_{\tilde{\gamma}\,\mu} \xi_\mu$.

Equations (\ref{eta-eqs-gen}), together with the gauge-fixing conditions
(\ref{eta-gf-2}), (\ref{eta-gf-1}) give 5+4+3 equations for 12 unknowns $\eta^i$.
To convince oneself that the can be solved let us describe a procedure that
works at least perturbatively. As the first step, let us multiply the relation (\ref{B-split})
by $\tilde{\Sigma}^{j\,\mu\nu}$ and use the fact that $\tilde{\Sigma}^i$ is metric
and so $\tilde{\Sigma}^{j\,\mu\nu}\tilde{\Sigma}^i_{\mu\nu}=4\delta^{ij}$. 
According to our gauge-fixing conditions (\ref{eta-gf-2}) the $ij$-matrix
$\tilde{\Sigma}^{j\,\mu\nu} D_{\tilde{\gamma}\,\mu}\eta^i_\nu$ is
symmetric and traceless. According to (\ref{B-split}) it is equal to
\be\label{X-ij}
2\tilde{\Sigma}^{j\,\mu\nu} D_{\tilde{\gamma}\,\mu}\eta^i_\nu=
\tilde{\Sigma}^{j\,\mu\nu} B^i_{\mu\nu} +
\frac{1}{2}\epsilon^{ijk} \epsilon^{klm} \tilde{\Sigma}^{l\,\mu\nu}B^m_{\mu\nu}
-\frac{1}{3}\delta^{ij} \tilde{\Sigma}^{k\,\mu\nu}B^k_{\mu\nu}\equiv 2X^{ij}.
\ee
Let us multiply this equation by $\tilde{\Sigma}^{j\,\rho\sigma}$ and then apply
the operator $D_{\tilde{\gamma}\,\rho}$ to both parts. In order to be able
to raise and lower spacetime indices under the covariant derivative it
is convenient to extend the operator $D_{\tilde{\gamma}}$ to $\tilde{\nabla}$.
On the left we get:
\be
8 \tilde{\nabla}_\rho \tilde{P}^{+\,\rho\sigma\mu\nu} \tilde{\nabla}_\mu\eta^i_\nu
= 2 \tilde{\nabla}^\rho \tilde{\nabla}_\rho \eta^i_\sigma - 
2 \tilde{\nabla}^\rho \tilde{\nabla}_\sigma \eta^i_\rho + (2/i) 
\epsilon^{\rho\sigma\mu\nu} \tilde{\nabla}_\rho \tilde{\nabla}_\mu \eta^i_\nu.
\ee
Using (\ref{eta-gf-1}) we can replace:
\be
\tilde{\nabla}^\rho \tilde{\nabla}_\sigma \eta^i_\rho = 
(\tilde{\nabla}^\rho \tilde{\nabla}_\sigma- \tilde{\nabla}_\sigma\tilde{\nabla}^\rho )\eta^i_\rho
= -\epsilon^{ijk} \tilde{F}^{j\,\rho}_{\sigma} \eta^k_\rho + \tilde{R}_\sigma^{\,\,\rho}\eta^i_\rho,
\ee
where $\tilde{F}^i$ is the curvature of $\tilde{\gamma}^i$ and $\tilde{R}_{\mu\nu}$ is the
Ricci tensor of $\tilde{g}_{\mu\nu}$. Similarly,
\be
\epsilon^{\rho\sigma\mu\nu} \tilde{\nabla}_\rho \tilde{\nabla}_\mu \eta^i_\nu
= \frac{1}{2}\epsilon^{ijk} \epsilon^{\rho\sigma\mu\nu} \tilde{F}^j_{\rho\mu}\eta^k_\nu.
\ee
Combining things together we get the sought differential equation for $\eta^i$:
\be\label{eta-eqn*}
\Delta \eta^i_\mu = \tilde{\Sigma}^{j}_{\mu\rho} \tilde{\nabla}^\rho X^{ij},
\ee
where the "Laplacian" $\Delta$ is defined as:
\be
\Delta \eta^i_\mu
:=-\tilde{\nabla}^\rho \tilde{\nabla}_\rho \eta^i_\mu
-2\epsilon^{ijk} \tilde{P}_\mu^{-\,\rho\alpha\beta} \tilde{F}^j_{\alpha\beta} \eta^k_\rho
+\tilde{R}_\mu^{\,\,\rho}\eta^i_\rho,
\ee
and the traceless, symmetric matrix $X^{ij}$ is as in (\ref{X-ij}).
The first and last terms in this second-order differential operator are exactly as in the 
Lichnerowicz Laplacian on one-forms, so it is an extension of this Laplacian to Lie-algebra-valued
one-forms. Note that for an Einstein metric $\tilde{g}$ the curvature
of $\tilde{\gamma}$ is self-dual, so the projection in the second term is zero. The Ricci
tensor is then proportional to the scalar curvature $\tilde{R}_{\mu\nu}=g_{\mu\nu}
\tilde{R}/4$, and so the non-derivative terms in the Riemannian signature 
Einstein case are positive on manifolds of positive scalar curvature. 
Thus, there are no non-trivial solutions to $\Delta \eta^i_\mu=0$ on compact
Riemannian signature Einstein manifolds of non-negative scalar curvature. 
This implies that if a solution to (\ref{eta-eqn*}) exists, then it is unique in this case.
For general metrics $\tilde{g}_{\mu\nu}$ the question of uniqueness has to be
investigated separately. 

One now has to solve for $\eta^i_a$ given $B^i$. The complication lies in the fact that
the metric $\tilde{g}_{\mu\nu}$ which is used to write the equation (\ref{eta-eqn*})
is itself unknown, and is determined via (\ref{B-split}) only after $\eta^i_\mu$ is
found. This complicated prescription makes it not obvious that a solution to this
problem exists. However, below we shall see that at least a perturbative solution
(when all quantities are expanded around the Minkowski spacetime background)
does exist. In the Riemannian signature case the problem reduces to an elliptic
equation of which there is a unique solution, order by order in perturbation. 
In the Lorentzian case we will find a formal solution that involves an inverse of
the $\Box$ operator. Let us now postpone the problem of finding a solution to 
(\ref{B-split}) and see what the availability of such a shift implies about our theory.

\subsection{Effect on the action}

Now with the representation (\ref{B-split}) being available, we can see what it implies
about the action of our theory. Recall that this has two terms: One is just the usual
BF-term, while the other is the potential term for the B-field. Substituting (\ref{B-split})
into the first term and using the fact that it is invariant under the shift symmetry,
we get back the BF-term but this time for a metric two-form $\tilde{\Sigma}^i$.
However, this is just the Einstein-Hilbert action for the corresponding metric,
as we have seen above. 

It remains to see what happens with the other, potential term. Of course, the potential
term is not invariant under the shift symmetry, and this is why there are propagating
degrees of freedom in this theory. So, if we substitute (\ref{B-split}) into the potential term,
we get a very complicated functional involving all powers of the derivatives of $\eta^a_\mu$,
which is very hard to deal with. Thus, it is probably not a very good idea to treat 
$\eta^a_\mu$ as a fundamental field in the action. However, there is another possibility.
Indeed, as we shall explicitly see below, the field $\eta^a_\mu$ in (\ref{B-split})
carries precisely the same information as the "non-metric" part of the B-field.
Thus, we can keep describing this non-metric part as we did before, using
the scalars $b^i_a$, and only change the variables for the metric part, describing it
not with some metric in the conformal class of $B^i$, but with the metric that appears 
in (\ref{B-split}). In other words, the following mixed parameterization of the B-field
is possible:
\be\label{B-mixed}
B^i = (\tilde{\Sigma}^i, b^i_a),
\ee
where $\tilde{\Sigma}^i$ is as introduced via (\ref{B-split}) and $b^i_a$ is
as before appears in:
\be
B^i = b^i_a \Sigma^a,
\ee
where we in addition impose the condition that the volume forms of the metrics
$g, \tilde{g}$ are the same:
\be\label{conf-cond}
\Sigma^a\wedge \Sigma^a = \tilde{\Sigma}^i\wedge \tilde{\Sigma}^i.
\ee
This is always possible using the conformal freedom in the choice of the metric $g$.
Note that (\ref{conf-cond}) fixes a metric in the conformal class defined by $B^i$
uniquely, and that this metric is, in general, distinct from the one fixed by
the condition (\ref{mat-pot}).

It is then clear that in the mixed parameterization (\ref{B-mixed}) the action
of the theory as a functional of the metric $\tilde{g}$ and the scalars $b^i_a$
has the following simple form:
\be\label{action-mixed}
S[\tilde{g}_{\mu\nu}, b^i_a] = \int d^4x \, \sqrt{-\tilde{g}} \left( \frac{1}{2} \tilde{R} - V(m^{ij})\right),
\ee
where $\tilde{g}$ is the determinant of the metric $\tilde{g}_{\mu\nu}$, $\tilde{R}$
is its Ricci scalar, $m^{ij}= b^i_a b^j_a$, and $V(\cdot)$ is the homogeneous order one
potential function. The action (\ref{action-mixed}) now has a form of the Einstein-Hilbert
action plus a potential term for the non-propagating scalars $m^{ij}$. Their field equations
set them to sit at the minimum of their potential, and the value of this minimum becomes
the cosmological constant of the usual Einstein general relativity for the metric 
$\tilde{g}_{\mu\nu}$. This discussion establishes that there exists a field parameterization
in which our theory is the usual (complex) general relativity. The only subtlety here is
the fact that the gauge-fixing condition (\ref{mat-pot}) that selects the "physical" metric from
a given conformal class is in general not the same as the condition (\ref{conf-cond}) used 
above. Thus, in general, before a transformation to the form (\ref{action-mixed}) can be applied,
one first has to apply to the metric $g_{\mu\nu}$ a conformal transformation.

At the same time, as we have seen previously, in a different field parameterization, namely
that in terms of the metric $g_{\mu\nu}$ and the scalars $b^i_a$ (or $m_{ab}$), the
resulting metric theory is quite non-trivial, containing an infinite number of higher curvature
invariants of the type familiar from the studies of renormalization in perturbative GR. 
We thus arrive at our central conclusion: there exists a field redefinition 
$g_{\mu\nu}\to \tilde{g}_{\mu\nu}$ of the metric tensor that maps an infinite expansion in 
curvature invariants of the metric $g_{\mu\nu}$ into the usual
Einstein-Hilbert action for the metric $\tilde{g}_{\mu\nu}$. This field redefinition is
the composition of a conformal transformation that makes (\ref{conf-cond}) satisfied, 
and then the topological shift symmetry discussed above. The central point is that this
field redefinition is, in general, non-local, for its determination involves, at the BF-level,
a solution of the differential equation (\ref{B-split}) for the $\eta^i_\mu$ field.
It is now our goal to see how all this works explicitly in first orders of the perturbation theory.

\subsection{First-order treatment: BF-level}

To first order in perturbations, a two-form field $B^i$ can be described as 
the Minkowski spacetime two-form field background $\delta^{ia} \Sigma^a_{0\,\mu\nu}$ plus
a perturbation that  can be decomposed into a metric and a non-metric part:
\be\label{pert-m-nm}
\delta^{ia} \delta^{(1)}\! B^i_{\mu\nu} = 
M^{(1)\,ab}\Sigma^b_{0\,\mu\nu} + \Sigma^{a\,\,\,\,\rho}_{0\,[\mu}\dot{g}_{\nu]\rho}.
\ee
Here $\dot{g}_{\mu\nu}$ is a perturbation of the metric described by the two-form field,
and $M^{(1)\,ab}$ is a perturbation of the non-metric part. Since the anti-symmetric part
of the matrix $M^{(1)\,ab}$ describes a perturbation that is a pure gauge, it is convenient
to gauge-fix this ${\rm SO}(3)$ freedom from the beginning by setting 
$M^{(1)\,ab}=M^{(1)\,(ab)}$.
There is also the conformal freedom ambiguity in (\ref{pert-m-nm}) for the change
in the conformal factor of the metric is described by the trace part 
$\dot{g}_{\mu\nu}\sim\eta_{\mu\nu}$, and this has the same form as the first non-metric
term in (\ref{pert-m-nm}). We can gauge-fix this freedom
by choosing the trace part $b^a_{\mu\nu}\sim \Sigma^a_{0\,\mu\nu}$,
where we have denoted $\delta^{ia} \delta^{(1)}\! B^i_{\mu\nu}\equiv b^a_{\mu\nu}$,
to correspond to a conformal transformation of the metric and not to a change in the 
non-metric part. In other words, we gauge-fix this freedom by setting the trace part of 
$M^{(1)\,ab}$ to zero. With these choices we have parametrized a perturbation $b^a_{\mu\nu}$ by
5 scalars $M^{(1)\,ab}, {\rm Tr}(M^{(1)})=0$, as well as 10 components of the metric perturbation
$\dot{g}_{\mu\nu}$, which overall gives us 15 scalars, as it should. 

We would now like to solve the equation (\ref{eta-eqn*}) for $\eta^a_\mu$ to
first order in perturbations. To first order the relevant Laplacian is simply 
$-\Box$, and all the terms apart from the first on the right-hand-side of (\ref{eta-eqn*})
are zero. We immediately get:
\be\label{eta-sol}
\eta^{(1)\,a}_\mu=\frac{2}{\Box} \partial^\rho M^{(1)\,ab}\Sigma^b_{0\,\rho\mu},
\ee
where $\Box=\partial^\mu\partial_\mu$. It is easy to verify that (\ref{eta-sol})
satisfies all gauge-fixing conditions (\ref{eta-gf-2}), (\ref{eta-gf-1})
to first order. 

Let us now find the corresponding relation between the two metric perturbations
$\dot{\tilde{g}}_{\mu\nu}$ and $\dot{g}_{\mu\nu}$. Extracting the metric part of the
two-form $d\eta^a$ (and taking into account that there is no trace part), we have:
\be
 \Sigma^{a\,\rho}_{0\,\,\,\,\mu} (\partial_\rho \eta^{(1)\,a}_\nu - \partial_\nu \eta^{(1)\,a}_\rho) =
 \dot{g}_{\mu\nu}-\dot{\tilde{g}}_{\mu\nu},
 \ee
 Substituting (\ref{eta-sol}), and using the fact that $M^{(1)\,ab}$ is symmetric and traceless we
 get:
 \be\label{2-metr-rel}
 \dot{g}_{\mu\nu}-\dot{\tilde{g}}_{\mu\nu}
 = \frac{2}{\Box} \Sigma^{a\,\rho}_{0\,\,\,\,\mu}\partial_\rho
 \Sigma^{b\,\sigma}_{0\,\,\,\,\nu}\partial_\sigma M^{(1)\,ab}\equiv
 \frac{2}{\Box} \partial^a_\mu \partial^b_\nu M^{(1)\,ab},
 \ee
 where we have introduced a notation:
 \be\label{a-der}
 \partial^a_\mu = \Sigma^{a\,\nu}_{0\,\,\,\,\mu} \partial_\nu.
 \ee
 This solves the problem of finding a relation between the metric perturbations in two
 different parameterizations of the two-form field perturbation. The relation
 (\ref{2-metr-rel}) has already been noted in \cite{Freidel:2008ku}. In this reference
 it was shown that this is the field redefinition that maps the quadratic part of
 the action into the usual EH form (plus a potential for the $M^{(1)\,ab}$-matrix), 
 and was noted that this field redefinition is
 related to the topological shift symmetry $B^i \to B^i + D_A \eta^i$ of the BF
 part of the action, but no explicit derivation was given. The above discussion fills
 this gap. Note that the formula (\ref{2-metr-rel}) we have obtained is precisely the
 one we have previously encountered in the section on degenerate Lagrangians,
 see (\ref{h-redef-H}).

\subsection{First-order treatment: Metric level}

We would now like to see what the transformation (\ref{2-metr-rel}) becomes at the
metric level, after the scalars $M^{ab}$ have been solved for in terms of the metric.
We have done it to a large extent in section \ref{sec:deg}, but here we repeat
the derivative in the current notations.

We have obtained a solution for $M^{ab}$ to first order in curvature in (\ref{H-1*}). At this
order the matrix $2M^{(1)\,ab}=H^{(1)\,ab}=(l^2/4g_2)(F^{ab})_{tf}$. We then have:
\be
\Sigma^{a\,\mu\nu} \Sigma^{b\,\rho\sigma} 2M^{ab} = \frac{l^2}{g_2}\left(
2P^{+\,\mu\nu\alpha\beta}R_{\alpha\beta\gamma\delta} P^{+\,\gamma\delta\rho\sigma}
-\frac{R}{3} P^{+\,\mu\nu\rho\sigma}\right),
\ee
where we have used (\ref{F-ab}) and (\ref{proj-s}). Expanding the projectors we get:
\be\label{SMS}
\Sigma^{a\,\mu\nu} \Sigma^{b\,\rho\sigma} 2M^{ab}=
\frac{l^2}{g_2}\left( C^{\mu\nu\rho\sigma} + \frac{1}{4i}\epsilon^{\mu\nu\alpha\beta}
R_{\alpha\beta}^{\quad\rho\sigma} + \frac{1}{4i}R^{\mu\nu\alpha\beta}
\epsilon_{\alpha\beta}^{\quad\rho\sigma} -\frac{R}{12i}\epsilon^{\mu\nu\rho\sigma}\right),
\ee
where
\be
C^{\mu\nu\rho\sigma}=R^{\mu\nu\rho\sigma}-g^{\mu[\rho}R^{\sigma]\nu}+
g^{\nu[\rho}R^{\sigma]\mu}+\frac{R}{3}g^{\mu[\rho}g^{\sigma]\nu}
\ee
is the Weyl tensor. Applying $\partial_\rho\partial_\sigma$ to (\ref{SMS})
we see that the imaginary terms drop out due to the Bianchi identity
$\epsilon^{\mu\nu\alpha\beta}\nabla_\nu R_{\alpha\beta\rho\sigma}=0$
and we have:
\be
\dot{g}_{\mu\nu}-\dot{\tilde{g}}_{\mu\nu}=\frac{l^2}{g_2}
\frac{1}{\Box}\partial^\rho\partial^\sigma C_{\mu\rho\nu\sigma}^{(1)}.
\ee
Here $C^{(1)}_{\mu\nu\rho\sigma}$ is the Weul tensor to first order in the metric
perturbation $\dot{g}_{\mu\nu}$. We have already computed the quantity
on the right-hand-side in section \ref{sec:deg}, see formula (\ref{dd-C}).
Thus, we get:
\be\label{del-g-1}
\dot{g}_{\mu\nu}-\dot{\tilde{g}}_{\mu\nu}=\frac{l^2}{2g_2}\left( R_{\mu\nu}^{(1)}
-\frac{1}{6} \eta_{\mu\nu} R^{(1)} \right) - \frac{l^2}{6g_2} \partial_\mu\partial_\nu 
\frac{1}{\Box} R^{(1)}.
\ee
Thus, the $\Box$ operator has cancelled in all but the last term. However, it
is clear that the last term is a diffeomorphism, even though non-local, and is of no
importance. The interesting part at this first order is given by the first term, which
we see to be local. However, this is not at all surprising, for it is well-known that
the counterterms at one-loop level are removable by a {\it local} redefinition
of the metric variable. It is also easy to see that what we have found for this
local field redefinition is precisely what we would obtain for the one-loop
action (\ref{act-1-loop}) with (\ref{curv-1-loop}) via the usual argument. Indeed,
we have:
\be
{\cal L}^{(1)}=
\frac{l^2}{8g_2}\left( R^{\mu\nu} R_{\mu\nu} - \frac{1}{3}R^2\right)=
\frac{1}{2}\left(R^{\mu\nu}- \frac{1}{2}g^{\mu\nu} R\right) \frac{l^2}{4g_2}\left( R_{\mu\nu}
-\frac{1}{6} \eta_{\mu\nu} R\right) .
\ee
The first quantity on the right-hand-side is the variation of the Einstein-Hilbert action,
while the second is precisely the local part of (\ref{del-g-1}). Thus, we get 
full agreement at the one-loop level, with the standard local field redefinition that
is used in this context to remove the one-loop counterterms being interpreted as a particular case
of the field redefinition (\ref{B-split}) that has its origin in the topological shift symmetry
of BF theory.

\subsection{Second-order treatment: BF-level}

Here we would like to extend the above analysis to the next order in perturbation theory.
To this end we have to expand the equation (\ref{eta-eqn*}) to second order in the perturbations. 
Recall that we are trying to relate two different parameterizations of a general two-form field.
In one of them one is representing the field $B^i$ as a metric two-form field $\tilde{\Sigma}^i$
for some metric $\tilde{g}_{\mu\nu}$ plus the covariant derivative (with respect to the 
$\tilde{\Sigma}$-compatible connection) of a Lie-algebra-valued one-form. In another
$B^i$ is represented as a set of metric two-forms $\Sigma^a$ for a metric $g_{\mu\nu}$ in
the conformal class defined by $B^i$ "twisted" by a ${\rm GL}(3)$ matrix $b^i_a$. 
Now each of these quantities must be expanded till second order in perturbations. 

Let us first describe an expansion for the metric two-forms. We have seen that to 
first order the perturbation of a metric two-form is given by 
$\Sigma^{a\,\,\,\,\rho}_{0\,[\mu} \dot{g}_{\nu]\rho}$, where $\dot{g}_{\mu\nu}$ is the
metric perturbation. The trace part of the metric perturbation gives rise to a self-dual
two-form perturbation proportional to $\Sigma^a_{0\,\mu\nu}$, while the tracefree
part of $\dot{g}_{\mu\nu}$ gives rise to an anti-self-dual two-form perturbation. Let us
describe what happens to second order in $\dot{g}_{\mu\nu}$. In order for
\be\label{Sigm-expans}
\Sigma^a_{\mu\nu}=\Sigma^a_{0\,\mu\nu}+\Sigma^{a\,\,\,\,\rho}_{0\,[\mu} \dot{g}_{\nu]\rho}+
\delta^{(2)}\!\Sigma^a_{\mu\nu}+\ldots\, ,
\ee
where $\delta^{(2)}\!\Sigma^a_{\mu\nu}$ is the second-order perturbation,
to remain a metric two-form it must satisfy the metricity equation:
\be\label{pert-metricity}
\epsilon^{\mu\nu\rho\sigma}
(\Sigma^a_{0\,\mu\nu}+\Sigma^{a\,\,\,\,\alpha}_{0\,[\mu} \dot{g}_{\nu]\alpha}+
\delta^{(2)}\!\Sigma^a_{\mu\nu})
(\Sigma^b_{0\,\rho\sigma}+\Sigma^{b\,\,\,\,\beta}_{0\,[\rho} \dot{g}_{\sigma]\beta}+
\delta^{(2)}\!\Sigma^b_{\rho\sigma} )\sim \delta^{ab}.
\ee
To first order, using the self-duality of the background we have on the left-hand-side:
\be
4i \Sigma^{(a\,\mu\nu}_0 \Sigma^{b)\,\,\,\,\alpha}_{0\,\mu}\dot{g}_{\nu\alpha}=4i \delta^{ab}
\eta^{\alpha\beta} \dot{g}_{\alpha\beta},
\ee
so the metricity holds. To second order the left-hand-side of (\ref{pert-metricity})
gives:
\be\label{metr-O2}
4i \Sigma^{(a\,\mu\nu}_0 \delta^{(2)}\!\Sigma^{b)}_{\mu\nu} +
\epsilon^{\mu\nu\rho\sigma} \Sigma^{a\,\,\,\,\alpha}_{0\,\mu} \dot{g}_{\nu\alpha}
\Sigma^{b\,\,\,\,\beta}_{0\,\rho} \dot{g}_{\sigma\beta}.
\ee
We can now rewrite the second term as $\epsilon^{\mu\nu\rho\sigma} 
\Sigma^{a\,\,\,\,[\alpha}_{0\,[\mu} \Sigma^{b\,\,\,\,\beta]}_{0\,\rho]} 
\dot{g}_{\nu\alpha}\dot{g}_{\sigma\beta}$ and use (\ref{2s-antisymm}) to rewrite
(\ref{metr-O2}) as 
\be
4i \Sigma^{(a\,\mu\nu}_0 \delta^{(2)}\!\Sigma^{b)}_{\mu\nu} 
-i \left(\Sigma^{(a\,\nu\sigma}_0 \Sigma^{b)\,\alpha\beta}_0 +
2\delta^{ab} \eta^{\alpha[\nu}\eta^{\sigma]\beta}\right) \dot{g}_{\nu\alpha}\dot{g}_{\sigma\beta}.
\ee
The last term is proportional to $\delta^{ab}$, so overall we get the following equation:
\be
4 \Sigma^{(a\,\mu\nu}_0 \delta^{(2)}\!\Sigma^{b)}_{\mu\nu} \Big|_{tf}
= \Sigma^{(a\,\nu\sigma}_0 \Sigma^{b)\,\alpha\beta}_0 \dot{g}_{\nu\alpha}\dot{g}_{\sigma\beta}
\Big|_{tf},
\ee
where, as before, $tf$ stands for the tracefree parts. This is an equation for the tracefree
symmetric part of the matrix $X^{(2)\,ab}$ of coefficients in the decomposition 
$\delta^{(2)}\!\Sigma^a_{\mu\nu}=
X^{(2)\,ab}\Sigma^b_{0\,\mu\nu} + Y^{(2)\,ab}\bar{\Sigma}^b_{0\,\mu\nu}$.
The equation we obtained leaves unconstrained the anti-symmetric and trace parts of
$X^{(2)\,ab}$, as well as the matrix $Y^{(2)\,ab}$ of anti-self-dual coefficients. This
is as expected, for the anti-symmetric part of the self-dual matrix $X^{(2)\,ab}$ is
pure gauge, while the trace part and the anti-self-dual matrix $Y^{(2)\,ab}$ describe
the metric part of the perturbation and cannot be constrained by the requirement of
metricity. Allowing for a convenient trace part, which at second order of the perturbation
is at our disposal, and fixing the anti-symmetric self-dual and anti-self-dual parts to be absent,
we can finally write a convenient expression for the second-order perturbation of the
metric two-forms:
\be\label{Sigm-O2}
\delta^{(2)}\!\Sigma^a_{\mu\nu} = \frac{1}{4} \Sigma^{a\,\rho\sigma}_0 \dot{g}_{\rho\alpha}
\dot{g}_{\sigma\beta} P^{+\,\alpha\beta}_{\quad\mu\nu}.
\ee

We will also need an expression for the metric two-forms with both indices raised. This can be
computed using $\Sigma^{a\,\mu\nu}=g^{\mu\alpha}g^{\nu\beta}\Sigma_{\alpha\beta}$
and expanding $g^{\mu\nu}=\eta^{\mu\nu}-\dot{g}^{\mu\nu}+ \dot{g}^{\mu\rho}\dot{g}^{\rho\nu}$.
After some algebra we get an expression to second order:
\be\label{sigm-inv-O2}
\Sigma^{a\,\mu\nu} = \Sigma^{a\,\mu\nu}_0 - \Sigma^{a\,[\mu|\rho|}_0 \dot{g}^{\nu]}_{\,\,\rho}
+\frac{1}{4} \Sigma^{a\,\alpha\beta}_0 P^{+\,\mu\nu\rho\sigma} 
\dot{g}_{\alpha\rho}\dot{g}_{\beta\sigma}+ \Sigma^{a\,[\mu}_{0\,\,\,\,\alpha}
\dot{g}^{\alpha\beta} \dot{g}^{\nu]}_{\,\,\beta}+\ldots\, .
\ee

Let us now discuss the second-order perturbation of the ${\rm GL}(3)$ matrix $b^i_a$.
We have:
\be\label{b-expans}
b^i_a = \delta^{ib}(\delta_{ab} + M^{(1)}_{ab} + M^{(2)}_{ab}+\ldots),
\ee
where at first order we have fixed the matrix $M^{(1)}_{ab}$ to be symmetric and tracefree. 
A convenient choice of the second-order perturbation is:
\be\label{M-O2}
M^{(2)}_{ab}=
\tilde{M}^{(2)}_{ab}-\frac{1}{2}M^{(1)}_{ac} M^{(1)}_{cb} + 2\kappa \delta_{ab} {\rm Tr}(M^{(1)})^2,
\ee
where $\tilde{M}^{(2)\,ab}$ is symmetric and tracefree, $\kappa$ is a 
parameter that determines which precisely metric in the conformal
class of $B^i$ one is using in the representation $B^i=b^i_a\Sigma^a$. Note that at
this stage we have not yet imposed the condition (\ref{conf-cond}). We will take
care of it later by an appropriate conformal transformation. The choice
of the second term in (\ref{M-O2}) is motivated by the fact that the internal metric $m_{ab}$
in this case has the expansion:
\be
m_{ab}=b^i_a b^i_b = \delta_{ab} + 2M^{(1)}_{ab} + 
2\tilde{M}^{(2)}_{ab}+ 4\kappa \delta_{ab} {\rm Tr}(M^{(1)})^2,
\ee
and so its tracefree part on which the potential function depends is just 
$H_{ab}=H_{ab}^{(1)}+H_{ab}^{(2)}+\ldots$, with 
$H^{(1)}_{ab} = 2M^{(1)}_{ab}$ and $H^{(2)}_{ab} = 2\tilde{M}^{(2)}_{ab}$.
The expansion is then the same as we have used above, see (\ref{m-expans}).

Collecting (\ref{Sigm-expans}) with (\ref{Sigm-O2}) and (\ref{b-expans}) with 
(\ref{M-O2}) we can write an expression for $B^i=b^i_a\Sigma^a$ to second order:
\be\label{B-O2}
\delta^{ia}B^i = \Sigma^a_{0\,\mu\nu} + M^{(1)\,ab}\Sigma^b_{0\,\mu\nu}
+ \Sigma^{a\,\,\,\,\rho}_{0\,[\mu}\dot{g}_{\nu]\rho} 
\\ \nonumber
+M^{(2)\,ab}\Sigma^b_{0\,\mu\nu}
+ \frac{1}{4} \Sigma^{a\,\rho\sigma}_0 \dot{g}_{\rho\alpha}
\dot{g}_{\sigma\beta} P^{+\,\alpha\beta}_{\quad\mu\nu}
+ M^{(1)\,ab} \Sigma^{b\,\,\,\,\rho}_{0\,[\mu}\dot{g}_{\nu]\rho}.
\ee
The second line here contains terms of the second order in the perturbations. We note that
the first two terms in the second line are self-dual (with respect to the background metric), while
only the last term is anti-self-dual. 

We can now expand the equation (\ref{eta-eqn*}) to second order. Let us first
work out the right-hand-side. To this end, we need to expand 
$2X^{ab}=\tilde{\Sigma}^{a\,\mu\nu}B^b_{\mu\nu}$ to second order and extract the
symmetric tracefree part of this matrix. The first of the quantities is given in
(\ref{sigm-inv-O2}), where one has to put $\dot{\tilde{g}}_{\mu\nu}$ 
everywhere. The second is given in (\ref{B-O2}). To second order the result is:
\be\label{X-2-1}
X^{ab}= 2M^{(1)\,ab} + 2M^{(2)\,ab}\Big|_{tf} + \frac{1}{2}M^{(1)\,ab}
(\dot{g}-\dot{\tilde{g}}) + \frac{1}{8}\Sigma^{a\,\mu\nu}_0\Sigma^{b\,\rho\sigma}_0
(\dot{g}_{\mu\rho}-\dot{\tilde{g}}_{\mu\rho})(\dot{g}_{\nu\sigma}-\dot{\tilde{g}}_{\nu\sigma})
\Big|_{tf}.
\ee
We can now express this in terms of $M^{(1)}, M^{(2)}$ only since we have, to first
order 
\be
\dot{g}_{\mu\nu}-\dot{\tilde{g}}_{\mu\nu}= \frac{2}{\Box} \partial_\mu^a\partial_\nu^b M^{(1)\,ab}.
\ee
The trace part of this expression vanishes, so there is no third term in (\ref{X-2-1}).
Thus, we get:
\be
X^{ab}= 2M^{(1)\,ab} + 2\tilde{M}^{(2)\,ab}+ \frac{1}{2}
\left( \Sigma^{a\,\mu\nu}_0\Sigma^{b\,\rho\sigma}_0 \frac{1}{\Box}\partial_\mu^c\partial_\rho^d
M^{(1)\,cd} \frac{1}{\Box}\partial_\nu^e\partial_\sigma^f M^{(1)\,ef} - 2M^{(1)\,ac}
M^{(1)\,cb}\right)_{tf}
\ee

We should now also expand (\ref{eta-eqn*}) to second order in perturbations.
Equating the second-order terms we get:
\be\label{eta-2}
\eta^{(2)\,a}_\mu = \frac{2}{\Box} \partial^\rho \left(M^{(2)\,ab} - 
\frac{1}{2} M^{(1)\, ac} M^{(1)\, cb}\right)_{tf} \Sigma^{b}_{0\, \rho\mu} 
+ \ldots\, ,
\ee
where the dots denote contributions involving other first-order terms. However, we do
not need to compute these. Indeed, the contribution from $X^{ab}$ that
contains $\dot{f}_{\mu\nu}-\dot{\tilde{g}}_{\mu\nu}$, as we already know, see 
(\ref{del-g-1}), depends only on $R_{\mu\nu}$ and $R$ that vanish on shell.
Thus, we are not interested in this contribution. The second-order contributions that come
by expanding the Laplacian in (\ref{eta-eqn*}) are all proportional to $l^2$,
and are not interesting since our aim is to find terms that cancel the $l^4$ terms
(\ref{act-2-loop}).

We can now repeat the same steps as in the one-loop case to find that the
metric redefinition is given by:
\be\label{del-g-2}
\dot{g}^{(2)}_{\mu\nu}-\dot{\tilde{g}}^{(2)}_{\mu\nu}
 =  \frac{2}{\Box} \partial^a_\mu \partial^b_\nu \left(M^{(2)\,ab} - 
\frac{1}{2} M^{(1)\, ac} M^{(1)\, cb}\right)_{tf} + \ldots \, ,
 \ee
 where the dots again stand for either on-shell vanishing or $l^2$ order terms. Note
 that $\dot{g}^{(2)}_{\mu\nu}$ is in fact zero at the second order considered, 
 but we kept it in the formula to make it look similar to (\ref{2-metr-rel}).
 
 \subsection{Second-order treatment: Metric level}
 
 We can now substitute into (\ref{del-g-2}) the solution (\ref{H-2*}) and compute the
 field redefinition at the metric level. Since $F^2$ and $\cD^2 F$ terms have 
 independent coefficients in front of them, they can be treated separately. The
 corresponding field redefinitions separately cancel their own terms in the effective 
 metric Lagrangian. We shall only consider the $F^2$ term. Expanding
 \be
 \frac{1}{\Box}\Sigma^a_{\alpha\mu}\Sigma^b_{\beta\nu} \partial^\alpha\partial^\beta
 (F^{ac}F^{cb})_{tf},
 \ee
taking the real part of the result, and dropping all on-shell vanishing, terms of the type
$\partial_{(\mu}\xi_{\nu)}$ that describe a diffeomorphism, as well as terms 
proportional to $\eta_{\mu\nu}$ that describe a conformal transformation that we shall
not attempt to reproduce, we get a multiple of 
 \be
 \frac{1}{\Box}\partial^\alpha\partial^\beta R_{\mu\alpha}^{\quad\gamma\delta}
 R_{\nu\beta\gamma\delta},
 \ee
 which is exactly the quantity that appears in the field redefinition (\ref{h-non-loc}).
 This completes the circle and shows how the field redefinition that removes
 the $(Riemann)^3$ term has its origin in the (non-local) topological shift symmetry
 that maps our theory to GR.
 
\section{Discussion}

We have considered an infinite-parametric class of effective metric Lagrangians that
arise from an underlying theory with two propagating DOF. In its simplest formulation
(\ref{action-A}) the underlying theory is given by just the most general gauge-invariant 
Lagrangian for an $\SOC$ connection that can be written without any background structure 
such as a metric. We have seen that
the low energy limit of any of the theories (\ref{action-A}), i.e. for any generic
choice of the defining potential, is general relativity. Thus, the theories (\ref{action-A})
provide particular UV completions of GR with rather appealing "minimal" property of no new 
propagating DOF being introduced. Moreover, the class (\ref{action-A}) consisting of all
generally-covariant theories of connection, it may be closed under the 
renormalization. If this is the case, then the class (\ref{action-A}), after it is
quantized (e.g. perturbatively), could be seriously considered as a candidate 
quantum theory of (pure) gravity.

We have also described the two-form field formulation that makes the 
spacetime metric of the theory (almost) explicit. In this formulation 
the theory is (\ref{action-AB}) the topological BF theory with a potential for the two-form
field. However, in addition to the metric the BF-formulation introduces certain non-propagating 
auxiliary fields that have to be integrated out to arrive at a purely metric description.

A certain complicated non-local field redefinition that has its origin in the topological symmetry of
BF theory can map any one of the effective metric Lagrangians to any other. In particular,
any of our effective Lagrangians can be mapped to the Einstein-Hilbert one,
which gives another explanation for why the theories we have studied have just two 
propagating DOF. The Lagrangian that is required to renormalize divergences
of perturbative quantum GR up to two loops lies within our class, which suggests that the
theory underlying the effective metric Lagrangians of gravity may be the one studied in
this paper. 

Importantly, we have seen that, if one enlarges the class of allowable field redefinitions to
those that are non-local but map local theory to a local one (non-trivial assumption),
then all our effective metric theories are equivalent. In particular, as we have shown in this paper,
the two-loop divergence of quantum gravity \cite{Goroff:1985th} is removable
by a field redefinition of this new type. The most intriguing question that arises is what all 
this means for the problem of quantum gravity. We will not attempt to provide an answer in 
this purely classical paper. However, some suggestive remarks can be made. 

In general, non-local field redefinitions (unlike the local ones)
do change the S-matrix of the theory, see \cite{Marcus:1984ei}, section 2,
as well as \cite{'tHooft:1973pz}, section 10 for good discussions of this point. 
The reason for this is that the determinant of the arising Jacobian contains factors of 
$1/\Box$ operator, which makes the corresponding ghost action non-trivial. 
However, our non-local field redefinitions are certainly of
a very special type and the conclusion about S-matrix being changed
needs to be re-examined. 

One intriguing possibility
can then be as follows. It is clear that what makes our non-local field redefinitions
possible is the topological symmetry of the BF part of the action of our theory.
An interesting, and potential deep way to understand this is to view our class
of theories as the topological BF theory in which the topological symmetry has
been gauge-fixed by the potential term. Then different gauge-fixings lead to
different effective metric theories. It is then not surprising that non-local field
redefinitions of the type described are possible. Indeed, it is known that one
can change the gauge-fixing term by a non-local gauge transformation, see the
example in section 11.3 of \cite{'tHooft:1973pz}. Since our different metric
theories correspond to different gauge-fixings of the same underlying 
theory (topological BF), it is possible that they give rise to equivalent S-matrices.
As the cited example in \cite{'tHooft:1973pz} shows, the way this must happen is that the 
ghost action that arises from the Jacobian of the field redefinition is precisely 
cancelled by the Faddeev-Popov ghost action correcting the
integration measure for the fact that the second-class constraints are present.
Here we will not attempt to demonstrate that this mechanism is indeed at play
in our class of theories, leaving it to future research. But the arguments given
do suggest that the theories described may be quantum-equivalent.
If so, and if the class of theories described is closed under renormalization,
then quantum gravity would be a finite theory, for all its divergences would
be removable by field redefinitions of the new type described in this paper.
This is certainly an exciting prospect, but certainly much more work is needed
before these ideas can be made concrete.

It is important to emphasize that the fact that all our effective metric theories are 
related in the sense explained above does not mean that they are in any natural way 
equivalent as classical theories, for a non-local field redefinition is involved. One
way to see it is to note that while solutions of GR are Einstein metrics, solutions
of any of our modified theories are not. 

An important question that we have not touched upon in this paper 
is how much of what we have described survives
when we couple gravity to matter. It is clear that if the above field redefinition
ideas are to work the coupling cannot be arbitrary -- one needs to continue to
have the same underlying topological symmetry of BF theory at play. One way to
introduce matter in a way that satisfies this requirement is to simply enlarge
the gauge group in question, and consider the theories of the type 
(\ref{action-A}) for a different gauge group. This is the unification proposal
first put forward in \cite{Peldan:1992iw}, further studied in \cite{Chakraborty:1994vx} and
\cite{Chakraborty:1994yr}, and recently revisited in \cite{Smolin:2007rx} and 
\cite{TorresGomez:2009gs}. It seems likely that Yang-Mills fields coupled to gravity 
(and Higgs fields) can be described via
this proposal. If so, then much of what we have said above about quantum
gravity applies to gravity plus Yang-Mills-Higgs system. It is not impossible that
fermions can also be introduced in a similar fashion by an appropriate
Grassmann-valued extension of the connection, but this is much more
speculative. Overall, we feel that there is reasonable hope that at least some
types of matter can be coupled to gravity in a way that keeps the non-local
field redefinitions acting on effective Lagrangians intact. Then, whatever the
story is for the pure gravity case, it will extend with very little changes to
gravity coupled to matter.

Let us conclude this paper with a list of open problems on the set of ideas described.
First and foremost, it is important to quantize our class of theories to see whether
our hopes of closeness under renormalization and possibly even finiteness have
any chance of being realized. Work on the perturbative quantization is in progress, 
with the theory linearized about the Minkowski background having been worked 
out in \cite{TorresGomez:2009gs}. It would also be important to try to find an explicitly real
formulation of this class of theories, as the prospect of having to work with 
holomorphic Lagrangians is bound to make some uneasy. It is also very important
to continue the work \cite{Smolin:2007rx}, \cite{TorresGomez:2009gs}
on unification by enlarging the gauge group to see what
types of matter can be realistically coupled to gravity in this form. 

We close with expressing a feeling/hope that the class of theories envisaged already 
two decades ago in \cite{Capovilla:1989ac}, \cite{Capovilla:1991kx}, \cite{Capovilla:1992ep},
\cite{Bengtsson:1990qg}, \cite{Peldan:1992iw} contains still many more surprises
waiting to be uncovered.

\section*{Acknowledgments} The author is supported by an EPSRC Advanced Fellowship. 
I am indebted to Oleksii Boyarskyi for his deep and stimulating questions about 
the theory described here. I am also grateful to my collaborator Yuri Shtanov for 
numerous discussions on the subject of this paper. I would like to thank Jorma Louko for a 
discussion about holomorphic Lagrangians and Ingemar Bengtsson for remarks on the draft.

\section*{Appendix}

\subsection*{Conventions}

Our conventions are as follows. We work in the signature $(-,+,+,+)$ that
is standard in the GR literature. We define the volume form so that the
object $\tilde{\epsilon}^{\mu\nu\rho\sigma}$ of density plus one has
in any coordinate system components $\tilde{\epsilon}^{0123}=-1$. Then
we have: 
\be
dx^\mu\wedge dx^\nu\wedge dx^\rho\wedge dx^\sigma = (-1)\tilde{\epsilon}^{\mu\nu\rho\sigma} d^4x.
\ee
A similar formula in terms of the tetrads reads:
\be
\theta^I\wedge \theta^J\wedge \theta^K\wedge \theta^L=(-1)\, \epsilon^{IJKL} \sqrt{-g} \, d^4x,
\ee
where $\epsilon^{IJKL}$ is the ``internal'' completely anti-symmetric tensor for which
our convention is that $\epsilon^{0123}=-1$, and $\sqrt{-g}$ is the square root of (minus) the determinant
of the metric 
\be
g_{\mu\nu}=\theta_\mu^I \theta_\nu^J \eta_{IJ},
\ee 
with $\eta_{IJ}$ being the
Minkowski metric. Here and everywhere the capital Latin letters are ``internal'' 
indices $I=0,1,2,3$. Our conventions on forms are:
\be
X^{(n)} = \frac{1}{n!} X_{\mu_1 \ldots \mu_n} dx^{\mu_1} \wedge dx^{\mu_n}.
\ee

\subsection*{Algebra of $\Sigma$-matrices}

Introducing:
\be
\Sigma^a=\im dt\wedge dx^a-\frac{1}{2}\epsilon^{abc}dx^b\wedge dx^c,
\ee
the following relation can be verified:
\be\label{algebra}
\Sigma^a_{\mu\rho}\Sigma^{b\,\rho}_{\quad\nu} = - \delta^{ab}  \eta_{\mu\nu} 
+ \epsilon^{abc}\Sigma^c_{\mu\nu}.
\ee
Using these it is not hard to get:
\be\label{prod-2s}
\Sigma^{a\,\mu\nu}\Sigma^b_{\mu\nu} = 4\delta^{ab}, \\
\label{proj-s}
\Sigma^a_{\mu\nu} \Sigma^a_{\rho\sigma} = 4 P^+_{\mu\nu\rho\sigma}: =
\eta_{\mu\rho}\eta_{\nu\sigma} - \eta_{\mu\sigma}\eta_{\nu\rho} 
+ (1/i) \epsilon_{\mu\nu\rho\sigma}, \\
\label{det-s}
\epsilon^{abc} \Sigma^{a\,\,\,\nu}_\mu \Sigma^{b\,\,\,\rho}_\nu \Sigma^{c\,\,\,\mu}_\rho = -4!, \\
\label{ident-s}
\epsilon^{abc} \Sigma^a_{\mu\nu}\Sigma^b_{\rho\sigma} \Sigma^{d\,\mu\rho} = -2 \delta^{cd}
\eta_{\nu\sigma}.
\ee
In (\ref{proj-s}) the tensor $P^+$ is the projector on self-dual two-forms. Sometimes we shall
also use the following more involved relation:
\be\label{2s-antisymm}
\Sigma^{a\,\,\,[\rho}_{[\mu} \Sigma^{b\,\,\,\sigma]}_{\nu]} = \frac{1}{2} \Sigma^{(a}_{\mu\nu}
\Sigma^{b)\,\rho\sigma} + \frac{1}{2i}\delta^{ab} \epsilon^{\quad\rho\sigma}_{\mu\nu}.
\ee

One can also introduce the anti-self-dual matrices
\be
\bar{\Sigma}=\im dt\wedge dx^a + \frac{1}{2}\epsilon^{abc}dx^b\wedge dx^c.
\ee
Their algebra is similar to that of self-dual quantities (\ref{algebra}):
\be\label{algebra-asd}
\bar{\Sigma}^a_{\mu\rho}\bar{\Sigma}^{b\,\rho}_{\quad\nu} = - \delta^{ab}  \eta_{\mu\nu} 
- \epsilon^{abc}\bar{\Sigma}^c_{\mu\nu}.
\ee
Thus, the only difference is the sign in the last term. Correspondingly, there will be a different
sign on the right-hand-side of analogs of relations (\ref{det-s}) and (\ref{ident-s}).

It is more non-trivial to compute the algebra between the self-dual and anti-self-dual matrices.
It can be computed case by case, but we were not able to find a simple closed formula. 
However, the result of a product of a self- and anti-self-dual matrix is always a symmetric tensor.
Thus, the following identity holds:
\be\label{id-sd-asd}
\Sigma^{a\,\,\,\rho}_{[\mu} \bar{\Sigma}^b_{|\rho|\nu]} = 0.
\ee

\subsection*{Curvature}

According to the definition of the rotation coefficients:
\be\label{def-Gamma}
\nabla_\mu \theta^I_\nu = - \Gamma_\mu^{IJ} \theta_{\nu\, J}.
\ee
Here $\nabla_\mu$ is the metric-compatible $\nabla_\mu g_{\rho\sigma}=0$ 
derivative operator that acts only on the spacetime indices.
We can use this equation to compute the ${\rm SO}(3)$-connection $\gamma_\mu^a$
in terms of the rotation coefficients $\Gamma_\mu^{IJ}$. Thus, it is not hard to check
that the expression:
\be\label{app-gamma}
\gamma^a= i\Gamma^{0a}-\frac{1}{2}\epsilon^{abc}\Gamma^{bc}
\ee
solves the compatibility equation:
\be
\nabla \Sigma^a + \epsilon^{abc}\gamma^b\wedge \Sigma^c=0,
\ee
where 
\be
\Sigma^a=i\theta^0\wedge \theta^a - \frac{1}{2}\epsilon^{abc}\theta^b\wedge \theta^c.
\ee

It is similarly not hard to check that the expression:
\be
F^{0a}-\frac{1}{2}\epsilon^{abc}F^{bc},
\ee
where
\be
F^{IJ} = d\Gamma^{IJ}+\Gamma^{IK}\wedge \Gamma_K^{\,\,\,\,J},
\ee
coincides with the curvature
\be
F^a = d\gamma^a +\frac{1}{2}\epsilon^{abc}\gamma^b\wedge\gamma^c
\ee
of the ${\rm SO}(3)$ connection (\ref{app-gamma}). 

The Riemann curvature tensor can be expressed in terms of $F^{IJ}$. We have:
\be\label{Riemm-F}
R_{\mu\nu\rho\sigma} = F_{\mu\nu}^{IJ} \theta_{\rho\,I}\theta_{\sigma\,J},
\ee
where our conventions for forms are $F^{IJ}=(1/2)F_{\mu\nu}^{IJ}dx^\mu dx^\nu$.
Indeed, the usual definition of the Riemann curvature is: 
$2\nabla_{[\mu}\nabla_{\nu]} X^\rho = -R_{\mu\nu\sigma}^{\quad\rho} X^\sigma$.
We can now introduce a derivative operator $D_\Gamma$
that acts on spacetime as well as on the internal indices, with the action on spacetime
indices being that of $\nabla$, and $D_\Gamma X^I = dX^I + \Gamma^{IJ} X_J$. Then,
from the definition (\ref{def-Gamma}) of $\Gamma^{IJ}$ it follows that 
$D_\Gamma \theta_\mu^I=0$. Let us now replace the metric-compatible derivative
operator $\nabla$ in the commutator of the Riemann curvature by the operator
$D_\Gamma$. This is legitimate, as it acts on an object without internal indices.
We can compute the same commutator in a different way by decomposing 
$X^\mu  =\theta^\mu_I X^I$. Since $\theta^\mu_I$ is preserved by $D_\Gamma$, it
can be taken outside of the derivatives and we have:
\be
2\nabla_{[\mu}\nabla_{\nu]} X^\rho = \theta^\rho_I 2D_{\Gamma\,[\mu}D_{\Gamma\,\nu]} X^I
= \theta^\rho_I F_{\mu\nu}^{IJ} X_J,
\ee
where we have used the definition of the curvature $F^{IJ}$. Now writing 
$X_J = \theta_{\sigma\,J} X^\sigma$ we get (\ref{Riemm-F}).

It is now easy to see that (\ref{Riemm-F}) can be rewritten as:
\be\nonumber
R_{\mu\nu\rho\sigma} = (iF^{0a}_{\mu\nu}-\frac{1}{2}\epsilon^{abc}F^{bc}_{\mu\nu})
(i\theta^0_{[\rho}\theta^a_{\sigma]}- \frac{1}{2}\epsilon^{aef}\theta^e_{[\rho}
\theta^f_{\sigma]}) +(iF^{0a}_{\mu\nu}+\frac{1}{2}\epsilon^{abc}F^{bc}_{\mu\nu})
(i\theta^0_{[\rho}\theta^a_{\sigma]}+ \frac{1}{2}\epsilon^{aef}\theta^e_{[\rho}
\theta^f_{\sigma]})
\\ \label{Riemm-decomp}
= \frac{1}{2} F^a_{\mu\nu} \Sigma^a_{\rho\sigma} 
+ \frac{1}{2} \bar{F}^a_{\mu\nu} \bar{\Sigma}^a_{\rho\sigma},\quad 
\ee
where $\bar{F}^a$ is the curvature of the anti-self-dual connection 
$\bar{\gamma}^a=i\Gamma^{0a}+(1/2)\epsilon^{abc}\Gamma^{bc}$.

Some consequences of (\ref{Riemm-decomp}) are easy to derive. First, we have
the first Bianchi identity $R_{\mu\nu\rho\sigma}\epsilon^{\mu\nu\rho\sigma}=0$.
Since $\Sigma^a_{\mu\nu}$ is self- and $\bar{\Sigma}^a_{\mu\nu}$ anti-self-dual
this gives:
\be
\Sigma^{a\,\mu\nu} F^a_{\mu\nu} - \bar{\Sigma}^{a\,\mu\nu} \bar{F}^a_{\mu\nu}=0,
\ee
or, in other words, the quantity $\Sigma^{a\,\mu\nu}F^a_{\mu\nu}$ is real. The relation
(\ref{Riemm-decomp}) then shows that it is equal to the Ricci scalar:
\be\label{R-scalar}
R=\Sigma^{a\,\mu\nu}F^a_{\mu\nu}.
\ee

Let us also write down the inverse relation that allows to find $F^a$ from the Riemann
curvature. We have:
\be\label{F-Riemm}
F^a_{\mu\nu} = \frac{1}{2} R_{\mu\nu\rho\sigma}\Sigma^{a\,\rho\sigma},
\ee
where we have used (\ref{prod-2s}).

\end{document}